\documentclass[12pt,hidelinks]{article}
\usepackage[body={17.5cm, 22cm},right=2cm]{geometry}
\usepackage{color}
\usepackage{amssymb,graphicx}
\usepackage{amsmath}
\usepackage{epsfig}
\usepackage[bookmarks=true,pdfborder={0 0 0}]{hyperref} 
\usepackage{cite}
\pdfoutput=1

\usepackage{pifont}

\usepackage[utf8]{inputenc}
\usepackage{verbatim}   
\usepackage{amsfonts}
\usepackage{amssymb}
\usepackage{graphicx}
\usepackage{slashed}
\usepackage[normalem]{ulem}
\usepackage[multiple]{footmisc}

\newcommand{\Mpc}{{\, {\rm Mpc}}}

\newcommand{\eV}{{\, {\rm eV}}}
\newcommand{\keV}{{\, {\rm keV}}}
\newcommand{\MeV}{{\, {\rm MeV}}}
\newcommand{\GeV}{{\, {\rm GeV}}}
\newcommand{\TeV}{{\, {\rm TeV}}}

\def\beq{\begin{equation}}
\def\eeq{\end{equation}}
\def\bea{\begin{eqnarray}}
\def\eea{\end{eqnarray}}
\def\bitem{\begin{itemize}}
	\def\eitem{\end{itemize}}
\newcommand{\bec}{\begin{center}}
	\newcommand{\eec}{\end{center}}
\newcommand{\ba}{\begin{array}}
	\newcommand{\ea}{\end{array}}

\def\bar#1{\overline{#1}}

\def\abs#1{\left| #1\right|}

\def\inv{^{\raise.15ex\hbox{${\scriptscriptstyle -}$}\kern-.05em 1}}
\def\lbar{{\lower.35ex\hbox{$\mathchar'26$}\mkern-10mu\lambda}} 

\def\to{\rightarrow}

\let\<=\langle
\let\>=\rangle

\let\+=\uparrow

\begin{document}

\pagenumbering{Alph}

\begin{titlepage}
	~\vspace{1cm}
	\begin{center}

		{\LARGE \bf Observing Invisible Axions with
	\\
	\vspace{3.1mm}
		Gravitational Waves}

		\vspace{0.7cm}

		{\large
			Marco~Gorghetto$^a$,
			Edward~Hardy$^b$, and Horia Nicolaescu$^b$}
		\\
		\vspace{.6cm}
		{\normalsize { \sl $^{a}$ 
				Department of Particle Physics and Astrophysics, Weizmann Institute of Science,\\
				Herzl St 234, Rehovot 761001, Israel }}
		
		\vspace{.3cm}
		{\normalsize { \sl $^{b}$ Department of Mathematical Sciences, University of Liverpool, \\ Liverpool, L69 7ZL, United Kingdom}}

	\end{center}
	\vspace{.8cm}
	\begin{abstract}

If the Peccei-Quinn symmetry associated to an axion has ever been restored after inflation, axion strings inevitably produce a contribution to the stochastic gravitational wave background. Combining effective field theory analysis with numerical simulations, we show that the resulting gravitational wave spectrum  has logarithmic deviations from a scale invariant form with an amplitude that is significantly enhanced at low frequencies. As a result, a single ultralight axion-like particle with a decay constant larger than $10^{14}~{\rm GeV}$  and any mass between $10^{-18}~{\rm eV}$ and $10^{-28}~{\rm eV}$ leads to an observable gravitational wave spectrum and is compatible with constraints on the post-inflationary scenario from dark matter overproduction, isocurvature and dark radiation. Since the spectrum extends over a wide range of frequencies, the resulting signal could be detected by multiple experiments. We describe straightforward ways in which the Peccei-Quinn symmetry can be restored  after inflation for such decay constants. We also comment on the recent possible NANOgrav signal in light of our results.

	\end{abstract}

\end{titlepage}

\pagenumbering{arabic}

	\thispagestyle{empty}
\newgeometry{height=22cm, width=17.5cm, top=2.3cm, bottom=2cm}

{\fontsize{11}{10}
\tableofcontents
}
\restoregeometry

\newpage


\section{Introduction}

The discovery of gravitational waves (GWs) from binary mergers by the Ligo/Virgo collaboration \cite{Abbott:2016blz} has opened a new window through which our Universe can be observed. This has already led to discoveries in astrophysics \cite{Abbott:2020tfl,Abbott:2020mjq} and many others will inevitably follow. Moreover, GWs also have the potential to provide invaluable   insights into fundamental particle physics. This is a particularly promising avenue  since there are plans for numerous new detectors, which will  have access to a much wider range of frequencies and much greater sensitivity than the current generation. It is therefore worth understanding whether the simplest and most motivated  models of new physics predict  GW backgrounds left over from the early Universe, and, if they do, whether these are in reach of future detectors.

In this regard, many extensions of the Standard Model (SM) predict the existence of additional U(1) (global or local) symmetry factors that are spontaneously broken. For instance, local U(1)s appear in grand unified theories, and theories of leptogenesis and neutrino masses. On the other hand, the QCD axion, introduced to solve the strong CP problem, is the pseudo Nambu-Goldstone boson (PNGB) of a spontaneously broken global U(1) Peccei-Quinn (PQ) symmetry. Similarly, axion-like-particles are PNGBs of new global U(1) symmetries, are common in well motivated phenomenological models and appear ubiquitous in typical string theory constructions \cite{Svrcek:2006yi,Arvanitaki:2009fg,Acharya:2010zx,Cicoli:2012sz,Demirtas:2018akl}.\footnote{Particularly relevant for our present work are axions from the closed string sector \cite{Ibanez:1999it}, for which global string defects can form due to symmetry restoration in the early Universe.} 
Additionally, the QCD axion and axion-like-particles (both of which we refer to just as axions) are compelling candidates to comprise some or all of the dark matter (DM), since they are automatically produced in the early Universe and are usually cosmologically stable.  Nevertheless, over large parts of their parameter space the detection of axions is challenging due to their extremely weak (or possibly vanishing) couplings to the SM, see e.g. \cite{Irastorza:2018dyq}.

If it has ever been restored in the early Universe, a spontaneously broken U(1) symmetry leads to the formation of cosmic strings. 
 Thanks to their topological nature, after they form, a network of 
such objects typically persists as the Universe expands. They therefore provide a sustained, possibly substantial, contribution to the (transverse-traceless component of the) energy momentum tensor of the Universe, sourcing GWs for an extended period of time. The resulting GW spectrum could span a wide range of frequencies, so is potentially relevant to numerous proposed detectors, including pulsar timing arrays such as SKA~\cite{janssen2014gravitational}; space based laser interferometers such as LISA~\cite{amaroseoane2017laser} as well as terrestrial laser interferometers including LIGO~\cite{Aasi:2014mqd} and ET~\cite{Hild:2010id,Punturo:2010zz}; and searches utilising novel approaches including atom interferometery such as AEDGE~\cite{Bertoldi:2019tck}.

In this paper we study the GWs produced by cosmic strings in a generic axion model in which the U(1) global symmetry has been restored after inflation, known as the post-inflationary scenario. Our analysis includes the QCD axion and axion-like particles as it only relies on the universal coupling to gravity. The signals are also independent of possible couplings of these particles to the SM and they do not depend on the possible local DM axion abundance, unlike many other detection strategies.\footnote{We will see that, regardless of the axion's couplings, friction from the thermal bath does not affect the string dynamics at the times when GWs with observable frequencies are emitted for axion decay constants that lead to an amplitude that could be detected.} Consequently our results apply to a remarkably wide class of models, most of which are otherwise not presently under experimental scrutiny.

After a network of axion strings first forms (e.g. when the temperature of the Universe drops sufficiently that the PQ symmetry spontaneously breaks) it evolves towards an attractor solution that is independent of the initial conditions~\cite{Gorghetto:2018myk}. Such a regime -- known as scaling -- is the result of the competing effects of string recombination and Hubble expansion.  As we discuss in Section~\ref{ss:Review}, during scaling the network's statistical properties drastically simplify and their time-dependence is fixed only by one scale, the Hubble parameter, up to crucial corrections that are logarithmic in the UV physics scale. For instance, since the strings arise from a global symmetry their tension manifestly has such a dependence.\footnote{Such corrections to naive scaling laws are common in a wide range of physical systems whenever a UV cutoff is present~\cite{Mussardo:1281256}.}  In particular, in this regime the number of strings per Hubble volume is driven to a critical value (also subject to logarithmic corrections~\cite{Gorghetto:2018myk,Kawasaki:2018bzv,Vaquero_2019}) and to maintain this the network releases energy, dominantly in the form of axions.

The motion and recombination of strings during scaling also sources GWs, which propagate freely until today and therefore contribute to the stochastic background. 
The crucial ingredients needed to determine the resulting spectrum are: 1) the energy emitted instantaneously from the string network in GWs as a function of time, and 2) the momentum distribution with which this energy is emitted. In fact, we will be able to derive these quantities (up to order one coefficients) analytically from energy conservation and  
the Nambu--Goto effective theory with the Kalb--Ramond term, which describes strings coupled to the axion field in the limit of small string thickness (and captures the logarithmic dependence of the tension mentioned above). We will then see that these  predictions are reproduced remarkably well by first principles numerical simulations of the physical system, which confirm the validity of the theoretical assumptions and allow us to extract the unknown coefficients. As will be clear in what follows, numerical simulations can only access a relatively small time range, and it is impossible to directly extract the GW energy and its momentum distribution at the physically relevant time, so a careful extrapolation is essential. However, the existence of the scaling solution, in combination with our analytic understanding on the GW emission, makes this extrapolation reliable.

We will show that the energy emitted in GWs at later times during the scaling regime is logarithmically enhanced, primarily as a result of the logarithmic increase of the string tension, and the energy is always produced with a momentum distribution localised at frequencies of order the Hubble scale. When the emission from the entire scaling regime is taken into account, this leads to  logarithmic deviations from  a scale invariant GW spectrum, which increase the amplitude of the spectrum at low frequencies (indeed, these frequencies are emitted the latest, when the enhancement is largest). As we will show in Section~\ref{sec:GWs}, the deviation can be approximated by a spectrum $d\Omega_{\rm gw}/d\log f\propto\log^4(f_a/H_f)$, where $f_a$ is the axion decay constant and $H_f\propto f^2$ is the Hubble parameter at the time when GWs of present day frequency $f$ are emitted. 
Given the large value of this logarithm (up to $10^2$ for the relevant axion masses), the deviation from scale invariance is substantial and means that axions with $f_a\gtrsim 10^{14}$ GeV lead to GW spectra that are observable in multiple upcoming experiments.

To better understand the range of decay constants and masses that could be discovered via GW observations, in Section~\ref{sec:Constraints} we study additional properties of the post-inflationary scenario, which give constraints on the axion parameter space. 
 In particular, we derive a lower bound on the relic abundance of axions from strings. We also calculate the spectrum of density perturbations in the axion field, which leads to isocurvature perturbations measurable for instance in the cosmic microwave background (CMB) that are potentially in conflict with observations. Finally, we discuss how the axions emitted during the scaling regime contribute to dark radiation, which is constrained by big bang nucleosynthesis (BBN) and CMB measurements.

We will see there is a significant region of allowed and observable masses and decay constants for ultralight axions, i.e. with a mass $\lesssim 10^{-17}~\eV$. However, the GWs from QCD axion strings are not observable due to the bound $f_a\lesssim 10^{10}$ GeV from DM overproduction in this case~\cite{moreaxions}.  GW searches are particularly useful since they are complementary to other approaches, such as astrophysical observations and DM direct detection experiments, with their sensitivity strongest for large decay constants for which  the axion couplings are typically suppressed. Our work will enable limits from GW observations, or even possible future discoveries, to be related to physics at energy scales far beyond any that could be explored directly.  It will also allow complementary progress (e.g. potential improvements in searches for isocurvature perturbations and of the measurement of $N_{\rm eff}$) to be interpreted in terms of the post-inflationary scenario.

The paper is structured as follows. After reviewing the properties of global strings in Section~\ref{ss:Review}, we begin our new work in Section~\ref{sec:GWs} by calculating the GW emission from the string network. In Section~\ref{sec:Constraints} we analyse additional properties and constraints on axions in the post-inflationary scenario. Following this, in Section~\ref{sec:restore} we study ways in which the U(1) symmetry can be restored in the early universe for large $f_a$. Finally, we conclude and discuss directions for future work in Section~\ref{sec:Conclusion}. Further details and supporting analysis is given in Appendices where we also compare our work to the previous literature.

\section{Properties of Axion String Networks} \label{ss:Review}

We consider a single axion, i.e. a PNGB of a spontaneously broken global U(1) symmetry, softly broken by the axion potential $V$, which is a periodic function of period $2\pi f_a$ and leads to an axion mass $m_a$. We remain agnostic about the origin of the potential (either by UV or IR physics) and its particular form. The QCD axion is a particular case, with a mass that arises due to an anomalous coupling to the gluon field strength and which is related to the axion decay constant by $m_af_a\simeq m_\pi f_\pi$ (where $m_\pi$ and $f_\pi$ are the pion mass and decay constant)  \cite{Weinberg:1977ma}. A prototypical axion model comprises a complex scalar field $\phi$ with Lagrangian
\begin{equation} \label{eq:LPhi}
{\cal L}=|\partial_\mu \phi|^2-\frac{m_r^2}{2v^2}\left (|\phi|^2-\frac{v^2}2 \right )^2~,
\end{equation}
leading to the spontaneous U(1) symmetry breaking at the scale $v$. The axion $a(x)$ is associated to the phase of $\phi$ as $\phi(x)=\frac{v+r(x)}{\sqrt2} e^{i a(x)/v}$, while the radial mode $r(x)$ is a heavy field of mass $m_r$. The equations of motion of the Lagrangian in eq.~\eqref{eq:LPhi} admit solitonic string-like solutions, called axion strings~\cite{Sikivie:1982qv,Vilenkin:1982ks,Vilenkin:1984ib,Davis:1986xc}, which are topologically non-trivial configurations that contain loops in space around which the axion field  wraps the fundamental domain $[-\pi v,\pi v]$ with non-zero winding number. At the string centre, $r(x)$ acquires a value of the order $v$ over a distance of order $m_r^{-1}$, which sets the string core thickness.

A network of axion strings forms after the U(1) symmetry is broken, and this subsequently approaches the attractor solution. We assume that the Universe is in radiation domination with metric $ds^2=dt^2-R^2(t)dx^2$, where $R(t)\propto t^{1/2}$, and Hubble parameter $H\equiv\dot{R}/R=1/(2t)$. Extensive evidence for the attractor was given in~\cite{Gorghetto:2018myk,moreaxions}, where more details can be found. The attractor is independent of the network’s initial properties, allowing us to make predictions that do not depend on the details of the breaking of the U(1) symmetry and of the very early history of the Universe, i.e. at times when $H\gg m_a$.\footnote{We will see that the GW spectrum at the observationally relevant frequencies is also independent of the very early evolution.} The existence of the attractor can be understood as resulting from a balance between two opposing effects: the expansion of the Universe continually increases the number of strings per Hubble patch, but if the critical density is exceeded string interactions and recombinations become efficient enough that the number of strings decreases. Consequently, the system is held at a critical point at which the number of strings per Hubble patch is approximately constant.

As mentioned, on the attractor solution the statistical properties of the string network follow fixed scaling laws that are (approximately) determined only by the one evolving scale: the Hubble parameter. For instance, 
 the average energy density of the string network can be written as
\begin{equation} \label{eq:rhoscal}
\rho_s(t) = \xi \frac{ \mu_{\rm eff}}{t^2} \equiv 4\xi  \mu_{\rm eff} H^2\,,
\end{equation}
where $\xi$ is the number of strings per Hubble patch, which measures the total length $\ell$ of the strings inside a Hubble volume in units of Hubble length, namely $\xi \equiv \lim_{L\to \infty } {\ell}(L)\,t^2/L^3$, while $\mu_{\rm eff}$ is the effective tension of the strings, i.e. their energy per unit length. For local strings the string tension is constant as the energy density is localised on their cores, and the explicit factor of $H^2$ in eq.~\eqref{eq:rhoscal} might capture the full time dependence of $\rho_s$. 
 However, the situation is different for the global strings that we study. In this case, owing to a logarithmic divergence, the tension of a single long straight string in one Hubble patch is $\mu=\pi v^2 \log(m_r/H)$. Consequently, during the scaling regime $\mu_{\rm eff}$ is expected to take the form
\beq\label{muth}
\mu_{\rm th} = \pi v^2 \log\left(\frac{m_r}{H}\frac{\eta}{\sqrt{\xi}}\right) ~,
\eeq
where $\eta$ is a dimensionless quantity that parametrises the typical shape of the strings in the scaling regime.\footnote{In eq.~\eqref{muth}, the argument of the logarithm should capture the main time dependence on $t$ since the logarithm is cut-off by the average distance between strings $\propto t/\sqrt{\xi}$.} Given the self-similarity of the network during scaling, $\eta$ is expected to have, at most, a weak time-dependence, so $\mu_{\rm eff}$ increases logarithmically with time. In axion theories that are more general than eq.~\eqref{eq:LPhi}, eq.~\eqref{muth} still holds with $m_r$ a UV-dependent parameter representing the typical mass of the heavy degrees of freedom associated with the U(1) breaking. In Appendix~\ref{app:simscal} we will show that numerical simulations of eq.~\eqref{eq:LPhi} confirm the validity of eq.~\eqref{muth} with a fixed $\eta$. 

The linear dependence of the string tension on $\log(m_r/H)$ also implies that the effective coupling of the axion field to the string is proportional to $1/\log(m_r/H)$ (see e.g. the discussion around eq.~\eqref{NGeqs1} below). It is therefore plausible that other properties of network might also depend on the same factor (in the following we define $\log\equiv\log(m_r/H)$). Indeed, there is clear evidence from numerical simulation of eq.~\eqref{eq:LPhi} that such violations are present in a number of the network's properties. For example, $\xi$ itself grows linearly with log during the scaling regime, namely (up to $1/\log$ terms that depend on the initial conditions)
\begin{equation} \label{eq:xivslog}
\xi=c_1 \log +c_0 ~,
\end{equation}
where the coefficient $c_1=0.24(2)$ can be extracted numerically. Although numerical simulations can so far only simulate string networks, and hence confirm eq.~\eqref{eq:xivslog}, at small scale separations (with $\log\lesssim8$), the growth is likely to be an intrinsic property of the scaling solution and persist also at larger logs \cite{Gorghetto:2018myk}.\footnote{For instance, the logarithmic growth affects long strings at exactly the same rate as sub-horizon loops (which make up respectively $80\%$ and $20\%$ of the string network length).} The logarithmic dependence of $\mu_{\rm eff}$ and $\xi$ on $m_r/H$ is referred to as ‘scaling violation’, as it introduces an explicit dependence on the additional scale $m_r$ in the properties of the scaling regime and in eq.~\eqref{eq:rhoscal}.

Since the energy density in eq.~\eqref{eq:rhoscal} diminishes faster than the energy of a system of (long) free strings ($\rho_s^{\rm free}\propto R^{-2}$), energy must be continuously emitted from strings to maintain scaling. Conservation of energy and eq.~\eqref{eq:rhoscal} imply the energy density emission rate $\Gamma=\dot{\rho}_s^{\rm free}-\dot{\rho}_s$ is given by~\cite{Gorghetto:2018myk}
\begin{equation} \label{eq:gammaemi}
\Gamma=\rho_s \left[2H-\frac{\dot \xi}{\xi}-\frac{\pi v^2}{\mu_{\rm eff}} \left(H+\frac{\dot\eta}{\eta} 
-\frac12 \frac{\dot\xi}{\xi} \right)\right] \ \  \stackrel {\log\gg1}\longrightarrow \ \ 2H\rho_s=\frac{\xi\mu_{\rm eff}}{t^3}\,,
\end{equation}
where we assumed that $\mu_{\rm eff}$ defined by eq.~\eqref{eq:rhoscal} is indeed reproduced by $\mu_{\rm th}$  in eq.~\eqref{muth} once the parameter $\eta$ is fixed appropriately. The equality on the right hand side of eq.~\eqref{eq:gammaemi} holds in the large log limit, which as we will see is the regime relevant to the emission of observable GWs. In Appendix~\ref{app:simscal} we will show that numerical simulations confirm the validity of eq.~\eqref{eq:gammaemi} (and we give additional insights into the subtlety that eq.~\eqref{eq:gammaemi} applies only to the 80\% of the string length that is in long strings and the interplay of these with small loops).

The energy lost by the network is radiated into the degrees of freedom coupled to the string, which are axions, radial modes and GWs. We therefore split
\begin{equation}
\Gamma=\Gamma_a+\Gamma_r+\Gamma_g~, 
\end{equation}
to account for the respective emissions. At large enough value of log (but not too large), $\Gamma$ is dominated by $\Gamma_a$. Indeed, numerical simulations of eq.~\eqref{eq:LPhi} show that the radial mode decouples from the string, but only logarithmically with the ratio $m_r/H$, i.e. $\Gamma_r/\Gamma_a$ decreases as inverse powers of $\log$~\cite{moreaxions}. In particular, although some small fraction of the energy (about 10\%) is emitted into radial modes at $\log\lesssim8$, this is seen to reduce logarithmically and is expected to vanish in the large log limit. Meanwhile, as we will show in the next Section, $\Gamma_g/\Gamma_a\simeq G\mu_{\rm eff}^2/v^2=\pi/8(v \log/M_{\rm P})^2$, where $M_{\rm P}=1/\sqrt{8\pi G}$ is the reduced Planck Mass. Consequently the GW emission is suppressed with respect to that into axions until $\log\sim M_{\rm P}/v$. As a result, $\Gamma\simeq\Gamma_a$ for $1\ll\log\ll M_{\rm P}/v$. In this range of $\log$  eq.~\eqref{eq:gammaemi} fully fixes $\Gamma_a$ in terms of $\xi$ and $\mu_{\rm eff}$.

The string network and the scaling regime persist as the Universe expands until approximately $H = m_a \equiv H_\star$, when the axion potential $V$ becomes cosmologically relevant ($m_a$ may either be temperature dependent or independent). At this time  $\log(m_r/H_\star)=60\div70$ for the QCD axion and can be $\simeq100$ for ultralight axions. At $H=H_\star$ a network of domain walls forms, bounded by the strings. If the axion potential does not preserve any discrete subgroup of the U(1) symmetry, in which case $v=f_a$, the domain walls are unstable and decay destroying the string network in the process. Meanwhile, soon after $H=H_\star$, most of the axions emitted during the scaling regime turn nonrelativistic and contribute to the DM abundance. Additional axions are expected to be emitted as the domain walls annihilate, supplementing the relic abundance from axions produced during the scaling regime. 
In the following we will set $v=f_a$, as is the case for unstable domain walls. However, as discussed in more detail in Section~\ref{ss:genericN}, our derivation of the GW spectrum from strings can be easily generalised to axions with $v= N f_a$ by replacing $f_a\to v$ throughout (for the QCD axion $N$ is set by the anomaly coefficient between the PQ symmetry and QCD). Note that for $N>1$ additional explicit breaking of the remaining discrete symmetry is necessary to avoid the domain walls over-closing the Universe.

\section{Gravitational Waves from Strings} \label{sec:GWs}

During the scaling regime the motion and interactions of the strings act as a continual source of GWs. In this Section we study the resulting spectrum by combining the 
effective theory of global strings, and field theory simulations of the physical system in eq.~\eqref{eq:LPhi}. 
In particular, in Section~\ref{ss:theory} we use the Nambu--Goto effective theory coupled to the axion field via the Kalb--Ramond term, which captures the dynamics of the parts of the network with small curvature. Both the effective theory and numerical simulations will show that the GWs can be self-consistently treated as a perturbation of the string network if $G\mu^2/f_a^2\ll1$, which will be satisfied for all $f_a$ and $m_a$ of interest. In this case, the fact that GWs are produced does not significantly influence the evolution of the network, which follows the previously described attractor.\footnote{If instead $G\mu^2/f_a^2\gtrsim1$ gravity dramatically changes the evolution of the system, and affects the scaling regime in a way that is not known.} 

As mentioned,  for $G\mu^2/f_a^2\ll1$ energy conservation and the scaling regime fix the time dependence of $\Gamma_a$, via eq.~\eqref{eq:gammaemi}.  
However, this cannot be directly used to infer  $\Gamma_g$, which accounts for only a small fraction of the energy released. Nevertheless, in this Section we will show that we can still make use of eq.~\eqref{eq:gammaemi} thanks to a convenient relation between the rate of energy emission into GWs and that into axions. We will argue for this relation theoretically using the Nambu--Goto effective theory, and confirm it with numerical simulations of the physical system in the scaling regime.  This will allow us to have analytic control of $\Gamma_g$ at all times, except for an order one coefficient that will be directly extracted from the simulations. In combination with the momentum distribution of the instantaneous GW emission, whose general form can be easily guessed and will be confirmed in simulations, this will allow us to determine the total GW spectrum produced by the network up to $H=H_\star$ when it is destroyed. After being produced the GWs propagate freely, redshifting as the universe expands, so today they make up an irreducible contribution to the stochastic background. As we will see in Section~\ref{ss:presentday}, during the scaling regime the GWs at the observable frequencies are emitted when $\log\gg1$, and therefore in the following we will often refer to the large log limit.

\subsection{Theoretical Derivation of GW Emission} \label{ss:theory}

The required relation between $\Gamma_g$ and $\Gamma_a$ can be argued for via the low energy limit of eq.~\eqref{eq:LPhi}, which is the effective theory of Nambu--Goto strings coupled to the axion field~\cite{Lund:1976ze} by the Kalb--Ramond action~\cite{Kalb:1974yc}. 
In particular, this effective theory can be obtained from eq.~\eqref{eq:LPhi} on the background of a string and at energies smaller than $m_r$ (i.e. integrating out the radial mode, see \cite{Davis:1988rw} for the explicit derivation). It describes the evolution of an infinitely thin string, with a trajectory identified by the space-time coordinate $X^\mu(\tau,\sigma)$, where $\tau,\sigma$ are worldsheet coordinates. The string is coupled to the axion field, described by its (dual) antisymmetric tensor $A_{\mu\nu}$. The corresponding action is
\begin{equation}\label{NGaction}
S=-\mu\int{d\tau d\sigma\sqrt{-\gamma}}-\frac16\int{d^4x F^{\mu\nu\rho}F_{\mu\nu\rho}}-g\int{d\tau d\sigma}\epsilon^{ab}\partial_a X^{\mu}\partial_b X^{\nu}A_{\mu\nu}(X) ~,
\end{equation}
where  $F_{\mu\nu\rho}=\partial_\mu A_{\nu\rho}+\partial_\nu A_{\rho\mu}+\partial_\rho A_{\mu\nu}$ and $\gamma$ is the determinant of the induced metric on the worldsheet $\gamma_{ab}=\partial_a X^{\mu}\partial_b X_\mu$, with $a,b=\tau,\sigma$. The coupling $g$ defines the axion-string interaction, while $\mu$ is the string energy per unit length  (this is easily seen from the first component of the energy momentum tensor, see eq.~\eqref{gwsource}).  As we will see in the following, $\mu$ accounts for the energy in the axion gradients, as well as that localised in the core. 
 The axion is related to the only degree of freedom of $A_{\mu\nu}$ by $F^{\mu\nu\rho}=\epsilon^{\mu\nu\rho\sigma}\partial_\sigma a/\sqrt{2}$.\footnote{The normalisation is fixed by equating the energy momentum tensor of the second term in eq.~\eqref{NGaction} to that of a free axion.}

Since $a$ changes by multiples of $2\pi f_a$ around a string, the coupling $g$ is quantised in terms of $f_a$ as $g=2\pi n f_a/\sqrt{2}$, with $n$ integer.\footnote{This is a consequence of the fact that the commutator $[\partial_i,\partial_j]a$ is non-zero (and quantized) around a string \cite{Davis:1988rw} and is easily seen by imposing $2\pi n f_a=\oint_C{dx^\mu \partial_\mu a}$ where $C$ is a loop surrounding the string, and evaluating the right hand side of this equation via Gauss' theorem and using the equations of motion $\partial_\mu F^{\mu\nu\rho}=-g\int{d\tau d\sigma}\epsilon^{ab}\partial_aX^{\nu}\partial_bX^{\rho}\delta^4(x-X)$.} 
 The gauge invariance $A_{\mu\nu}\to A_{\mu\nu}+\partial_\mu \Lambda_\nu+\partial_\nu \Lambda_\mu$ and worldsheet reparametrization invariance of eq.~\eqref{NGaction} can be fixed by choosing the gauge $\partial_\mu A^{\mu\nu}=0$ and $\dot{X}\cdot X'=\dot{X}^2+ X'^2=0$ where $\dot{X}^\mu\equiv\partial_\tau X^\mu$ and $X'^\mu\equiv\partial_\sigma X^\mu$. In the frame $\tau=t$, the equations of motion for a string with winding $n=1$ are
\begin{align}\label{NGeqs1}
\mu(\ddot{X}^\mu-{X''}^\mu)&= 2\sqrt{2}\pi f_aF^{\mu\nu\rho}\dot{X}_\nu X'_\rho \ ,\\\label{NGeqs2}
\partial_\alpha\partial^\alpha A^{\mu\nu}&=\sqrt{2}\pi f_a\int d\sigma\left(\dot{X}^\mu X'^\nu-\dot{X}^\nu X'^\mu\right)\delta^3(\vec{x}-\vec{X}) \ .
\end{align}
This system of coupled equations determines the evolution of the string and the axion field. The axion is sourced from a moving string via eq.~\eqref{NGeqs2}, whose motion is itself influenced by the axion via~eq.~\eqref{NGeqs1}. 

Before proceeding, let us clarify a subtlety of this theory. As discussed in~\cite{Lund:1976ze,Dabholkar:1989ju} the action in eq.~\eqref{NGaction} is strictly speaking ill defined, since the solution of the equations of motion for $A_{\mu\nu}$  in eq.~\eqref{NGeqs2} is divergent as $x^\mu$ approaches $X^\mu$. This makes the interaction term in eq.~\eqref{NGaction} (logarithmically) divergent when evaluated on such solutions. This UV divergence can be regularized and completely reabsorbed in the redefinition of the (bare) string tension $\mu$. 
After reabsorbing the divergence, the equations of motion will have the same form as eqs.~\eqref{NGeqs1} and~\eqref{NGeqs2}, but (just like in the renormalization of quantum field theories) with finite $\mu(\Delta)$ and $A_{\mu\nu}(\Delta)$ depending on a new (unphysical) length scale $\Delta$, which can be interpreted as the length at which one probes the string core. In particular, under the change of this scale to $\Delta'$, $\mu(\Delta')=\mu(\Delta)+(g^2/2\pi)\log(\Delta'/\Delta)=\mu(\Delta)+\pi f_a^2\log(\Delta'/\Delta)$.\footnote{The same running has been studied in a generalisation of the theory we consider in the context of the effective string description of vortices in superfluids \cite{Horn:2015zna}.} 
As this scale is not physical, it can be chosen arbitrarily. If $\Delta$ is chosen as $m_r^{-1}\ll\Delta\lesssim L$, where $L$ is the IR cutoff ($\simeq H^{-1}$ for long strings), the interpretation of $\mu(\Delta)$ will be that of an effective tension that includes the energy in the axiostatic gradient (up to the IR cutoff $\Delta$), while $A_{\mu\nu}(\Delta)$ will include mostly the axion radiation.\footnote{The fact that changing $\Delta$ does not change the equations of motion implies that, as far as the dynamics of the string at small curvature is concerned, it does not matter whether the energy is localised in the string core or in the axion gradient. } In the following we will tacitly assume that the preceding regularization and subtraction has been performed, and that $\Delta$ has been always chosen in this way, so that $\mu(\Delta)$ (which we will call $\mu$ for simplicity) corresponds to the total energy per unit length, including the gradient energy (see~\cite{Dabholkar:1989ju} for a more complete treatment). From eq.~\eqref{NGeqs1} it follows that, as anticipated in Section~\ref{ss:Review}, the effective coupling of the axion to the string is determined by $f_a^2/\mu\propto 1/\log$.

Notice that the preceding discussion does not hold when the inverse core-size $m_r^{-1}$ is of the order of the IR cutoff $L$. Consequently, this effective theory describes the dynamics of the physical system in the parts of the network where the finite string thickness is smaller than the inverse string curvature, but it  will break down when strings intersect and reconnect, or when loops shrink (such processes are sensitive to the details of the structure of the potential of the field $\phi$ and will therefore need the full theory in eq.~\eqref{eq:LPhi}).

In the presence of gravity a moving string sources gravitational radiation, which can be determined by linearising Einstein's equations giving e.g. in the harmonic gauge $\partial^\mu h_{\mu\nu}=\frac12\partial^\mu h$ 
\begin{equation}\label{gwsource}
\partial_\alpha\partial^\alpha h^{\mu\nu}=16\pi G \left(T_s^{\mu\nu}-\frac12\eta^{\mu\nu} T_{s \, \lambda}^{   \lambda}\right) , \quad T_s^{\mu\nu}=\mu\int d\sigma\left(\dot{X}^\mu \dot{X}^\nu-X'^\mu X'^\nu\right)\delta^3(\vec{x}-\vec{X}) \, ,
\end{equation}
where $g_{\mu\nu}=\eta_{\mu\nu}+h_{\mu\nu}$ is the metric, $h\equiv h^\mu_{\ \mu}$ and $T_s^{\mu\nu}$ is the energy momentum tensor of the string from the first term of eq.~\eqref{NGaction}. It is straightforward to show that 
the energies radiated at infinity per unit time in axions and GWs from a string trajectory $X^\mu$ are respectively
\begin{equation}\label{EaEg}
\frac{dE_a}{dt}=r_a[X]\, f_a^2\ ,  \qquad \frac{dE_g}{dt}=r_g[X]\, G \mu^2 ~,
\end{equation}
where $r_a[X]$ and $r_g[X]$ are dimensionless functionals of the shape of the string trajectory (but independent of the string length). In more detail, for any string trajectory $X^\mu$ that is a solution of eqs.~\eqref{NGeqs1} and~\eqref{NGeqs2}, the axion and GW fields are determined by eqs.~\eqref{NGeqs2} and~\eqref{gwsource}. These are wave-like equations of the form $\partial_\alpha \partial^\alpha B=j$, with solution $B=\int d^3y j(t-|\vec{x}-\vec{y}|, \vec{y})/(4\pi|\vec{x}-\vec{y}|)$, and therefore $A_{\mu\nu}\propto f_a$ and $h_{\mu\nu}\propto G\mu$. The emitted energy is $dE/dt\equiv-\int d^3x \dot{T}^{00}$, where for the axion $T_a^{\mu\nu}\sim (\partial A)^2$ and for the GWs $T_g^{\mu\nu}\sim G^{-1}(\partial h)^2$. This fixes the dependence on $f_a^2$ and $G\mu^2$ of eq.~\eqref{EaEg}, while the remaining factors (called $r_a$ and $r_g$) must be dimensionless functionals of the string trajectory only.

The main conclusion from eq.~\eqref{EaEg} is that GWs are emitted proportionally to the (square of) the string tension, since they are sourced by the energy momentum tensor. Conversely, the axion coupling to the string is fixed by $f_a$ and the axion energy is proportional to $f_a^2$ only. We stress that eq.~\eqref{EaEg} is valid for any trajectory that is a solution of eqs.~\eqref{NGeqs1} and~\eqref{NGeqs2}, irrespective of the ratio $f_a^2/\mu$, i.e. regardless of the magnitude of the axion-string coupling. Therefore eq.~\eqref{EaEg} is expected to capture the energy emission from the pieces of the string network for which the string thickness can be neglected at all values of the log, including those accessible in simulations 
(related previous analysis in the literature has been carried out in the limit of zero coupling \cite{Vachaspati:1984gt,Vilenkin:1986ku}).

Since we will not need the functional form of $r_a[X]$ and $r_g[X]$, we give their expressions  
in Appendix~\ref{app:moretheory}, where we also give further details of the derivation of eq.~\eqref{EaEg}. From eqs.~\eqref{rg} in Appendix~\ref{app:moretheory} it can be seen explicitly that (as expected given that they are dimensionless) $r_a[X]$ and $r_g[X]$ are invariant under the rescaling of the length of the trajectory and of time, and therefore depend only on the shape of the trajectory. We note that the coefficients $r_a[X]$ and $r_g[X]$ have been calculated in~\cite{Vilenkin:1986ku} for particular trajectories in the limit of zero coupling.

Finally, we observe that, as mentioned, this effective field theory predicts that the GWs do not significantly influence the motion of the strings provided $G\mu^2/f_a^2\ll1$. Indeed, the inclusion of gravitational backreaction modifies eq.~\eqref{NGeqs1} by introducing, on the right hand side, the term  $-\mu \Gamma_{\nu\rho}^{\mu} (\dot{X}^{\nu}{X'}^{\rho}+\dot{X}^{\rho}{X'}^{\nu})$, where $\Gamma_{\nu\rho}^{\mu}$ are the Christoffel symbols (this was first studied in \cite{Copeland:1990qu,Quashnock:1990wv}). Since $\Gamma_{\nu\rho}^{\mu}\sim\partial h$, and $h$ is of order $G\mu$, this term is suppressed by $G \mu^2/f_a^2$ with respect to the one already present in eq.~\eqref{NGeqs1}. Similarly, the energy emitted in GWs from eq.~\eqref{EaEg} is suppressed with respect to that into axions by the same factor.

\subsection{GW Spectrum during the Scaling Regime} \label{ss:setup}

We now combine the results of Sections~\ref{ss:Review} and \ref{ss:theory}  to extract information on the emission of GWs during the scaling regime. As outlined, we use an approach that avoids having to calculate the GW emission directly from eq.~\eqref{EaEg}, which would require understanding the form of the trajectories $X^\mu_s$ of long strings and loops during scaling.

Given that eq.~\eqref{EaEg} holds for a generic string trajectory, the energy densities $\Gamma_a$ and $\Gamma_g$ emitted per unit time during the scaling regime are related by $rG \mu_{\rm eff}^2/f_a^2$, where $r\equiv r_g[X_s]/r_a[X_s]$. 
We can therefore use our knowledge of $\Gamma_a$ from energy conservation (i.e. eq.~\eqref{eq:gammaemi}) to infer the energy density emitted per unit time in GWs during the scaling regime. This reads
\begin{equation}\label{gammagw}
\Gamma_{g}=r\, \frac{G\mu_{\rm eff}^2}{f_a^2}\Gamma_a \  \stackrel {\log\gg1}\longrightarrow \ 8\xi rH^3\frac{G\mu_{\rm eff}^3}{f_a^2} \ ,
\end{equation}
where the second relation holds in the large log limit, and in that case $\Gamma_{g}\simeq8 \pi^3 r  G f_a^4   H^3 \xi \log^3$. In eq.~\eqref{gammagw} the dimensionless coefficient $r$ is a functional of the average shape of the string network (and expected to be of order $1$). The average shape of the strings is preserved throughout the scaling regime, and therefore we expect $r$ to be time-independent, or at most have a weak log dependence.  
The coefficient $r$ can be interpreted as a form factor of the string network that encodes how  efficiently the string trajectories during scaling emit GWs compared to axions. In particular it parametrises the string dynamics that are responsible for the GW emission (long strings, small loops, string reconnection, etc.). 

The validity of eq.~\eqref{gammagw} with a constant $r$ relies solely on energy conservation during the scaling regime and on the Nambu--Goto effective theory. While the latter must break down when strings reconnect and loops shrink, it is possible that most of the axion and GW energy is emitted in the regime where the effective theory is valid. Indeed, we will see in  Section~\ref{ss:gwsim} that eq.~\eqref{gammagw} is reproduced remarkably well with a constant $r$ in first principles field theory simulations, which will allow us to also directly extract its value (instead of calculating it from its definition). We will also see that $\mu_{\rm eff}$, which we defined in terms of the string energy, is well matched by the theoretical form eq.~\eqref{muth}. Given this, we will assume that eq.~\eqref{gammagw} holds in the remainder of our present analysis.\footnote{As mentioned in Section~\ref{ss:Review}, eq.~\eqref{gammagw} must break down when $\log\sim M_{\rm P}/f_a$. In Section~\ref{sec:Constraints} we will see that the values of $f_a$ allowed by existing constraints are always small enough for this to be true (for all $m_a$). With an abuse of language, we will therefore use the phrasing `large log' to indicate $\log\lesssim M_{\rm P}/f_a$.} In particular, assuming the growth of $\xi$ as in eq.~\eqref{eq:xivslog}, this implies that $\Gamma_g$ is proportional to $H^3\log^4$ at large log.

 Since GWs redshift freely, we can  straightforwardly obtain the GW energy density $\rho_g$ at a generic time during the scaling regime from $\dot{\rho}_g+4H\rho_g=\Gamma_g+\dots$, where the dots stand for possible additional GW sources, which we subsequently neglect. 
It immediately follows that $\rho_g(t)=\int_{t_1}^tdt'(R(t')/R(t))^4\Gamma_g(t')$, where $t_1$ is the time when the scaling regime starts. The remaining ingredient required to calculate the GW spectrum is the momentum distribution of $\Gamma_g$. 
 It is convenient to write $\Gamma_g$ as a function of the differential emission rate $\partial \Gamma_g/\partial k$ and to further express this in terms of the instantaneous emission spectrum $F_g$, i.e. 
\begin{equation}\label{Fdef}
\Gamma_g(t)=\int{dk  \frac{\partial\Gamma_g}{\partial k}[k,t]} \ , \qquad\quad \frac{\partial\Gamma_g}{\partial k}[k,t]=\frac{\Gamma_g(t)}{H(t)} F_g\left[\frac{k}{H},\frac{m_r}{H}\right] \, .
\end{equation}
The function $F_g[x,y]$ fully captures the momentum dependence of the instantaneous emission via the variable $x=k/H$ and its possible time dependence via the variable $m_r/H$, and is  normalised to one by definition, $\int dx F_g[x,y]=1$. Plugging eq.~\eqref{Fdef} into $\rho_g(t)$ we obtain the total GW energy density spectrum, defined by $\rho_g=\int dk \partial \rho_g/\partial k$,
\begin{equation} \label{eq:drhogdk}
\frac{\partial \rho_g}{\partial k}\left[k,t\right]=\int^t_{t_1} dt' \frac{\Gamma_g'}{H'}\left( \frac{R'}{R}\right)^3 F_g\left [\frac{k'}{H'},\frac{m_r}{H'}\right ]\,,
\end{equation}
where $k'=k R/R'$ is the redshifted momentum and all other the primed quantities are evaluated at $t'$. Eq.~\eqref{eq:drhogdk} is just the superposition of all the spectra emitted from $t_1$ to $t$, properly redshifted.

Given the existence of a scaling solution, we have some theoretical expectations for the form of the instantaneous GW spectrum $F_g$, which closely resembles the analogous axion spectrum studied in \cite{moreaxions}. First, since strings typically have a curvature of order Hubble, the spectrum of GWs emitted at each instant should be peaked at momenta of order the Hubble parameter at that time. Meanwhile, production of GWs with momentum below Hubble or above the string core scale is expected to be strongly suppressed. The absence of any scale between $H$ and $m_r$ suggests that between these two (IR and UV) cutoffs the spectrum follows a single power law $F_g\propto1/x^{q}$.  
It is also expected from the Nambu--Goto description  that most of the GW energy $\Gamma_g$ is contained in (IR) momenta of order Hubble, as opposed to (UV) momenta of order $m_r$ (see \cite{Dabholkar:1989ju}), which corresponds to $q>1$.

We will show in the next Section that all of the properties above are verified in numerical simulations of the scaling regime, where we will provide the exact form of $F_g$. To get a general picture of the resulting total GW spectrum, it is sufficient to approximate the instantaneous emission with sharp IR and UV cutoffs at momenta $x_0 H$ and $m_r$, i.e. $F_g[x,y]\propto 1/x^q$ for $x\in[x_0,y]$, and zero elsewhere. Inserting this $F_g$ into eq.~\eqref{eq:drhogdk} we obtain the GW spectrum. At a generic time during the scaling regime with $\log\gg1$, in the momentum range $k\gtrsim  x_0H$ but $k\lesssim x_0\sqrt{H H_1}$ (where $H_1=1/2t_1$ is the Hubble parameter at the start of the scaling regime) this is given by 
\begin{equation}\label{totalspectrum}
\frac{\partial \rho_g}{\partial \log k}\left[k,t\right]
=8\pi^3 c_1 r Gf_a^4 H^2
\log^4\left[\frac{m_r}{H}\left(\frac{x_0H}{k}\right)^2\right] \ ,  \ \ \ \ \text{for} \ \ x_0H\lesssim k \lesssim x_0\sqrt{H H_1} ~,
\end{equation}
where $c_1$ defined in eq.~\eqref{eq:xivslog} determines the growth rate of $\xi$, and we omitted terms proportional  
to additional inverse powers of log, which are negligible in the regime we are considering. 
The full expression for the spectrum is given in Appendix~\ref{app:strings}.

We observe that the spectrum is (approximately) scale invariant between the IR and UV momenta $x_0 H$ and $x_0\sqrt{H H_1}$, so the total GW energy is spread over a wide range of frequencies. This is because the rate at which energy in previously emitted GWs redshifts and the main decrease in $\Gamma_g \propto H^3$ in a radiation dominated universe combine to give $\partial \rho_g/\partial k \propto 1/k$. Such a spectrum accumulates only between the two extremes $x_0H$ and $x_0\sqrt{H H_1}$, which correspond to the peaks of the instantaneous emission at $H$ and at $H_1$ respectively (redshifted until $H$).  

However, as time progresses the GW emission is enhanced due to the (increasing) $\log^4$ factor in $\Gamma_g$. This leads to a violation of the spectrum's scale invariance, which consequently has larger values at smaller $k$. This is captured by the logarithmic factor in eq.~\eqref{totalspectrum}, which ranges from $\log^4(m_r/H)=\log^4$ at $k\simeq x_0H$ to $\log^4(m_r/H_1)$ at $k\simeq x_0\sqrt{HH_1}$. 
One $\log^2$ factor is due to the increase in the energy stored in the network (from $\xi$ and $\mu_{\rm eff}$), and the additional $\log^2$ factor to the efficiency at which this energy can be emitted in GWs, proportional to coupling of the GWs to the string $G\mu^2$. Although a single power of $\log$ relies on the extrapolation of the log growth of $\xi$ observed in simulations, the other three powers are inevitable. Since at the end of the scaling regime $\log\simeq100$, the violation of scale invariance is substantial and -- as we discuss in Section~\ref{ss:presentday} -- it could make the low frequencies of an otherwise invisible signal detectable. 
 It is also clear that it is essential to extrapolate to large $\log$ to make any sensible physical predictions, and results directly extracted from simulations, which can reach only $\log\lesssim8$, are guaranteed to be off by many orders of magnitude.

There are several other features of  the GW spectrum from scaling that are worth noting. First, the dependence of the spectrum in  eq.~\eqref{totalspectrum} on $x_0$ is only logarithmic, and the dependence on $q$ only comes in (neglected) terms proportional to $(q-1)^{-1}\log^{-1}$ (i.e. together with inverse powers of $\log$, see eq.~\eqref{eq:drhodlogkfull} in Appendix~\ref{app:strings}). Thus, as long as $q-1$ is definitely larger than $1/\log$ the dependence on $q$ is negligible for modes with $k>x_0 H$. This means that even an approximate determination of $x_0$ and $q$ from simulations will be sufficient to understand the spectrum in the momentum region of eq.~\eqref{totalspectrum}. Moreover, a possible dependence of $x_0$ and $q$ on $\log$ -- as long as it keeps $q>1$ -- would not significantly change the spectrum.

Second, if the effective number of degrees of freedom in thermal equilibrium $g$ is not constant, the scale factor away from particle thresholds is $R\propto t^{1/2} g^{-1/12}$ where $g$ is evaluated at the temperature corresponding to the time $t$. 
The spectrum in eq.~\eqref{totalspectrum} at time $t$ gets the ($k$-dependent) multiplicative correction $(g(t)/g(t_k))^{1/3}$. Here, $t_k$ is defined by $x_0 H(t_k)\equiv k R(t)/R(t_k)$, and is the time when most of the GWs that have momentum $k$ at time $t$ are emitted. We refer to Appendix~\ref{app:strings} for the explicit derivation. As we will see in more detail in Section~\ref{ss:presentday}, the change in $g$ distorts the $\log^4$ dependence at the momenta corresponding to the temperature when the degrees of freedom decouple from the thermal bath. 

Finally, as explained in Appendix~\ref{app:strings}, at UV momenta $k\gtrsim x_0\sqrt{ H H_1}$ the spectrum $\partial\rho_a/\partial\log k$  
 is suppressed as $1/k^{q-1}$ and rapidly falls. The critical Hubble $\sqrt{H H_1}$ is model dependent, since it depends on when the scaling regime began (and so when and how the string network forms). 
Meanwhile, at IR momenta $k\lesssim x_0H$ the spectrum is also power law suppressed and follows the same behaviour as $F_g$ for $x\lesssim x_0$. Contrary to the simplified case discussed above, we will see that $F_g[x,y]\propto x^3$ for $x\lesssim x_0$  and this implies $\partial\rho_g/ \partial k\propto k^3$ at $k\lesssim x_0H$. This last part of the spectrum is produced at the time when domain walls form, so will also get a contribution from domain walls, which is expected to change it by at least an order one amount (we discuss this contribution briefly in Appendix~\ref{app:massGW}).

We also note that in the large log limit the total energy in GWs emitted from $t_1$ to $t$, i.e. $\rho_g=\int_{t_1}^tdt'\Gamma_g'(R'/R)^4$, is $\rho_g=(4/5)H^2G\mu_{\rm th}^3\xi\log/f_a^2\propto H^2\log^5$. The additional log factor is related to the fact that this energy gets equal contributions from all the times from $t_1$ to $t$ on a logarithmic scale. It is straightforward to see that the approximately scale invariant spectrum in eq.~\eqref{totalspectrum} reproduces this formula.

\subsection{Comparison with Numerical Simulations} \label{ss:gwsim}

We now show that results from numerical simulations match the preceding analysis extremely well. This confirms our theoretical assumptions (i.e. that the Nambu--Goto EFT is valid at least for determining the relative emission into axions and GWs, and the general shape of the instantaneous emission spectrum) and allows us to extract the values of the parameters $r, x_0$ and $q$.

In the simulations we evolve the complex scalar $\phi$  by numerically integrating the equations of motion that follow from eq.~\eqref{eq:LPhi} on a discrete lattice. Starting from suitable initial conditions, a network of strings forms and evolves, and is driven to the attractor. Due to the competing requirements that the grid must contain a least a few Hubble patches (to capture the infinite-volume properties of the network), and must have at least a few lattice points per string core (to reproduce the string interactions correctly), such simulations can only access values of $\log\lesssim 8$.\footnote{In more detail, the maximum log is limited to $\sim\log N$, where $N$ is the number of gridpoints in the box side. In this paper we used grids of size to $N^3\sim2500^3$ (as opposed to $4500^3$ in previous work) given the additional computational cost in the evolution of the GWs. The resulting maximum $\log\sim 7.5$ is still sufficient for the properties and trends in the attractor solution to be reliably identified.} This is the origin of the previously mentioned required extrapolations (in our approach, we need to extrapolate $r,x_0$ and $q$, as well as $\xi$).

As well as the physical system of eq.~\eqref{eq:LPhi}, we also simulate the so-called `fat' string system, which is defined by the same Lagrangian as eq.~\eqref{eq:LPhi} but with $m_r\propto R^{-1}$ decreasing with time. In this way the string core size remains constant in comoving coordinates. In this system the same hierarchy in $\log$ corresponds to a larger ratio between final and initial cosmic times, and the string network therefore flows to the attractor faster, leading to cleaner results. Although the qualitative features are expected to be reproduced, the quantitative parameters of the scaling solution could differ from those of the physical string system (which we therefore use to extract the numerical values). For both systems we set the initial conditions as close as possible to the scaling solution, on which $\xi$ grows logarithmically (the evolution of $\xi$ is plotted in Figure~\ref{fig:xi} in Appendix~\ref{app:simscal}, where we give more details).
It is straightforward to evaluate the energy in axions and radial modes in simulations, and therefore the emission rates $\Gamma_a$ and $\Gamma_r$. 

The GWs produced during the evolution are obtained by numerically solving the linearised Einstein equations 
\beq\label{eq:gweom}
\ddot{h}_{ij}+3H\dot{h}_{ij}-R^{-2}\nabla^2 h_{ij}=16\pi G T^{\rm TT}_{\phi,ij} ~,
\eeq
where $T^{\rm TT}_{\phi,\mu\nu}$ is the transverse-traceless part of the energy momentum tensor of eq.~\eqref{eq:LPhi}.\footnote{Here $h$ is in the transverse-traceless gauge, which is convenient in the FRW background.} 
 At a generic time the energy density in GWs is $\rho_g\equiv T_g^{00}$, where $T_g^{\mu\nu}=(32\pi G R^2)^{-1}\langle \partial^\mu h_{\alpha\beta}\partial^\nu h^{\alpha\beta} \rangle$ is their energy momentum tensor 
and the brackets stand for the spatial average. From $\rho_g$ the emission rate of GWs can be calculated using $\Gamma_g=R^{-4}\frac{d}{dt}(R^4\rho_g)$. 
Further details concerning our numerical implementation can be found in Appendix~\ref{app:sim}. In Appendix~\ref{app:simUn} we analyse the systematic uncertainties in simulations, e.g. from the finite lattice spacing, and the results we show are with parameter choices such that these are negligible.

The gravitational backreaction (in the weak gravity regime) is represented in the equations of motion of the Lagrangian in eq.~\eqref{eq:LPhi} by the term  $R^{-2}h^{ij}\partial_i\partial_j\phi$. By carrying out simulations with this term included, in Appendix~\ref{app:back} we show that -- as expected -- the (evolving) effective parameter that controls the gravitational backreaction during the scaling regime is $G\mu_{\rm eff}^2/f_a^2$. This is shown in Figure~\ref{fig:xirhoback} of Appendix~\ref{app:back}, where the deviations in $\xi$ and $\rho_a$ due to the backreaction depend only on $G\mu_{\rm eff}^2/f_a^2$ for different choices of $f_a/M_{\rm P}$. In particular, provided $G\mu_{\rm eff}^2/f_a^2\lesssim0.5$, corresponding to $\log  \lesssim M_{\rm P}/f_a$: (1) gravity is self-consistently in the weak regime and (2) it does not alter the dynamics of the string network.\footnote{This is expected, since the backreaction term is negligible with respect to the gradient term $-R^{-2}\nabla^2\phi$ for such $f_a$, as $h\sim Gf_a^2$ from the linearised Einstein equations.} The decay constants (and logs) of interest, studied in Section~\ref{ss:presentday}, are all safely within this limit, so we do not include backreaction in the remainder of our simulations.

We now turn to our main results from simulations. We first observe that the total energy emitted per unit time from the network ($\Gamma_a+\Gamma_r$) in simulations matches the theoretical formula for $\Gamma$ in eq.~\eqref{eq:gammaemi} with the theoretically expected form of the tension $\mu_{\rm th}$ given by eq.~\eqref{muth}. In more detail, although eq.~\eqref{muth} determines uniquely $\mu_{\rm th}$ at large log, at the small log relevant in the simulations $\mu_{\rm th}$ (and consequently $\Gamma$) is sensitive to the (only) free parameter $\eta$. 
We extract the value of $\eta$ by requiring $\Gamma=\Gamma_a+\Gamma_r$ (with $(\Gamma_a+\Gamma_r)$ measured in simulations) in the range $\log >4.5$. 
For the fat string system, this determines $\eta_{\rm fat}\simeq0.27$ quite precisely, while for the physical system $\eta_{\rm phys}\simeq 0.20$ with a larger (about $50\%$) uncertainty. The fact that the ratio $(\Gamma_a+\Gamma_r)/\Gamma$ is 
 close to unity at all times for such time-independent choice of $\eta$ (this is shown in Figure~\ref{fig:Gamma_ratio} of Appendix~\ref{app:simscal}) is a non-trivial check that eq.~\eqref{eq:gammaemi} captures the emission rate at the logs accessible in simulations.\footnote{In fact, the expression for $\Gamma$ in eq.~\eqref{eq:gammaemi} applies only to long nonrelativistic strings, which make up at any time a constant $80\%$ fraction of the string network~\cite{Gorghetto:2018myk}. As explained in Appendix~\ref{app:simscal}, the ratio $(\Gamma_a+\Gamma_r)/\Gamma$ reproduces this percentage remarkably well.} This also (indirectly) shows that $\mu_{\rm eff}$ is well reproduced by the theoretical expectation $\mu_{\rm th}$ of eq.~\eqref{muth} and is growing logarithmically as expected, so we can use the latter expression in the analysis that follows. 
Although this check can be performed only at the small logs available in simulations, given the theoretical discussion of Section~\ref{ss:Review}, eqs.~\eqref{muth} and~\eqref{eq:gammaemi} will hold also at large log.\footnote{\label{ft1}As also explained in Appendix~\ref{app:simscal}, eq.~\eqref{muth} strictly speaking holds for nonrelativistic strings (since there is no boost factor), but any corrections from this are absorbed in the definition of $r$.}

Having confirmed the validity of $\Gamma$ and $\mu_{\rm th}$, we are  ready to study the GW emission $\Gamma_g$ and extract the coefficient $r$. As discussed, we do so using $r$'s relation to $\Gamma_g$ and $\Gamma_a$ of eq.~\eqref{gammagw}, i.e. $r=f_a^2\Gamma_g/(\Gamma_a G \mu_{\rm eff}^2)$.  
This leads to a small ambiguity in determination of $r$ at small log, because the small (and decreasing) proportion of energy emitted into radial modes  
is by construction not captured by the Nambu--Goto description, and could be included along with $\Gamma_a$ in eq.~\eqref{gammagw}. Since the radial modes take a proportion of the energy that would otherwise go into axions, it is natural to expect that the inclusion of $\Gamma_r$ leads to the quantity conserved during scaling.  
 In the following we therefore consider $r_{\rm sim}\equiv f_a^2\Gamma_g/(\Gamma G \mu_{\rm th}^2)$, where $\Gamma$ and $\mu_{\rm th}$ are calculated from eqs.~\eqref{muth} and~\eqref{eq:gammaemi} with the same constant value of $\eta$ mentioned before (indeed $\Gamma$ reproduces $\Gamma_a+\Gamma_r$ measured in simulations), while $\Gamma_g$ is evaluate in the simulations. 
Note that using the theoretical $\Gamma$ in the extraction of $r$ instead of the result of $\Gamma_a+\Gamma_r$ from simulations reduces possible systematic uncertainties related to fluctuations of the last two quantities, which especially at small logs are affected by parametric resonance effects between axion and radial modes especially for the physical system.\footnote{On the other hand $\Gamma_g$ fluctuates far less since the coupling of the GWs to the radial modes is much weaker.}

\begin{figure}[t]
	\begin{center}
		\includegraphics[width=0.48\textwidth]{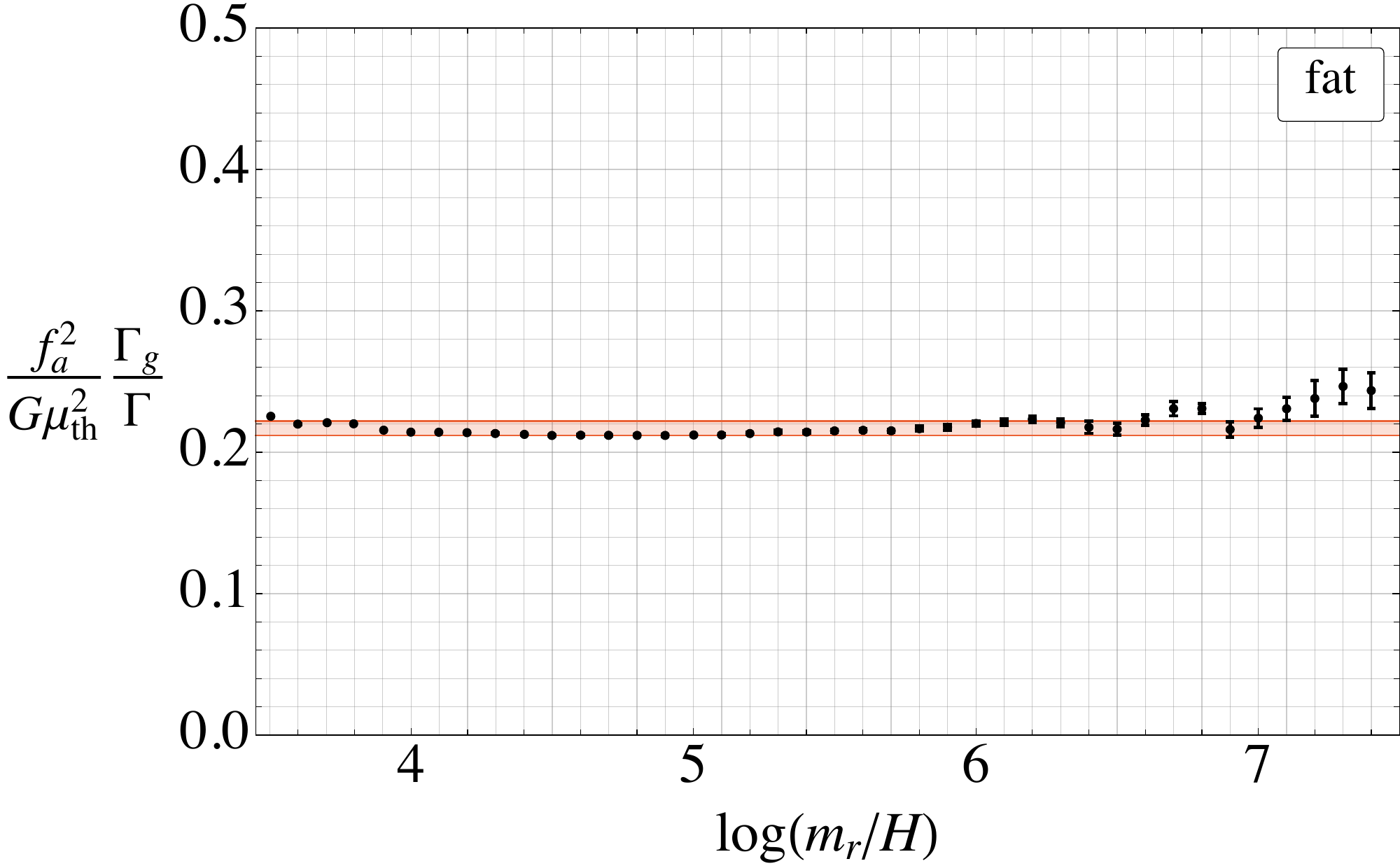}
		~~~				\includegraphics[width=0.48\textwidth]{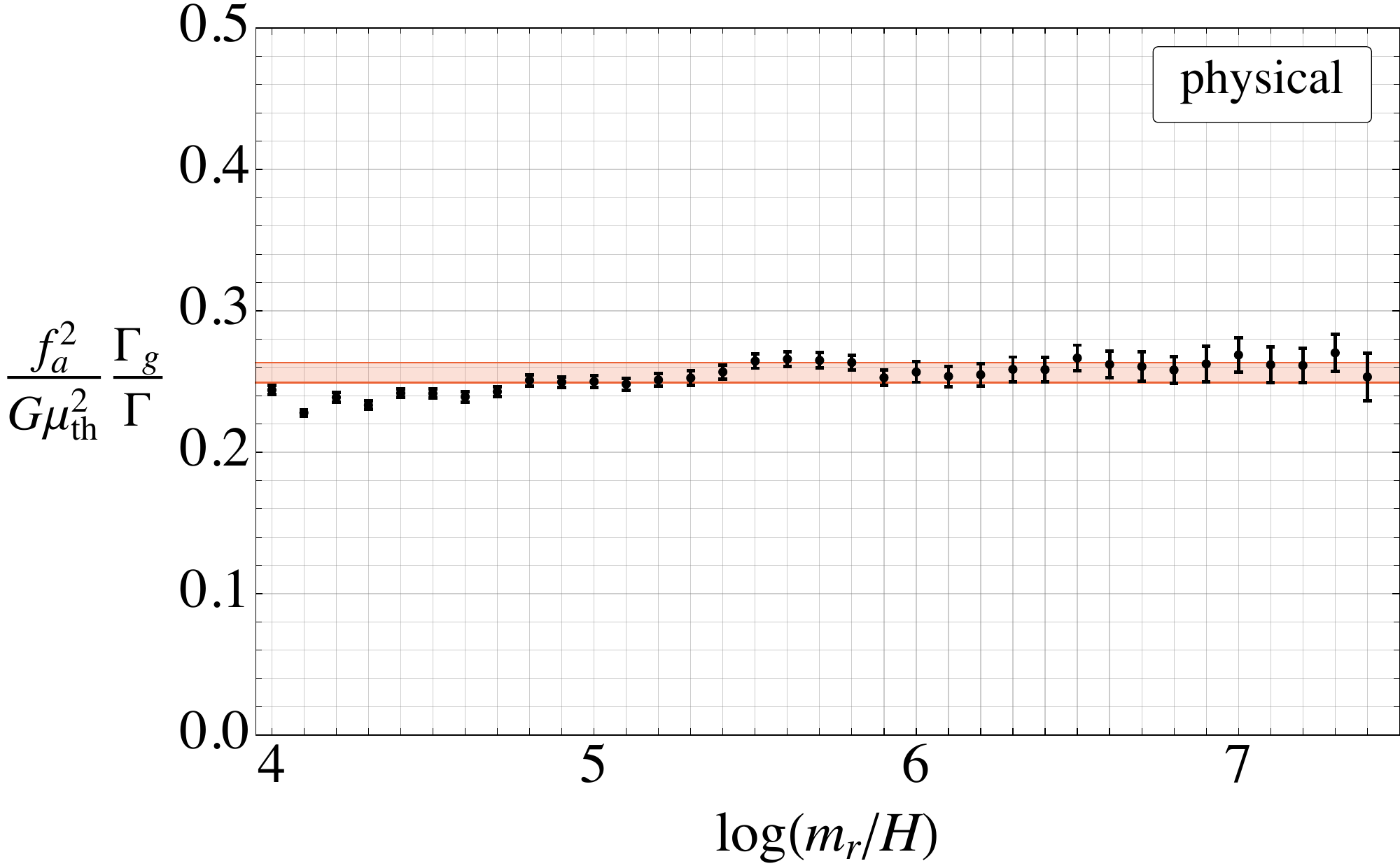}
	\end{center}
	\caption{The evolution of energy density $\Gamma_g$ emitted in GWs per unit time during the scaling regime  normalised to $\Gamma G \mu_{\rm th}^2/f_a^2$, where $\Gamma$ is the total energy emission rate and $\mu_{\rm th}$ is the theoretical expectation for the string tension, for the fat string system (left) and the physical system (right).	
	Error bars on the data points represent the statistical error over a set of $30$ simulations.		 The constant value of this ratio is in agreement with the theoretical expectation of Section~\ref{ss:theory}, and indicates that the relative emission to GWs grows as $\Gamma_g / \Gamma_a \propto \log^2(m_r/H)$ at large log. The red bands indicate the uncertainty we assign to the value of the ratio extracted from the data. 		 
		\label{fig:rsim}} 
\end{figure} 

In Figure~\ref{fig:rsim} we plot the time evolution of $r_{\rm sim}$ for the fat and the physical systems. The uncertainties on the data points represent the statistical error over a set of $30$ simulations with initial conditions with the same initial string density $\xi$. 
In both cases  $r_{\rm sim}$ is of order one and its time-independence is manifest over more than three $e$-foldings. This corresponds to a verification of eq.~\eqref{gammagw} and provides a remarkably consistent picture of the dynamics of the string system. First, it confirms the validity of the effective Nambu--Goto description at least for the emission of GWs, which therefore also allows analytic control of $\Gamma_g$ 
 beyond the range that can be simulated. 
The constant form of $r_{\rm sim}$ even holds well from $\log = 4$ when there is only a mild hierarchy between the string core scale and the Hubble parameter. In fact it is not unreasonable that $r$ is constant even at such early times, since the theoretical expectation in eq.~\eqref{gammagw} does not rely on the string system being in the pure Nambu--Goto limit (namely it should hold even at small log when the strings are strongly coupled to the axion). Moreover, as anticipated in Section~\ref{ss:setup}, the constant value of $r_{\rm sim}$ is strong evidence that the string configuration is self-similar. This in turn confirms that the scaling regime has been achieved in the simulation and that the logarithmic increase in $\xi$ is a part of it.\footnote{If the logarithmic increase were a transient, we would expect the configuration not to be self-similar. This is also consistent with the observation that the fraction of $\xi$ in strings of different lengths remains constant \cite{Gorghetto:2018myk}.} 
Although we cannot exclude a qualitative change in the evolution of $r_{\rm sim}$ after $\log=7.5$, both theory and simulations suggest that such a change is unlikely. The extrapolation of $r$ after $\log=7.5$ therefore seems robust despite the difference in $\log$ between the simulations and the physically relevant system. On the other hand, simulations can never exclude the possibility that there is a small logarithmic running of $r$ but only bound its value, and we comment on this when discussing the uncertainties on our predictions of GW spectra in Section~\ref{ss:presentday}. 

In Figure~\ref{fig:rsim} we also show the fit of this observable with a constant function.  
For the physical system we start the fit at $\log>4.5$, since the first few data points at small log seem to deviate slightly, as the system has not yet fully reached the scaling regime at such early times (see also the not completely linear behaviour of $\xi$ at those log in Figure~\ref{fig:xi}).\footnote{The small disagreement of the last two data points of the fat system is due to a statistical fluctuation.}

We now estimate the uncertainty on $r=f_a^2\Gamma_g/(\Gamma_a G \mu_{\rm eff}^2)$ for physical string network. A first source of uncertainty comes from the precise value of the string tension $\mu_{\rm eff}$ at small log (for which, as mentioned, the value of $\eta$ is relevant), which feeds into the value of $r$ extracted from $\Gamma_g$ measured in simulations. As mentioned, $\mu_{\rm eff}$ is well reproduced by $\mu_{\rm th}$ for $\eta_{\rm phys}\simeq 0.20$. However, given the large fluctuations of $\Gamma_a+\Gamma_r$ in the physical system, values of $\eta_{\rm phys}=0.1\div0.3$ still reproduce  $\Gamma=\Gamma_a+\Gamma_r$ acceptably well (this range is obtained by varying $\eta$ such that $(\Gamma_a+\Gamma_r)/\Gamma$ remains approximately constant, see Figure~\ref{fig:Gammaeta} and Appendix~\ref{app:simscal}). This  translates into an uncertainty on $r_{\rm sim}$ in the range $0.17\div0.34$ (estimated by evaluating $r_{\rm sim}$ at the final time for $\eta_{\rm phys}$ in the range $0.1\div0.3$, see Appendix~\ref{app:simscal} and Figure~\ref{fig:reta} for more details). 
Moreover, the previously mentioned ambiguity in using $\Gamma$ or $\Gamma_a$ introduces a theoretical uncertainty in the value of $r$  (although $r_{\rm sim}$ seems to be the conserved quantity). We can conservatively quantify the associated error as about $10\%$, corresponding to the contribution that $\Gamma_r$ provides provide to $\Gamma$ at the largest simulated log of Figure~\ref{fig:rsim}. The actual statistical uncertainty on data of $\Gamma_g$ is negligible with respect to the mentioned ones. As a result, we obtain a value for $r$ with a conservative error estimate of 
\beq\label{eq:rr}
r = 0.26(11) \ .
\eeq
Since $r$ is of order one, as expected for $f_a \ll M_{\rm P}$ only a small fraction of energy is emitted in GWs. Moreover, as well as qualitative agreement, in Appendix~\ref{app:back} we show that our value of $r$ is quantitatively consistent with the decrease in axion energy, for a fixed $f_a/M_{\rm{P}}$, when backreaction is included.

\begin{figure}[t]
	\begin{center}
		\includegraphics[width=0.485\textwidth]{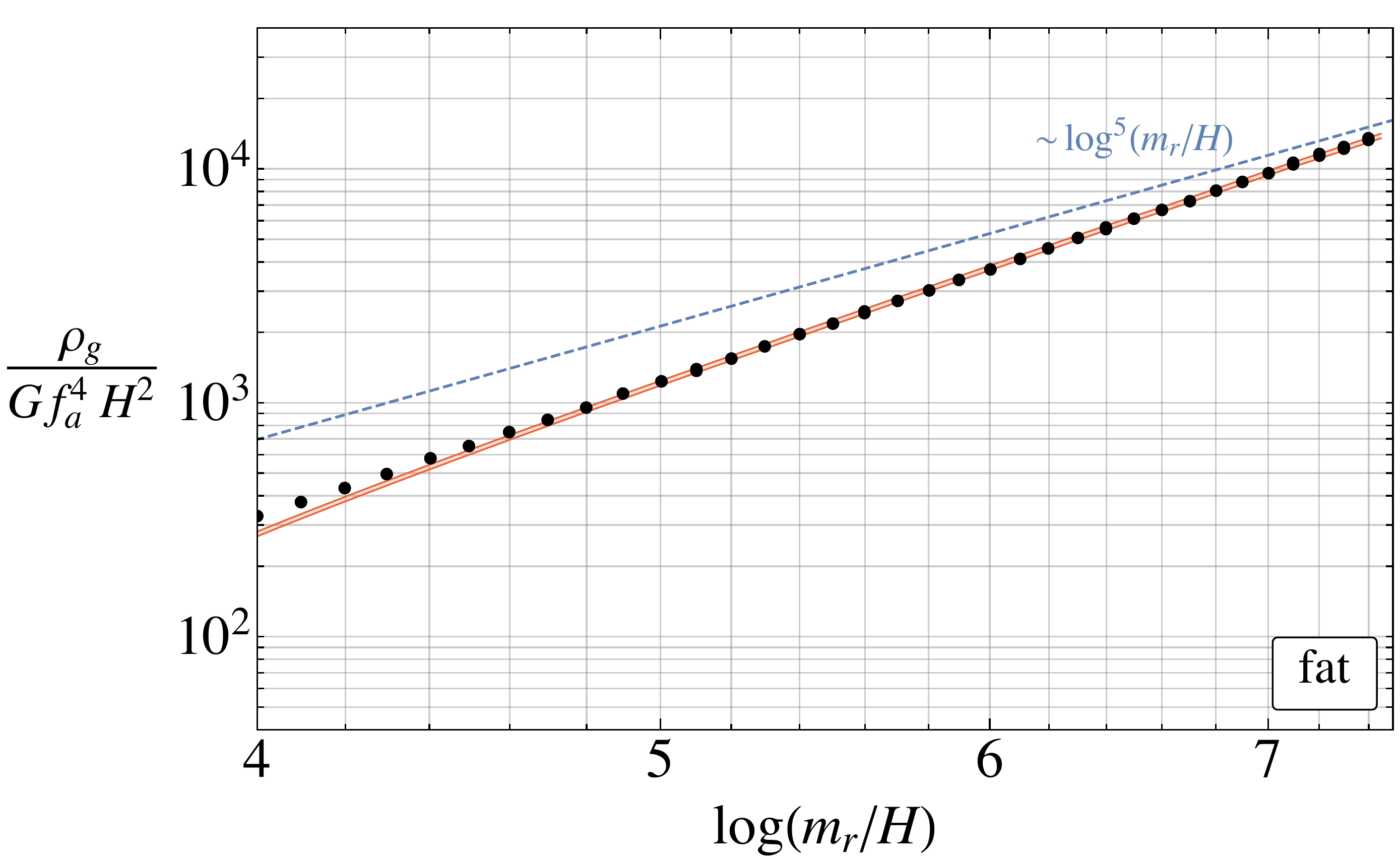}
		~				\includegraphics[width=0.485\textwidth]{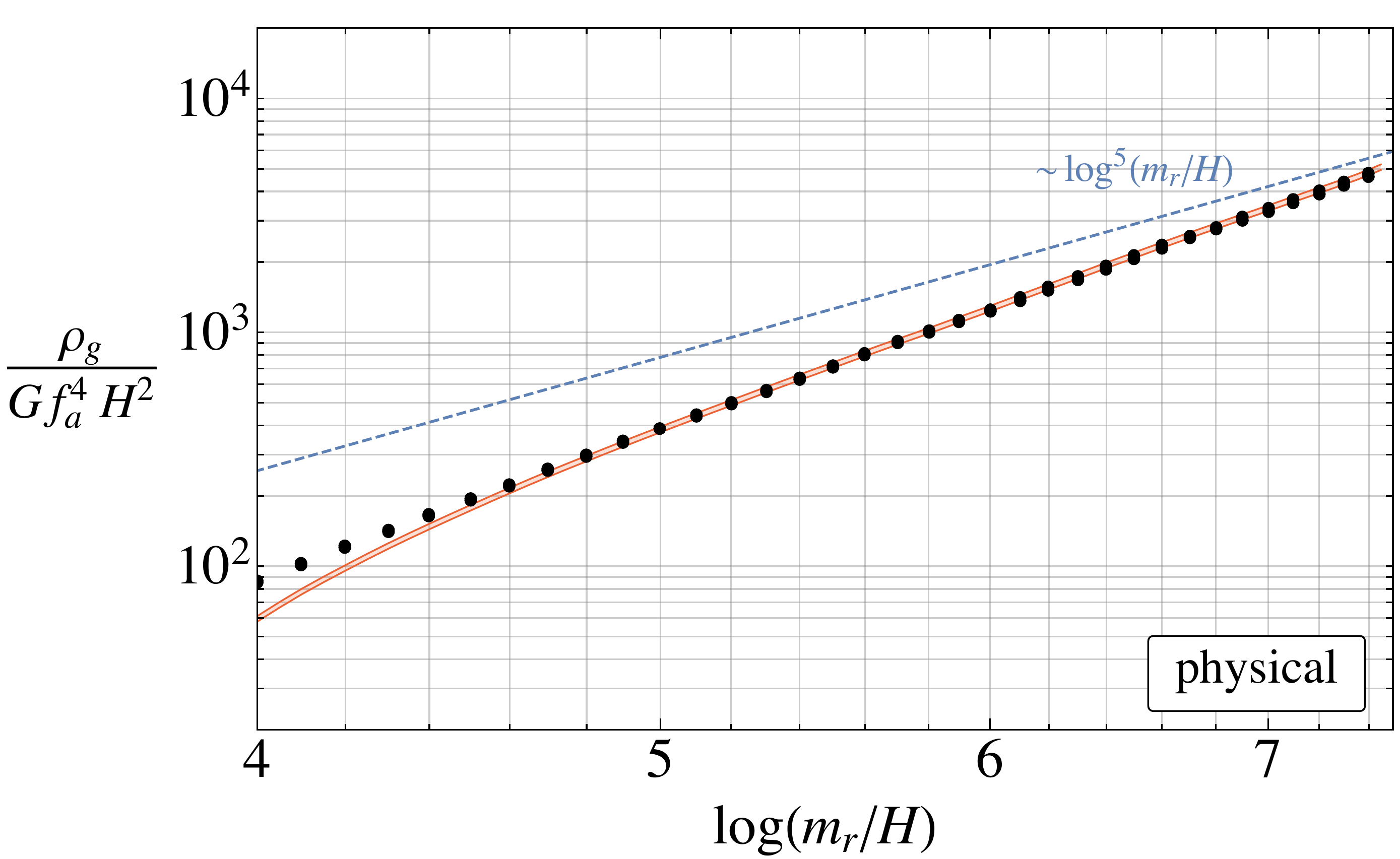}
	\end{center}
	\caption{The evolution of the total energy density in GWs $\rho_g$	(black points, with statistical errors smaller than the plotted data points), and the theoretical prediction for this quantity with the range of $r$ extracted from Figure~\ref{fig:rsim} (red bands). 		
		The data are in full agreement with the predictions. We also plot $\rho_g\propto H^2\log^5$, which is valid at $\log \gg 1$, and which the  prediction and data asymptote to.	\label{fig:rhog}} 
\end{figure} 

In Figure~\ref{fig:rhog} we also show the evolution of the total GW energy density $\rho_g$ measured in the simulations (black points). In the same plot we show the theoretical prediction for this quantity, i.e. $\rho_g=\int_{t_1}^tdt'\Gamma_g'(R'/R)^4$, with $\Gamma_g=r G\mu_{\rm th}^2\Gamma/f_a^2$ where the constant $r$ (together with its error) is extracted from Figure~\ref{fig:rsim} and $\Gamma$ is as in eq.~\eqref{eq:gammaemi} and is evaluated as before (in doing the integration we take $t_1$ such that $\log(m_r/H_1)=3.5$). As expected given Figure~\ref{fig:rsim}, the agreement between the prediction and the data is excellent. We also note that the data approach the expected large log behaviour $\rho_g\propto H^2\log^5$,  discussed in Section~\ref{ss:setup}.

\begin{figure}[t] 
	\begin{center}
		\includegraphics[width=0.445\textwidth]{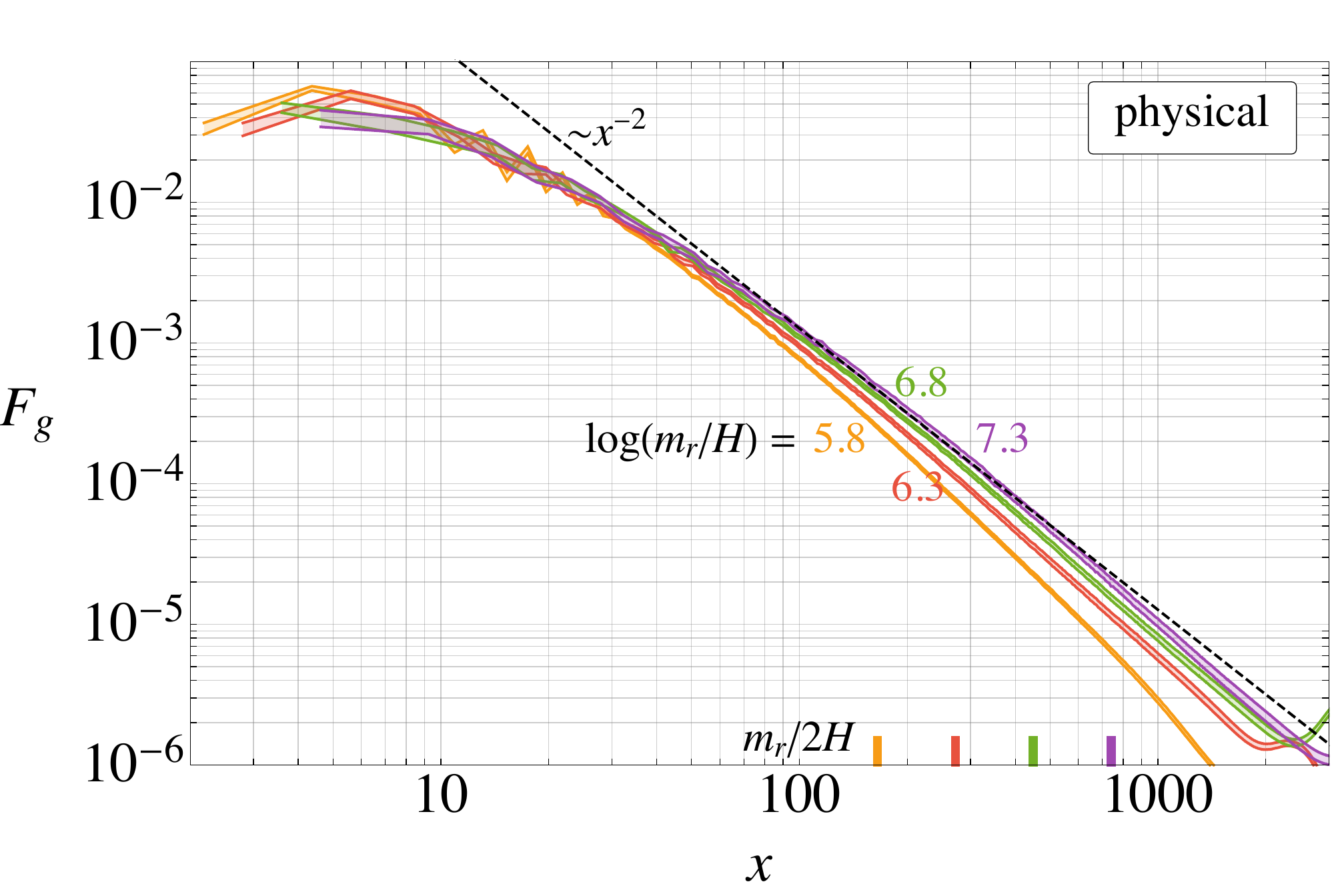}
		\qquad				\includegraphics[width=0.5\textwidth]{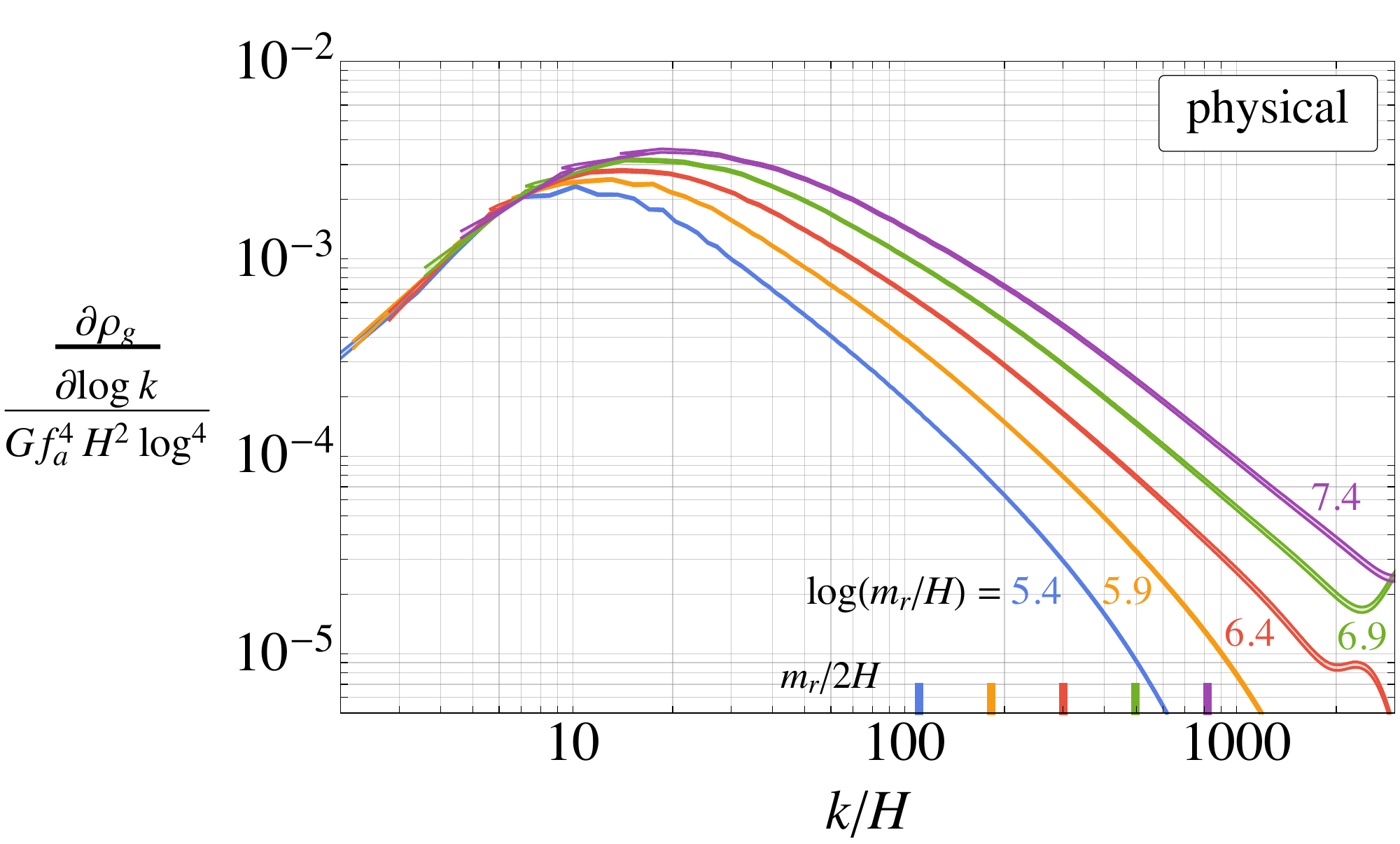}
	\end{center}
	\caption{\emph{Left:} The momentum distribution $F_g[x,y]$ of the GWs emitted instantaneously from the string network, as a function of their physical momentum normalised to Hubble, $x=k/H$. Different lines indicate different times i.e. changing $y=m_r/H$, which are labelled by $\log(m_r/H)$. At all times the distribution is dominated by (IR) momenta of order Hubble, and decays as $x^{-2}$ at higher momenta. \emph{Right:} The evolution of total spectrum of GWs produced by the network, plotted at different times. We factor out the expected time dependence of the amplitude $\propto H^2 \log^4$.
	\label{fig:Fg} } 
\end{figure}

The remaining input required to calculate the GW spectrum is the momentum distribution of the instantaneous GW emission $F_g$.  
It is straightforward to extract this from simulations, and results for $F_g[x,y]$ are plotted for the physical system in Figure~\ref{fig:Fg} (left) at different times, i.e. different $y=m_r/H$ (labelled by $\log(m_r/H)$). As expected, the distribution has an IR cutoff corresponding to $x_0\sim 2\pi$ and a UV cutoff at around the string core scale, corresponding to $x\sim m_r/(2H)$, which are the same as for the axion spectrum.\footnote{The value $k/H\simeq 2\pi$ is also motivated since for such momentum the wavelength equals the Hubble distance.} The spectrum has a somewhat broad peak around the IR cutoff, from $x\simeq 5$ to $20$. Above this, an intermediate power law is visible and compatible with $q=2$, which implies an IR dominated emission. Thus, the numerical results fully confirm (at least at small log) our assumptions of Section~\ref{ss:setup}.  

The parameters $x_0$ and $q$ appear to be time-independent in the range of log accessible to simulations. This can be seen from the fact that the $F_g$ overlap for all $y$ (we carry out a more detailed analysis in Appendix~\ref{app:simR}). It is interesting to note that, although the IR and UV cutoffs of $F_g$ are similar to those of the corresponding  axion instantaneous emission spectrum $F_a$, the spectral indices of the two, $q$ and $q_a$ respectively, are dramatically different. For the axion $q_a$ increases with log (from $q_a=0.75$ to $0.95$ between $\log=6$ and $8$ indicating a UV dominated spectrum that is gradually becoming IR dominated).\footnote{This could be due to the fact that the coupling of the radial modes to axions is much larger than the coupling of radial modes to GWs. Therefore the radial modes which are partly emitted by the strings at small log could produce energetic axions~\cite{moreaxions}, contributing to the UV part of the axion spectrum (but not to the GW one). The emission of radial modes diminishes logarithmically, and the spectrum therefore decreases in the UV.}  We also note that an emission with $q=2$ is characteristic of kink-kink collisions~\cite{Auclair:2019wcv}, as opposed to cusps and kinks which instead provide $q=4/3$ and $q=5/3$ respectively. Although this could be an indication that kink-kink collisions are the dominant source of the emission, our results do not rely on the modelling of the particular process sourcing the GWs, as the evolution of the field equations captures the full dynamics of the system. 

Despite these results, we cannot exclude a slow logarithmic running of $x_0$ and $q$. However, as discussed in Section~\ref{ss:setup}, a slow increase in $q$ would have little effect on the integrated GW spectrum (and a decrease in $q$ until $q<1$ would be extremely surprising). Likewise, a slow change in $x_0$ would only have a very minor effect, so we do not analyse this possibility in detail. Consequently, we can safely assume that the form of $F_g$ is preserved also at large $\log$.

In Figure~\ref{fig:Fg} (right) we also show the total GW spectrum $\partial\rho_g/\partial \log k$ at different times as a function of the physical momentum normalised to Hubble. The general features derived in Section~\ref{ss:setup} are reproduced. However, due to the small final $\log$, the approximately scale invariant region $x_0H\lesssim k\lesssim x_0\sqrt{HH_1}$ of eq.~\eqref{totalspectrum} corresponds only to a small portion of the spectrum.  Indeed, the first nontrivial emission in the scaling regime happens around $\log{(m_r/H_1)}\sim4$ and therefore eq.~\eqref{totalspectrum} holds in the restricted range $x_0\lesssim k/H\lesssim x_0\sqrt{HH_1}/H\sim40\div50$ at the final time $\log=7.4$.\footnote{$H_1$ corresponds to when the UV and IR cutoffs of the spectrum $x_0H$ and $m_r/2$ coincide.} Nevertheless,  the key result from eq.~\eqref{totalspectrum} that at fixed $k/H\sim x_0$ the spectrum (once the leading $H^2$ is factored out) grows proportionally to $\log^4(m_r/H)$ is matched well in Figure~\ref{fig:Fg} (right), which is of course consistent with the extrapolation to large log being essential (a version of the plot without the $\log^4$ factored out is given in Figure~\ref{fig:GWnolog} of Appendix~\ref{app:simR}).\footnote{At small $\log$, $\partial\rho_a/\partial\log k$ actually grows faster than $\log^4$ due to the subleading $\log$ corrections of $\Gamma_g$ in eq.~\eqref{gammagw}, and indeed at $k/H\sim x_0$ the spectrum in Figure~\ref{fig:Fg} (right) shows a slight increase.}  Finally we observe that $\partial\rho_g/ \partial k\propto k^3$ for $k\lesssim x_0H$ (this is the same behaviour as $F_g$ at small $x$), while  $\partial\rho_g/ \partial k\propto k^{q-1}$ at momenta higher than $x_0\sqrt{HH_1}$. 
 In Appendix~\ref{app:sim} we show the results for $F_g$ and the total spectrum for the fat string system, which are qualitatively, and even quantitatively, similar to those of the physical system.

\subsection{The GW Spectrum Today} \label{ss:presentday}

The present day remnant of the GW spectrum from the scaling regime can be straightforwardly computed by combining the theoretical discussion of Section~\ref{ss:setup} and the results of Section~\ref{ss:gwsim}. For the sake of definiteness for now we assume a temperature-independent axion mass, and we discuss the temperature-dependent case later.

As mentioned, the scaling regime ends (approximately) when $H= m_a$, at which time the network is destroyed and GWs stop being produced at the rate in eq.~\eqref{gammagw}. The GW spectrum at $H=m_a$ follows from eq.~\eqref{eq:drhogdk}.  From $H=m_a$ on the GWs redshift freely until today, and their contribution to the present day total energy density of the Universe is $\Omega_{\rm gw}\equiv \rho_g/\rho_c$, where $\rho_c$ is the critical density.

\begin{figure}[t]
	\begin{center}
		\includegraphics[width=0.8\textwidth]{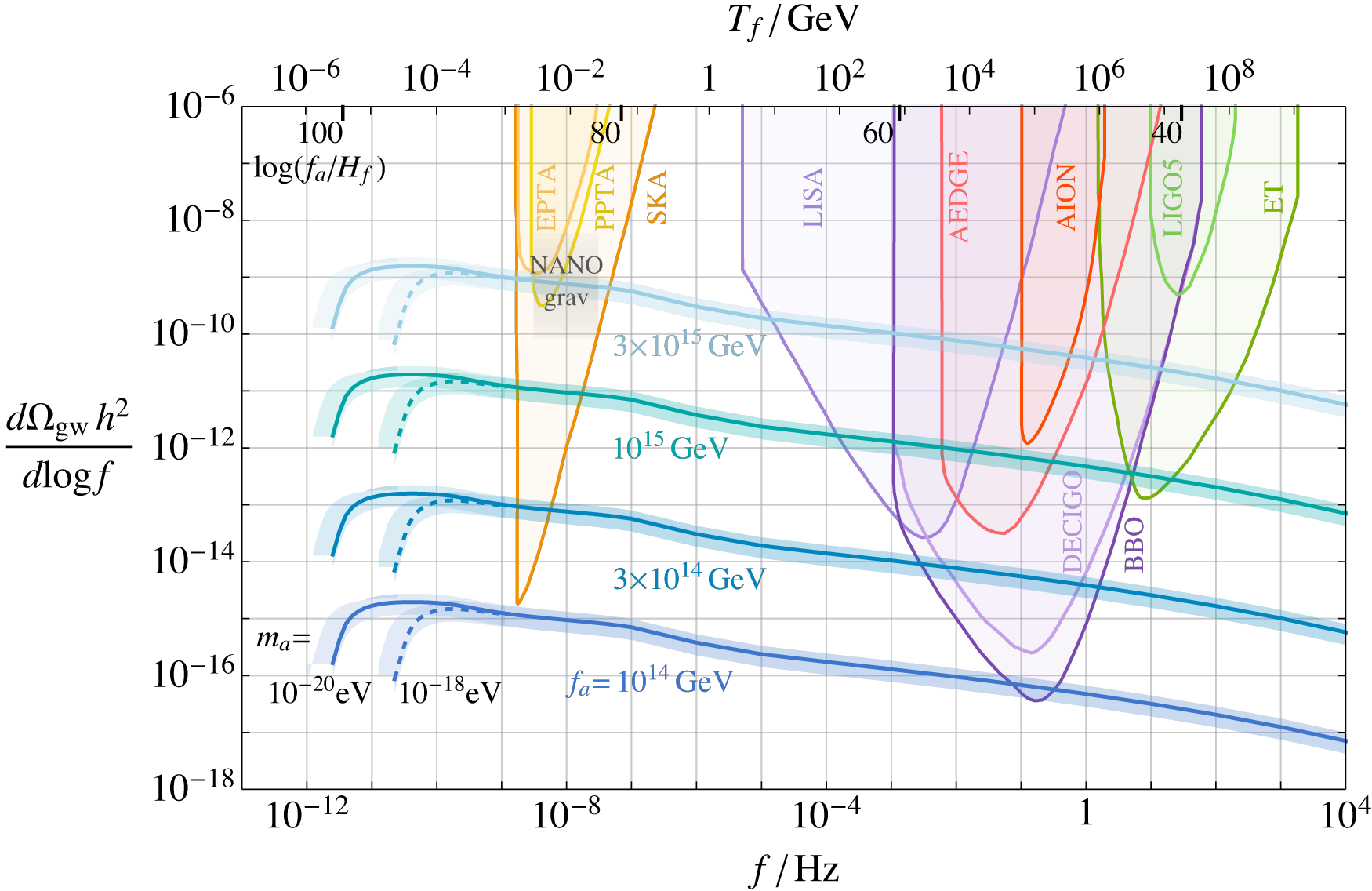}
	\end{center}
	\caption{
The contribution to the energy density of the Universe today from the GWs emitted by axion strings during the scaling regime, as a function of their present day frequency $f$. Different lines correspond to different values of the axion decay constant $f_a$ and mass $m_a$. We plot values of $m_a$ that are compatible with current constraints on the post-inflationary scenario. Values of $f_a\lesssim 10^{15}\GeV$ are possible, while larger values are  in tension with isocurvature and dark radiation bounds. GWs with higher frequencies are produced at earlier times, and we indicate the temperature of the Universe when GWs of a particular frequency are dominantly produced,   $T_f$, as well as the corresponding value of the log at this time $\log(m_r/H_f)$. All the GWs are emitted in radiation domination and the network decays before matter-radiation equality.
		  \label{fig:OmegaGW} } 
\end{figure}

In Figure~\ref{fig:OmegaGW} we plot the GW spectrum  $d \Omega_{\rm gw}/d \log f$ (from a numerical integration of eq.~\eqref{eq:drhogdk}), where $f$ is the frequency. We use the value of $r$ from eq.~\eqref{eq:rr} to evaluate $\Gamma_g$ in eq.~\eqref{gammagw}, and a functional form of $F_g$ that fits Figure~\ref{fig:Fg} (left) (see Appendix~\ref{app:strings} for more details). Results are shown for  different choices of axion decay constant and mass that are not excluded by other cosmological constraints (derived in Section~\ref{sec:Constraints}). For an axion that produces GWs in the detectable frequency range, these constraints require  $f_a\lesssim 10^{15} \GeV$ and $10^{-28} \eV\lesssim m_a\lesssim 10^{-17} \eV$ (and as a result the network is destroyed before matter-radiation equality).\footnote{This confirms that the condition for gravitational backreaction to be negligible $G\mu_{\rm eff}/f_a^2\ll1$ is satisfied.} We also show $f_a=3\times10^{15}\GeV$, which is likely to be in tension with dark radiation bounds, see Section~\ref{ss:darkradiation}. 

In this plot we assume that the Hubble parameter $H_1$ when the scaling regime starts is sufficiently large that the GW spectrum is in the approximately scale invariant region for the entire range of frequencies shown. Later in this Section we show that this assumption is highly plausible.  We also assume a standard cosmological history with radiation domination up to high temperatures and only the particle content of the SM plus the axion, and that $m_r=f_a$. We describe the errors that we include on the GW spectra and other sources of uncertainty in detail at the end of this Section.

A useful (and for most purposes accurate) analytic approximation for $\Omega_{\rm gw}$ in the $\log\gg1$ limit is derived from eq.~\eqref{totalspectrum} in Appendix~\ref{app:simscal} (see eq.~\eqref{eq:Omegaapproxanalytic}) and reads 
\small
 \begin{equation}\label{eq:Omegaapprox}
\frac{d \Omega_{\rm gw} h^2}{d \log f}  \simeq   0.80 \times 10^{-15} \left(\frac{c_1}{0.24}\right) \left(\frac{r}{0.26}\right) \left(\frac{f_a}{10^{14} {\rm GeV}}\right)^4 \left(\frac{10}{g_f} \right)^{\frac13} \\  \left\{1+ 0.12\log\left[\left(\frac{m_r}{10^{14} {\rm GeV}}\right)\left(\frac{10^{-8} {\rm Hz}}{f}\right)^2\right] \right\}^4 ,
\end{equation}
\normalsize
where $g_f$ is the effective number of degrees of freedom in thermal equilibrium at the temperature $T_f$, when most of the GWs with today's frequency $f$ are emitted.\footnote{More precisely, this temperature is defined by $x_0 H(T_f)\equiv f R(T_0)/R(T_f)$.} Eq.~\eqref{eq:Omegaapprox} holds in the frequency range $3\times10^{-12} (m_a/10^{-20} {\rm eV})^{1/2}\lesssim f/{\rm Hz}\lesssim 10^3(H_1/\GeV)^{1/2}$. This corresponds to the extremes in eq.~\eqref{totalspectrum} evaluated at $H=m_a$, redshifted to today. As stressed in Section~\ref{ss:setup}, at lower and higher frequencies than these IR and UV cutoffs, $d\Omega_{\rm gw}/d\log k$ is suppressed as $f^3$ and $f^{q-1}$ respectively.\footnote{The dependence on $f^3$ of the super-horizon modes is fixed by causality after such modes starts oscillating~\cite{Caprini:2009fx}.}

Several comments are in order. First, we observe that $f_a$ controls the overall amplitude of the spectrum, since it ultimately determines both the energy density of the string network and also the efficiency at which this is emitted into GWs. 
Conversely, $m_a$  only affects the position of the IR-cutoff $\propto m_a^{1/2}$ of the approximately scale-invariant part of $\Omega_{\rm gw}$ (which is $m_a$ independent). This is not surprising given that $m_a$ is unimportant during scaling, and its only role is in determining when the network is destroyed. The resulting IR cutoff is visible in Figure~\ref{fig:OmegaGW}.

The $\log^4$ dependence of the frequency is evident in Figure~\ref{fig:OmegaGW} and eq.~\eqref{eq:Omegaapprox}. In Figure~\ref{fig:OmegaGW}, we indicated the value of the log when most of the GWs with frequency $f$ are emitted, $\log_f\equiv\log(m_r/H_f)$. Since this varies by more than a factor of $2$ over the frequencies of observational interest (in the range $\log_f=30\div90$), there is a substantial effect on the spectral shape as well as the amplitude. 
We also note that the GWs in most of the observable range are emitted when  $\xi=10\div20$ from eq.~\eqref{eq:xivslog}. 
 As expected, the spectral shape is also modified by the changes in number of relativistic degrees of freedom in the Universe. In particular, this has an effect at frequencies that are dominantly emitted at temperatures $T_f$ (shown on the upper axis) at which such changes occur (the largest effect is around $T\simeq 100\MeV$ when a large number of the degrees of freedom decouple). 
As a result the spectrum at higher frequencies is suppressed by more that the $\log^4$ factor, since these are emitted at earlier times when $g_f$ is larger leading to increased expansion of the Universe. However, such effects are fairly weak, as $\Omega_{\rm gw}$ depends only on $g_f^{1/3}$.

In Figure~\ref{fig:OmegaGW} we also show the projected sensitivity curves for ongoing and proposed GW searches (EPTA~\cite{Lentati:2015qwp}, PPT~\cite{Shannon:2015ect}, SKA~\cite{janssen2014gravitational}, LISA~\cite{amaroseoane2017laser}, DECIGO/BBO~\cite{Yagi:2011wg}, AEDGE~\cite{Bertoldi:2019tck}, AION\cite{Badurina:2019hst}, LIGO~\cite{Aasi:2014mqd}, and ET~\cite{Hild:2010id,Punturo:2010zz}), as well as an extremely tentative possible signal by NANOgrav \cite{Arzoumanian:2020vkk}, which we comment on in the Conclusions. In particular, we plot the power-law-integrated sensitivity curves \cite{Thrane:2013oya,Moore:2014lga,Romano:2016dpx} as derived in \cite{Gouttenoire:2019kij}. Partly due to the enhancement of the signal at low frequencies, the near future detection prospects are best at Pulsar-Timing Arrays such as SKA, which scan the lowest frequencies and could be sensitive to all $f_a \gtrsim 10^{14}~\GeV$. Detection is also possible at space-based interferometers: although LISA could be sensitive only to $f_a \gtrsim 5\times10^{14}~\GeV$, its proposed successors could explore lower values of $f_a$. The wide range of axion masses and decay constants that lead to a measurable GW signal motivates the effort to develop such experiments.

As mentioned, complementary constraints require $m_a\lesssim 10^{-17}\eV$, and therefore in case of a detection the actual value of $m_a$ will not be inferred. This is because, for this mass range, all the detectable frequencies are emitted deep inside the scaling regime, so the IR cutoff of $\Omega_{\rm gw}$ is unobservable. Additionally, any temperature-dependence of the mass -- as long as is monotonically decreasing -- does not affect the detectable GWs, as it can only modify lower frequencies, emitted when the mass is relevant. 
We note that if the string network is destroyed before $T\simeq10^6$ GeV (e.g. for a heavy axion mass, $m_a\gtrsim$ MeV, which is still allowed by the observational constraints in Section~\ref{sec:Constraints}), the IR cutoff frequency is so large that the spectrum does not extend down to observable frequencies.

Although neglected in this discussion,  after $H=m_a$ additional GWs will be emitted by domain walls, which will supplement those from the scaling regime. As discussed in Appendix~\ref{app:massGW},  one calculable contribution to these GWs has frequencies (and amplitude) of the same order as that from the last $e$-folding of the scaling regime. 
Therefore, although it will modify the shape of the IR-cutoff of the GW spectrum from the scaling regime,  this contribution at least will not be observationally  relevant.

\subsubsection*{Uncertainties on the Spectrum}

Given its experimental importance, an understanding of the possible sources of uncertainty on the GW prediction is crucial. The error bands on the GW spectra plotted in Figure~\ref{fig:OmegaGW} are obtained by combining the uncertainties on the coefficient of the growth of $\xi$ (i.e. $c_1$ of eq.~\eqref{eq:xivslog}), on $r$, and by varying $x_0$ in the range $5\div10$ (which only has a visible effect on the location of the IR cutoff). We do not think it would be fair to associate a sharp numerical uncertainty to the extrapolations necessary to reconstruct the scaling regime at large log. Instead, we now summarise the assumptions needed to obtain eq.~\eqref{eq:Omegaapprox} and Figure~\ref{fig:OmegaGW}, and the corresponding possible uncertainties. In all cases we have made the most conservative extrapolations possible, and taken together a deviation of more than a factor of $2\div4$ from our predictions would be surprising.

\begin{itemize}
\item We assumed that $\xi$ continues to grow logarithmically as in eq.~\eqref{eq:xivslog} during the scaling regime beyond the range of simulations. While such an increase has not yet found a mathematical proof, it has been numerically demonstrated in simulations at $\log\lesssim8$, providing the best fit of the data (which disfavours any function that saturates soon after $\log=8$, as observed in~\cite{Gorghetto:2018myk,moreaxions} and evident in Figure~\ref{fig:xi} of Appendix~\ref{app:simR}).\footnote{This behaviour has been confirmed independently  \cite{Fleury:2015aca,Kawasaki:2018bzv,Buschmann:2019icd,Vaquero:2018tib}. An enhanced value of $\xi$ is also suggested by Nambu--Goto string simulations \cite{Bennett:1987vf,Allen:1990tv,Bennett:1989yp}, and in a system where the tension of the strings is increased via additional degrees of freedom, \cite{Klaer:2017qhr,Klaer:2019fxc}. Meanwhile \cite{Hindmarsh:2019csc} claims that simulation results indicate $\xi$ approaches a constant value $\sim 1$, however their results are fully consistent with a logarithmic increase and are not well fit by the saturating function the authors propose (in particular, the function suggested by the authors fits for less than a single $e$-folding, an interval over which practically any function can be fit).} 
Such a growth is theoretically plausible given the logarithmic sensitivity of the system to $m_r/H$, and given the excellent fit over the range of logs that can be simulated it is the most conservative assumption for the late time behaviour of $\xi$. The resulting string densities at the times relevant to the emission of observable GWs are $\xi = 10\div 20$. This is a factor of $10$ larger than is reached in simulations, correspondingly increasing the energy emitted into GWs, on top of the (much larger) enhancement from $\mu_{\rm eff}$.

Nevertheless, we cannot exclude the possibility that $\xi$ saturates (or its growth accelerates) at log far beyond the reach of simulations. In these cases, the emission would be damped (or increased) proportionally to the value of $\xi$. In particular, a different value of $\xi$ would modify the amplitude in Figure~\ref{fig:OmegaGW} at the frequency $f$ by the factor $\sim\xi/(c_1\log_f)$. Therefore, as long as $\xi$ does not saturate at a value smaller than $5$ (which seems highly unlikely given that such values are obtained in simulations that partly reproduce the dynamics of the system at large log by boosting the string tension \cite{Klaer:2019fxc}), in the case of a saturation the amplitude of $\Omega_{\rm gw}$ would only decrease by a factor of two at LISA frequencies and a factor of four at SKA frequencies. Consequently the uncertainty on $\xi$ does not qualitative change the prospects of detection, strengthening the robustness of our results (e.g.  $f_a \gtrsim 10^{14}\GeV$ remains just about detectable by SKA).

\item We assumed that the energy emission rate into GWs continues to follow the prediction in eq.~\eqref{gammagw} based on the Nambu--Goto strings effective theory, with $r$ constant, throughout the scaling regime. Numerical simulations confirmed this result for $\log\lesssim7.5$ with fixed $r=0.26(11)$, as in eq.~\eqref{eq:rr}. As mentioned, we cannot exclude a small logarithmic dependence in $r$ (for instance, due to a change in the average shape of the string trajectories). However such a running is bounded by Figure~\ref{fig:rsim} and (if present) would most likely give $\Gamma_g$ a  $\log$ dependence that is subleading to that from $\xi$. This possibility therefore only makes a small contribution to the overall uncertainty on $\Omega_{\rm gw}$. 

\item We assumed that the form of the instantaneous emission spectrum of GWs $F_g$ of Figure~\ref{fig:Fg} is preserved during the whole scaling regime, also after $\log\simeq7.5$. This assumption is motivated by the existence of the scaling regime (and $F_g$ is seen to be preserved during the whole simulation range). As noted in Section~\ref{ss:setup}, as long as the IR cutoff $x_0$ does not change exponentially and $q$ does not decreased below 1 (both of which are unlikely), the uncertainties on  $\Omega_{\rm gw}$ from these quantities are negligible.
\end{itemize}

As well as those from our reconstruction of the scaling regime, there are also uncertainties on $\Omega_{\rm gw}$ due to unknown features of the early Universe.
\begin{itemize}
	\item 
	In obtaining Figure~\ref{fig:OmegaGW} we have taken the Hubble parameter $H_1$ when the scaling regime starts sufficiently large that the GWs in the observable frequency range are in the approximately scale invariant part of the spectrum. This is actually a mild assumption requiring just that $H_1 \gtrsim \keV$ (corresponding to a temperature of the order of $100$ TeV), so the UV cutoff of the spectrum is at frequencies $f\gtrsim$ Hz. Values of $H_1$ much larger than this are expected in all of the models that lead to symmetry restoration at the relevant $f_a$, which we discuss in Section~\ref{sec:restore}.  It is also expected that there is sufficient time for the string network to reach the attractor prior to the Universe dropping to the temperature $10^7\GeV$ when the first GWs  in the observable frequency range are emitted. Moreover,  friction on the string due to interactions with the thermal bath, which is relevant at high temperatures  if the axion couples to standard model particles (but we do not include in our analysis of the scaling regime), will be negligible by this point for axion decay constants that lead to observable signals. In particular, the friction is expected to be irrelevant for temperatures $\lesssim 10^{11}\GeV (f_a/10^{14}\GeV)^2$ \cite{Battye:1994au}, which is safely far above those corresponding to observable frequencies (further analysis would be required to determine the temperature at which friction becomes irrelevant precisely, see also   \cite{Garriga:1993gj,Vilenkin:1991zk,Martins:1995tg,Agrawal:2020euj}). 

In contrast, the UV part of the spectrum $f\gtrsim (H_1/\keV)^{1/2}$~Hz  depends on how (and when) the network formed and reached the scaling regime, and possibly gets additional contributions e.g. from the U(1) phase transition. However, in practice this is not detectable in any motivated model.
\item

In our derivation we also assumed that the Universe is in radiation domination up to high temperatures (say, $10^{8}$ GeV, as show in Figure~\ref{fig:OmegaGW}). Different cosmological scenarios, e.g. matter domination or kination, would drastically modify the spectrum's shape. Indeed, it is easy to show from eq.~\eqref{eq:drhogdk} that, if $R\propto t^\alpha$, $\partial\rho_g/\partial\log k$ is proportional to a negative (positive) power of $k$ depending on $\alpha>1/2$ ($\alpha<1/2$) \cite{Giovannini:1998bp,Riazuelo:2000fc,Cui:2018rwi,Ramberg:2019dgi}. An accurate  determination of the spectrum in such scenarios would require recalculating $\xi$ and $r$ from simulations with such $\alpha$.  
\item
Finally, we assumed only SM degrees of freedom. If additional degrees of freedom are present the prediction will be modified as in eq.~\eqref{eq:Omegaapprox}. Indeed, as pointed out in~\cite{Battye:1997ji}, a precise measurement of the spectrum could in principle provide information about $g$ at high temperatures (it is plausible that a significant number of beyond SM degrees of freedom could enter at energies $\gtrsim \TeV$, and the possibility this could be detected by analysing the GW spectrum from strings has been considered in \cite{Cui:2018rwi}). However, given the uncertainties discussed above and the weak dependence of $\Omega_{\rm gw}$ on $g_f$, extracting the number of degrees of freedom would appear to be very challenging.\footnote{Strictly speaking, our results for $\xi$ and $r$ have been obtained from numerical simulations with $R\propto t^{1/2}$, that do not account for the change in the number of degrees of freedom. We expect however a very minimal change in such observables when one takes into account the time-dependence on $R$ from $g$.}
\end{itemize}

\section{Constraints on the Axion Mass and Decay Constant} \label{sec:Constraints}

Having shown that that post-inflationary axions lead to GWs with amplitude and frequency that could be accessible to proposed experiments, we now analyse some other phenomenological features of this scenario.  These provide direct and indirect constraints on $f_a$ and $m_a$, completing our understanding of the axions that could be discovered via GWs, and also giving complementary observational signatures.

In the following we will consider a single axion with a temperature-independent mass $m_a$.\footnote{Although we remain agnostic on how the mass in generated, we observe that this mass could arise from explicit breaking of the axion's shift symmetry in the ultra-violet (UV) theory.}  Axions with $f_a\gtrsim10^{14}$ and  $m_a \lesssim \MeV$ are cosmologically stable regardless of the details of their couplings to the SM~\cite{Arias:2012az}. In Section~\ref{ss:relic} we will see that, given this stability, only ultralight axions are not ruled out by dark matter overproduction for the relevant values of $f_a$, so we focus on this mass range (as described in Appendix~\ref{app:cosmostable}, axions with $m_a\gtrsim$ MeV can decay, but, as mentioned, for such values the scaling regime ends before observable GWs are produced). In Appendix~\ref{app:cosmostable} we show that only the ultra-light mass range is allowed for large $f_a$ also in the case of a temperature-dependent axion mass.

\begin{figure}[t]
	\begin{center}
		\includegraphics[width=0.75\textwidth]{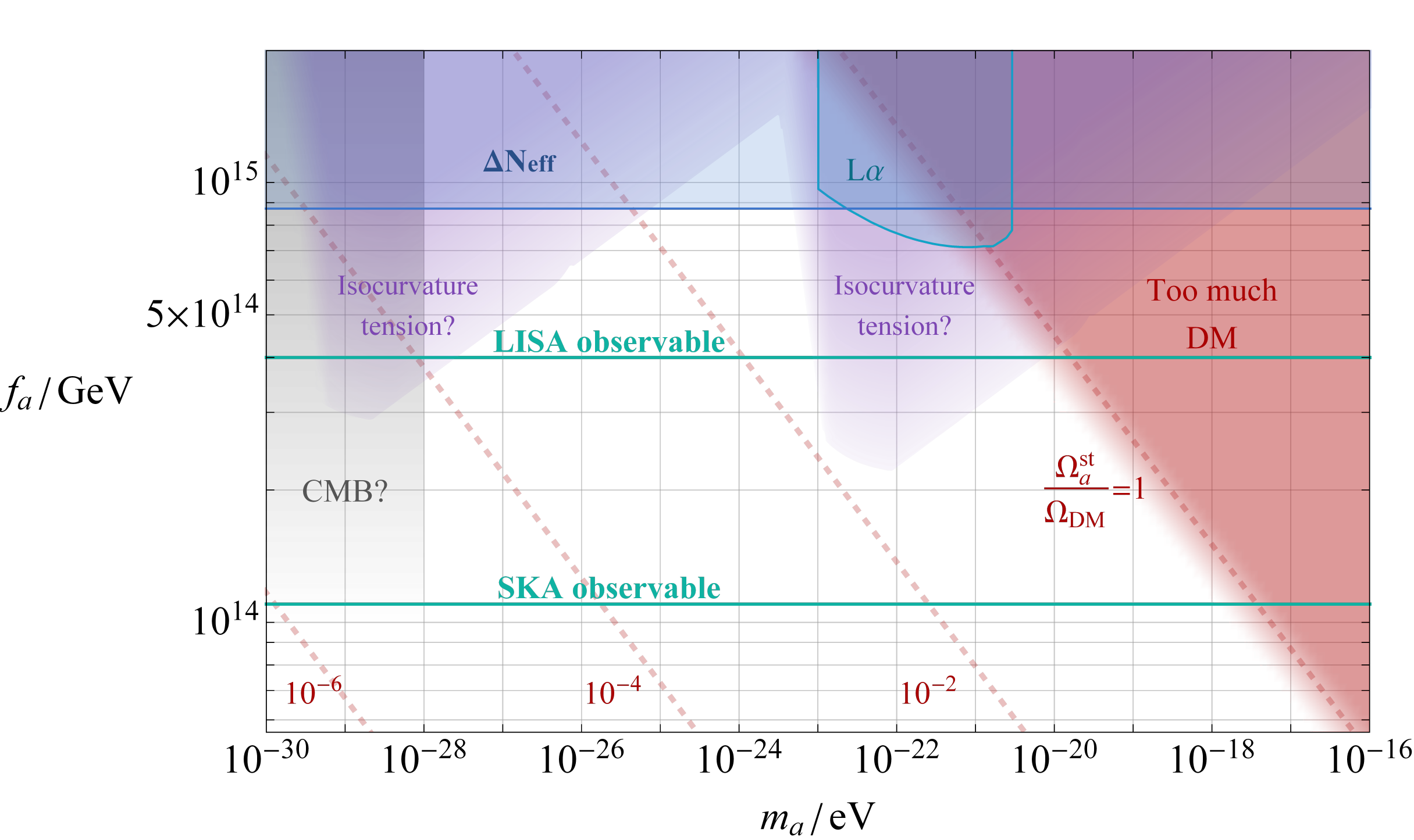}
	\end{center}
	\caption{Constraints on axion mass and decay constant in the post-inflationary scenario, with a temperature independent axion mass.  Limits from dark matter overproduction and dark radiation are shown in red and blue respectively. The bound from isocurvature perturbations is conservative but is still particularly uncertain (see the main text for details). Direct CMB observations rule out extremely light axions with large $f_a$ (in purple), since strings persist at the time of decoupling causing anisotropies. We also identify the parts of parameter space in which axion strings produce a GW spectrum that is detectable by SKA and LISA, and the fraction of DM that axions from the scaling regime comprise.
		\label{fig:param}} 
\end{figure}

In Figure~\ref{fig:param} we summarise the constraints, which we detail in the remainder of this Section. Given their stability, axions form a component of dark matter (potentially overproducing the observed abundance) and those that are relativistic at the time of BBN or decoupling act as dark radiation. The resulting limits are shown in red and blue respectively. DM axions from strings and domain walls have isocurvature perturbations, which are constrained by CMB and Lyman-$\alpha$ observations. As explained in Section~\ref{ss:iso}, it is challenging to determine the resulting limits precisely, however we will derive reasonable conservative bounds (which should still be treated with caution), shown in purple. We also show bounds on the fraction of dark matter that can be ultralight from Lyman-$\alpha$ observations (L$\alpha$).
 Finally, if $m_a$ is small enough, axion strings persist at the epoch of CMB decoupling, and are constrained by CMB observations. In Appendix~\ref{app:otherbounds} we summarise other constraints that are less strong than those shown, e.g. from from black hole superradiance.

Even though these limits are fairly restrictive and become stronger for larger $f_a$ (exactly when the GWs become detectable), there is about one order of magnitude of allowed values of $f_a$ and many orders of magnitude of $m_a$ that provide observable GW signals from strings. 
We note that these constraints do not depend on the possible axion couplings to the SM (which we did not specify). Moreover, although we considered a model with a single axion, the bounds are not expected to significantly change if additional light fields are included and coupled to the axion, so our conclusions apply to generic axion models in the post-inflationary scenario.\footnote{Some models with unusual features (such as the clockwork mechanism) might  avoid these constraints.}

\subsection{Dark Matter} \label{ss:relic}

As discussed in Section~\ref{ss:Review}, during the scaling regime energy is continuously radiated into axions at the rate $\Gamma_a$. Such axions are relativistic during the scaling regime (since $m_a\ll H$), and the majority of them become nonrelativistic soon after the axion potential $V$ becomes relevant, when $H\simeq m_a$.  Since these axions are stable they form a component of dark matter.

The number density of axions during the scaling regime can be obtained following a similar approach as for GWs in Section~\ref{ss:setup}. As mentioned, the momentum distribution of $\Gamma_a$ has the same form as that of $\Gamma_g$ in Figure~\ref{fig:Fg}. However, the spectral index $q_a$ for the axion emission changes in time and, although $q_a<1$ for $\log\lesssim 8$, its extrapolation indicates that $q_a>1$ at $\log\gtrsim 9$, which we will assume in the following (see~\cite{Gorghetto:2018myk,moreaxions} for more details). 
Once $q_a>1$, the axion number density during the scaling regime $n_a^{\rm st}\equiv \int dk/\omega_k\partial\rho_a/\partial k$ is approximately $8H\xi\mu_{\rm eff}/x_{0,a}$ (here $\omega_k^2\equiv k^2+m_a^2$; $\partial\rho_a/\partial k$ is the axion energy density spectrum, defined in the same way as the GW one in eq.~\eqref{eq:drhogdk}; and $x_{0,a}$ is the IR cutoff of the instantaneous axion emission spectrum).  
Consequently, the number density $n_{a,\star}^{\rm st}$ at $H=H_\star$ is enhanced by a factor of $\xi_\star\log_\star\gg1$ with respect to that in the pre-inflationary scenario, which is of order $\theta_0^2H_\star f_a^2$ at this time \cite{Preskill:1982cy} (the subscript `$\star$' refers to quantities evaluated at $H=H_\star\equiv m_a$).

We can obtain a conservative lower bound on the final axion abundance by considering only the axion waves emitted during the scaling regime up to $H_\star$ (i.e. $n_{a,\star}^{\rm st}$ only), which can be reliably determined (following the logic applied to QCD axion strings in \cite{moreaxions}). Additional axions will be produced during the destruction of the string network, however a reliable calculation of this component appears challenging as the system of strings and domain walls cannot be simulated at the physical value of $m_a/m_r$.\footnote{The final number density of axions arising from the axions emitted by the scaling regime is not expected to be affected by more than an order one factor by the presence of domain walls connected to strings \cite{moreaxions}.}

As shown in~\cite{moreaxions}, the waves emitted during the scaling regime redshift relativistically for some time even after $H_\star$, since their kinetic energy  a factor of $\xi_\star\log_\star\gg1$ larger than their potential energy (bounded by $V\lesssim m_a^2f_a^2$), which is therefore negligible.\footnote{The relevant kinetic energy is the one is IR modes, approximately given by $8\pi \xi_\star\log_\star H_\star^2f_a^2\gg m_a^2f_a^2$ at $H_\star$.} The relativistic redshift ends when the potential and kinetic energies become comparable. At this time, the waves experience a nonlinear transient (the main features of which can be understood analytically, up to order 1 coefficients that need to be obtained from numerical simulations). Soon after this the majority of the axions from the scaling regime become nonrelativistic and their number density per comoving volume is conserved. The net effect of the relativistic redshift and the nonlinear transient is a non-conservation of the comoving axion number density between $H_\star$ and the end of the transient. However, unlike the case of a temperature dependent mass, this only amounts to an up to $O(20\%)$ effect for a temperature-independent mass with $\xi_\star\log_\star=3\times10^3$ or smaller (see Appendix~\ref{app:destroy}  for the derivation). Given the much larger uncertainties involved, in what follows we will therefore make the approximation that $n_a^{\rm st}$ is conserved.

Number density conservation from $H_\star$ leads to the (nonrelativistic) axion energy density today $\rho_a^0=m_a (R_\star/R_0)^3n_{a,\star}^{\rm st}$, and the relic abundance of axions from strings during the scaling regime $\Omega_a^{\rm st}\equiv\rho^0_a/\rho_c$ is
\beq \label{eq:relic}
\Omega_a^{\rm st} \simeq 0.1\left(\frac{\xi_\star\log_\star}{3\times 10^3} \right) \left(\frac{f_a}{10^{14}\GeV}\right)^{2} \left( \frac{m_a}{10^{-18} \eV} \right)^\frac12 \left(\frac{10}{x_{0,a}} \right)\left(\frac{3.5}{g_\star (T_\star)} \right)^{\frac14}  ~,
\eeq
for $\xi_\star\log_\star\lesssim 10^3$. For larger values of $\xi_\star\log_\star$, eq.~\eqref{eq:relic} must be changed to take into account the non-conservation of the number density. For a temperature independent axion mass, this results in $\Omega_a^{\rm st}\propto (\xi_\star\log_\star)^{3/4}$ up to logarithmic corrections (see Appendix~\ref{app:destroy}). Note that, unlike the QCD axion, in eq.~\eqref{eq:relic} $f_a$ and $m_a$ can vary independently and, for a fixed $f_a$, smaller values of $m_a$ give smaller $\Omega_a^{\rm st}$.

As expected, for $\xi_\star\log_\star$ small enough that the transient is negligible,  
 the relic density of axions from strings is a factor of $\xi_\star\log_\star$ larger than that from misalignment  in the pre-inflationary scenario (with $\mathcal{O}(1)$ misalignment angle) \cite{Hui:2016ltb}, with the same leading parametric dependence on $f_a$ and $m_a$.   For ultralight axions $\log_\star=\log(m_r/m_a)\simeq10^2$ and, as discussed in Section~\ref{ss:presentday}, the extrapolation of $\xi$ suggests that $\xi_\star=c_1\log_\star\simeq25$. Therefore, the relic density leads to severe constraints for $f_a \gtrsim 10^{14}\GeV$, forcing  $m_a \lesssim 10^{-17}\eV$. 
 We show the bound $\Omega_a^{\rm st}<\Omega_{\rm DM}$ in Figure~\ref{fig:param} assuming that the IR cutoff of the spectrum is $x_{0,a}=10$ also at large log, and that the growth of $\xi$ in eq.~\eqref{eq:xivslog} continues (we also fix that the network persists until $3H=m_a$, as suggested by simulation results for a temperature independent axion mass in Appendix~\ref{app:destroy}, which leads to a constraint that is  stronger by a factor of $\sqrt{3}$).

Eq.~\eqref{eq:relic} is expected to hold at best up to an order one coefficient that encodes the effect of the system of strings and domain walls on the axion waves produced during the scaling regime. We finally note that eq.~\eqref{eq:relic} only provides a lower bound on the axion dark matter abundance, as it misses the unknown part from string-domain wall collapse. In any case, as shown in Figure~\ref{fig:param}, even if this contribution to the axion relic abundance is a few orders of magnitude larger than $\Omega_a^{\rm st}$, a wide range of axion masses are still allowed for $f_a \gtrsim 10^{14}\GeV$ (thanks to the dependence of $\Omega_a^{\rm st}$ on $m_a$ and $f_a$).

We also note that Lyman-$\alpha$ observations limit the fraction of the dark matter that can have a mass below $\sim 10^{-20}~\eV$ \cite{Irsic:2017yje,Kobayashi:2017jcf}. We plot the resulting constraints in Figure~\ref{fig:param}, labelled L$\alpha$. Even though they are weaker than those from isocurvature that we study later, these bounds have the advantage of not depending on the (uncertain) details of the axion density power spectrum.\footnote{There is an additional constraint from Lyman-$\alpha$ \cite{Rogers:2020ltq} observations, not plotted, which requires $m_a >2 \times 10^{-22}~\eV$ in the case the axion makes up the entirety of the dark matter. While this excludes some values of $m_a$ and $f_a$ such that the axion makes up the whole DM abundance, the extension of this bound to the case of an axion that makes up a fraction of the total DM is currently unknown.}

\subsection{Dark Radiation}\label{ss:darkradiation}

For the ultra-light axion masses allowed by the relic abundance constraint in eq.~\eqref{eq:relic} (plotted in Figure~\ref{fig:param}), the scaling regime ends when the Universe's temperature is less than an MeV (see also the upper axis of Figure~\ref{fig:OmegaGW}). Consequently all the axions emitted during the scaling regime up to that point are relativistic at the time of BBN. Additionally, due to the approximately scale-invariant form of the axion energy density spectrum $\partial\rho_a/\partial k$ (see eq.~\eqref{drhodk} in Appendix~\ref{app:destroy}),  
 an order one fraction of the axion energy is in modes that are relativistic at CMB decoupling even if the string network is destroyed prior to this. Constraints from current limits on dark radiation at these times, usually expressed in terms of the effective number of neutrinos $N_{\rm eff}$, are therefore potentially relevant.\footnote{For larger axion masses a fraction of $\rho_a$ is in modes that are relativistic at BBN, but  constraints from dark matter overproduction are stronger in this case.}

Similarly to gravitational waves, during the scaling regime the energy density in axions is $\rho_a(t)=\int_{t_1}^tdt'(R'/R)^4\Gamma_a'$, where $\Gamma_a$ takes the form in eq.~\eqref{eq:gammaemi}. As a result, up to $1/\log$ corrections and neglecting changes in the number of relativistic degrees of freedom $g$, and assuming $\xi= c_1 \log$ as in eq.~\eqref{eq:xivslog}, 
\begin{equation}\label{eq:rhoarel}
\rho_a= \frac43 H^2 c_1\pi f_a^2\left(\log^3-\log_1^3\right) \ ,
\end{equation}
where $\log_1\equiv\log(m_r/H_1)$ is the value of the log when the scaling regime starts.  From eq.~\eqref{eq:rhoarel} we see that the energy in axions could be sizeable as it is enhanced by a $\log^3$ factor, which ultimately comes from the fact that the axion spectrum deviates from scale invariance by a $\log^2$ correction. 
Moreover, the dependence on the initial condition, which sets $\log_1$, is not important as long as $\log_1\ll\log$ (in the following we therefore neglect $\log_1$, and the resulting bound is barely affected by varying this in its plausible range).

The axion energy corresponds to an effective number of neutrinos  $\Delta N_{\rm eff}\equiv (8/7)(11/4)^{4/3}\rho_a/\rho_\gamma$ where $\rho_\gamma$ is the energy density in photons.\footnote{The effective number of neutrinos is defined by $N_{\rm eff}\equiv (8/7)(11/4)^{4/3}(\rho_\nu+\rho_X)/\rho_a$. In principle  the energy density in strings and gravitational waves also contributes to $\rho_X$, however this energy is much smaller than $\rho_a$.} At a temperature of $1 \MeV$ eq.~\eqref{eq:rhoarel} leads to
\beq \label{eq:DNeff}
\Delta N_{\rm eff} = 0.6 \left( \frac{c_{1}}{0.24}\right) \left(\frac{f_a}{10^{15}~\GeV}\right)^2\left(\frac{\log}{90}\right)^3 \ ,
\eeq
where $\log(m_r/H_{\rm BBN})\simeq 90$ if $m_r=10^{14}\div10^{15}\GeV$. To determine the overall coefficient in eq.~\eqref{eq:DNeff} precisely, we improved the calculation of $\rho_a$ in eq.~\eqref{eq:rhoarel} by numerically integrating $\Gamma_a$ to account for the changing number of relativistic degrees of freedom. Owing to the increased expansion of the Universe this suppresses the result by about $30\%$.\footnote{In particular, the contribution to $\rho_a$ from earlier times, already suppressed by $\log^3$, is further suppressed by $g^{1/3}$.}

Although bounds that we will study in Section~\ref{ss:cmb} require that the string network is destroyed prior to the formation of the CMB for the $f_a\gtrsim10^{14}\GeV$, limits on dark radiation from the CMB are still potentially relevant. 
In this case, the energy in relativistic axions at the time of decoupling is simply obtained by redshifting the value of eq.~\eqref{eq:rhoarel} from the time of the network destruction, accounting for the proportion of the axions that become non-relativistic.

We use the $95\%$ limits from \cite{Aghanim:2018eyx}, which correspond to $\Delta N_{\rm eff} < 0.46$ and $0.28$ at BBN and CMB times respectively (imposing constraints derived from different analyses or the $68\%$ limits, e.g. \cite{Pitrou:2018cgg} only changes the constraint on $f_a$ relatively mildly). As can be seen from eq.~\eqref{eq:DNeff}, the limits from BBN constrain $f_a \lesssim 10^{15}~\GeV$ as shown in Figure~\ref{fig:param}. Those from the CMB lead to a comparable bound, although the maximum allowed $\Delta N_{\rm eff}$ is more uncertain given tensions between different determinations of the Hubble parameter. As we will see shortly, all such values of $f_a$ are in tension with limits from isocurvature. However, the constraints from dark radiation are still useful since they are subject to fewer uncertainties.

We finally note that the detection of GWs from strings would  predict a non-vanishing $\Delta N_{\rm eff}$. Conversely, given the quadratic dependence on $f_a$, 
a plausible improvement in the measurement of $\Delta N_{\rm eff}$ to an uncertainty of $\sigma(\Delta N_{\rm eff}) \simeq 0.02$ \cite{Abazajian:2013oma} could rule out $f_a\gtrsim 2\times 10^{14}$, which is a large part of the range that gives GWs that could be observed in the near future.\footnote{Additionally, the possibility that the background of relativistic axions left over today could be directly detected (for axion-to-photon coupling $g_{a\gamma\gamma}$ larger than $f_a^{-1}$) has recently been studied \cite{Dror:2021nyr}.}

\subsection{Isocurvature Perturbations} \label{ss:iso}

During the scaling regime, and after domain walls form and annihilate, the axion field contains inhomogeneities. Since these involve only the axion energy density, they correspond to isocurvature perturbations~\cite{Hogan:1988mp}, which 
are potentially in conflict with cosmological observations, even if the axion makes up only a fraction of the total dark matter. An analysis of such perturbation has previously been carried out for the QCD axion \cite{Enander:2017ogx} and axion-like particles in \cite{Feix:2019lpo,Feix:2020txt}, and we comment in Appendix~\ref{app:compare} on the differences with our approach.

The perturbations in the axion energy density $\rho_a$ are most easily studied using the overdensity field $\delta_a(x)\equiv(\rho_a(x)-\langle\rho_a\rangle)/\langle\rho_a\rangle$, where the brackets indicate the spatial average. It is useful to express $\delta_a$ in terms of the (dimensionless) power spectrum $\Delta_a^2(k)$, defined by
\begin{equation}\label{eq:delta2k}
\langle\tilde{\delta}_a(\mathbf{k})\tilde{\delta}_a(\mathbf{k}')\rangle=\frac{2\pi^2}{k^3}\Delta^2_a(k)\delta^3(\mathbf{k}-\mathbf{k}') \ ,
\end{equation}
where $\tilde{\delta}_a$ is the Fourier transform of $\delta_a$ and we assumed statistical homogeneity and isotropy. It is straightforward to show that $\langle\delta_a^2\rangle=\int \Delta_a^2(k)/k~ dk$.

After the string network is destroyed, $\Delta_a^2$ is expected to have a non-zero (and order one) value at momenta $k\simeq H_\star$, since $H_\star$ sets the typical size of fluctuations, and therefore the typical correlations between momentum modes. At these scales, the precise shape of $\Delta_a^2$ depends on the dynamics of the string network at $H\simeq H_\star$. Conversely, fluctuations at scales larger than the horizon at that time, i.e. $k\ll H_\star$, are expected to be uncorrelated. From eq.~\eqref{eq:delta2k}, this fixes the form of $\Delta_a^2\propto k^3$, which is the prediction from a white noise spectrum, so $\Delta_a^2$ is suppressed at large scales. We therefore parametrise $\Delta_a^2(k)=C(k_{\rm com}/H_\star)^3$ at comoving momenta $k_{\rm com}\equiv k R/R_\star \lesssim H_\star$, where $C$ is a dimensionless coefficient. Finally, the power spectrum at $k\gtrsim H_\star$ depends on the details of the destruction of strings and domain walls, and it could be affected by oscillons \cite{Vaquero:2018tib}, 
 and is beyond the scope of our present work.

During the scaling regime, and as the string network is destroyed, $\Delta^2_a$ changes. However, once all the strings and domain walls have vanished and the axion field has mostly settled down to have amplitude $\ll f_a$, and the axion energy density is redshifting non-relativistically, the $k^3$ part of $\Delta_a^2$ is constant leading to a time-independent $C$.\footnote{We neglect the effect of gravity, which is expected to not affect the far IR-momenta \cite{Feix:2020txt}.} 

CMB observations dominantly constrain $\Delta_a^2$  at momenta that are of order the Hubble parameter at decoupling. For $m_a\gtrsim 10^{-28}~\eV$, the network decays before decoupling, and such scales lie in the IR $k^3$ tail, on which we will therefore focus (for smaller $m_a$, strings are present at decoupling, and gravitational effects lead to more stringent bounds, discussed in Section~\ref{ss:cmb}). As we discuss shortly the situation is more complicated for isocurvature constraints from Lyman $\alpha$ observations.

Given the highly nonlinear dynamics of the scaling regime and the destruction of the string network, the only way to determine $\Delta^2_a$ accurately would be to numerically simulate the system until the strings and domain walls have all vanished. Unfortunately, as discussed, it is extremely hard to study the network's destruction reliably.  Despite this, similarly to the dark matter abundance, we can obtain a conservative isocurvature constraint by  considering only the axion waves produced during the scaling regime up to $H_\star$. Provided the presence of domain walls does not affect these waves by more than an order one amount,  the axions from the scaling regime can be approximated as a separate component of DM (distinct from that produced by domain walls) with its own power spectrum and a relic abundance given by eq.~\eqref{eq:relic}. Neglecting the effects of domain walls, we will be able to determine the power spectrum for this component and in doing so obtain an isocurvature bound. The DM axions  from domain walls will also have density perturbations so are expected to only strengthen the constraint.\footnote{A large weakening of the bound would require that the domain walls absorb a large fraction of low momentum axion waves emitted during scaling and homogeneously remit them as high momentum modes.}

To determine $\Delta_a^2$, we solve the equations of motion $\ddot{a}+3H\dot{a}-R^{-2}\nabla^2a+V'(a)=0$ (with $V=m_a^2f_a^2(1-\cos(a/f_a))$ and $m_a$ temperature independent as before) numerically. As discussed in Appendix~\ref{app:density} we expect very similar results from any other potential bounded to be $\lesssim m_a^2f_a^2$. We start at $H=H_\star$ with initial conditions given by a superposition of waves with the energy density spectrum $\partial\rho_a/\partial k$ predicted by the scaling regime at $\xi_\star \log_\star=10^3$ (see Appendix~\ref{app:density} and \cite{moreaxions} for more details). As mentioned in Section~\ref{ss:relic}, the field undergoes a period of relativistic redshift and a nonlinear transient, after which the nonrelativistic regime is rapidly reached with (at least the IR part of) $\Delta^2_a$ constant.\footnote{Unlike simulations of the string network, the axion only simulations do not need to probe length scales $f_a^{-1}$ (since it is the IR part of the axion spectrum that contains the majority of the axion number density so is relevant to the constraint), therefore the physical dynamics are reproduced directly.}

In Figure~\ref{fig:denpow} (left) of Appendix~\ref{app:density} we plot the resulting $\Delta^2_a$, defined as in eq.~\eqref{eq:delta2k}, as a function of the momentum and time. As expected, the spectrum reaches an order one value at momenta corresponding to the scale $H_\star$. The peak is somewhat above $H_\star$ since $x_{0,a}\simeq10$ and correlations at smaller scales are likely since the average axion momentum is larger than $x_0$ (see~\cite{moreaxions}). 
At smaller momenta the expected $k^3$ behaviour is reproduced, with a constant coefficient $C\approx 2\times 10^{-5}$ (the nonlinearities turn out to be important, as this coefficient is about a factor of $4$ smaller than if the evolution of these waves were linear, i.e. with $V=1/2m_a^2a^2$, see the green line in Figure~\ref{fig:denpow} (left) of Appendix~\ref{app:density}). In the following we will derive constraints assuming $\Delta_a^2$ and $\Omega_a^{\rm st}$ as described above. To get a feel for how much the DM axions from domain walls might potentially strengthen the constraint, in Appendix~\ref{app:density} we also plot and discuss results for the power spectrum obtained after the destruction of the full string-domain wall network at the unphysical value $\log_\star=5$.

\subsubsection*{\emph{Isocurvature constraints from the CMB and Lyman-$\alpha$}}

The CMB is the longest standing source of constraints on isocurvature perturbations, with the best currently available data coming from Planck \cite{collaboration2018planck}, from modes close to the pivot scale $k_{\rm CMB}=0.05~{\rm MPc}^{-1}$. Such observations bound the fraction of isocurvature fluctuations $f_{\rm iso}$ relative to the curvature perturbations at this scale, defined by $f^2_{\rm iso}\equiv\Delta_{\rm iso}^2(k)/\Delta_\mathcal{R}^2(k)$, where $\Delta_{\rm iso}^2$ is the spectrum of isocurvature perturbations and $\Delta_\mathcal{R}^2(k)=A_s(k/k_{\rm CMB})^{n_s-1}$ is the (almost scale invariant) spectrum of curvature perturbations.

For $m_a\gtrsim10^{-28}$ eV,  $k_{\rm CMB}$ is inside the $k^3$ part of $\Delta^{2}_a(k)$ and we can directly apply the limit $f_{\rm iso}<0.64$ obtained in~\cite{Feix:2020txt}.\footnote{The precise numerical bound depends on the cosmological data that is combined. Interestingly the analysis in \cite{Feix:2020txt} shows a mild preference for a non-zero isocurvature component.} In general the axions from scaling only comprise a fraction of the total dark matter, so there is a factor of $\Omega_a^{\rm st}/\Omega_{\rm DM}$ in their contribution to $f_{\rm iso}$. Consequently, assuming there are no isocurvature fluctuations in the remainder of the DM,  at $k=k_{\rm CMB}$
\beq
\begin{aligned} \label{eq:cmbiso}
f_{\rm iso} &= \frac{\Omega_a^{\rm st}}{\Omega_{\rm DM}} \sqrt{\frac{C\, k_{\rm CMB}^3}{A_s k_\star^3} } \\
&\simeq 0.2 \left(\frac{\xi_\star\log_\star}{3\times10^3}\right)\left(\frac{f_a}{5\times 10^{14}~\GeV} \right)^2 \left(\frac{m_a}{10^{-28}~\eV} \right)^{-1/4}\left(\frac{C}{2\times10^{-5}}\right)^{1/2}~,
\end{aligned}
\eeq
where $A_s = 2.2\times 10^{-9}$, and $k_\star\equiv H_\star R_\star/R_0$ is the comoving momentum corresponding to $H$ at $H=H_\star=m_a$ redshifted to today. As discussed, we expect eq.~\eqref{eq:cmbiso} to give a lower bound on $f_{\rm iso}$, and to be valid up to an order one factor. We note that the isocurvature constraint is potentially important even when the axions produced by the scaling regime make up a small fraction of the abundance. This is because, for sufficiently light axions, the relative density fluctuations reach values close to one at the observationally constrained scales.

In addition to CMB bounds, it has recently been shown that isocurvature fluctuations at smaller scales are constrained by Lyman-$\alpha$ observations~\cite{Murgia_2019}, and that these can be important in post-inflationary axion dark matter models~\cite{Irsic:2019iff}. We use the constraint from~\cite{Murgia_2019}, which assumes a $k^3$ power spectrum. Expressed in terms of the isocurvature fraction at the Planck pivot scale, this requires $f_{\rm iso}<0.004$. 
For axion masses such that the observed modes are in the $k^3$ part of $\Delta_a^2$, this can be immediately converted into a bound on $f_a$ similarly to eq.~\eqref{eq:cmbiso}. However, the dominant scales in Lyman-$\alpha$ studies correspond to momenta of order $k_{{\rm L}\alpha} \sim 10~{\rm Mpc}^{-1}$ much larger than those relevant to CMB observations \cite{Irsic:2019iff} (and more sophisticated analyses are potentially sensitive to even smaller scales). 
For sufficiently small masses, $m_a \lesssim 10^{-22}~\eV$, such modes are not in the $k^3$ region. Consequently the results of \cite{Murgia_2019} cannot be applied and instead a complete reanalysis is needed. We do not attempt in our present work. However, we note that given the density power spectrum remains of order 1 up to $k/H\sim 100$ and drops relatively slowly above this, there are likely to be relevant constraints in this mass range as well.

\subsubsection*{\emph{Impact on the  allowed axion masses and decay constants}}

The bounds on the axion mass and decay constant from CMB and Lyman-$\alpha$ isocurvature constraints are plotted in Figure~\ref{fig:param}. The limits at higher axion masses come from Lyman-$\alpha$ observations, while those at lower masses are due to the CMB observations. As discussed, we do not give bounds from Lyman-$\alpha$ for axion masses smaller than about $10^{-22}~\eV$.

Although we expect the domain walls not to affect the power spectrum associated to pre-existing axions waves by more than an order one factor, we blur the resulting constraints in Figure~\ref{fig:param} to reflect this uncertainty. Moreover, the constraints from isocurvature would strengthen if the axions from domain walls have large density perturbations, i.e. if their density power spectrum corresponds to a larger coefficient $C$, if their relic abundance is comparable to that of the axions from scaling (though the bound on $f_a$ would increase only proportionally to $C^{1/4}$). The bounds would also strengthen if the domain walls produced more DM axions than the scaling regime, even if they have the same density power spectrum as those from scaling, because $f_{\rm iso}$ includes a factor of  $\Omega_a/\Omega_{\rm DM}$, as in eq.~\eqref{eq:cmbiso}.\footnote{We also note that there are other possible constraints on isocurvature from spatial scales between the CMB and Lyman-$\alpha$ observations \cite{Irsic:2019iff}, which could matter for intermediate axion masses $\sim 10^{-25}~\eV$. However, these are more sensitive to astrophysical assumptions, and would require detailed analysis.}

We note that it is clear from eq.~\eqref{eq:cmbiso} that the isocurvature constraints are not relevant for the QCD axion in the post-inflationary scenario. Indeed, for the QCD axion $H_\star$ corresponds to the Hubble parameter shortly before the QCD crossover, which is much larger than that for ultralight axions, and so $k_\star$ is larger and $f_{\rm iso}$ more suppressed. 
Finally, for both the QCD axion and ultralight axions, isocurvature perturbations sourced during inflation (if the axion was present at this point) are averaged out by the dynamics of the strings and are simply incorporated into the coefficient $C$ of the white noise spectrum.

\subsection{Strings and CMB anisotropies} \label{ss:cmb}

If strings are present at the time of decoupling their gravitational interactions introduce additional anisotropies in the CMB \cite{Zeldovich:1980gh,Vilenkin:1981iu}. In particular, a long string induces a `deficit angle' $\delta = 8\pi G \mu$ in the locally flat metric around it, giving the metric a global conical structure. As a result, two particles moving towards the string in parallel  acquire a nonzero relative velocity as they pass the string, and eventually collide.\footnote{On the other hand, loops are expected to oscillate many times before disappearing and at large distances the gravitational field averaged over one oscillation is the usual Newtonian potential, which affects the motion of bodies like conventional matter.} If this is applied to an observer and a source of photons, the observer will see a discontinuous Doppler shift of the photons as the string is passed. Consequently, strings lead to discontinuous temperature fluctuations,  of order $\delta T/T \propto G \mu$, in the CMB photons around them \cite{Kaiser:1984iv}.  
Moreover, a string that moves in the primordial plasma produces a wake behind it by pure gravitational interactions, and therefore additional density perturbations, which are again potentially measurable in the CMB. Since the strings act as a random source, such perturbations are not coherent and do not result in the observed acoustic oscillations~\cite{Albrecht:1995bg,Magueijo:1995xj,Pen:1997ae}. Consequently they can contribute, at most, a relatively small fraction of the total anisotropy (see e.g.~\cite{Pogosian:2003mz,Wyman:2005tu,Fraisse:2006xc,Ade:2013xla}).

A statistical analysis comparing the effects mentioned above  with CMB observations gives an upper limit on the string tension $\mu$ and consequently on the allowed $f_a$ for an axion network that is not destroyed prior to decoupling. These constraints are therefore relevant for axion masses $m_a \lesssim 10^{-28}~\eV$.  
As for the GW spectrum and axion DM abundance, it is challenging to determine the string induced anisotropy spectrum at the relevant time, which cannot be studied directly in simulations.  Reliable constraints can only be obtained by fully accounting for the effects of scaling violations on the string induced anisotropy spectrum, and we do not attempt this in our present work.\footnote{It would be particularly challenging to accurately determine constraints around $m_a\sim 10^{-28}~\eV$, given that the string network is being destroyed by domain walls around the time of decoupling in this case.} Instead, we simply note that, based on the previous literature (e.g. \cite{Battye:2010xz,Charnock:2016nzm,Lizarraga:2016onn} for local strings and \cite{Lopez-Eiguren:2017dmc} for global strings), a bound very roughly in the range of $G \mu \lesssim 10^{-7}$ is plausible, corresponding to $f_a \lesssim 2 \times 10^{14}~\GeV$. This is in possible tension with any axion with $m_a\lesssim 10^{-28}~\eV$ and $f_a$ large enough for observable GWs. We however stress that there is significant uncertainty remaining and blur the bound in Figure~\ref{fig:param} to indicate this. In particular, if the bound on $G\mu$ turns out to be significantly weaker than the quoted limit, observable GWs from axion strings would be allowed for arbitrarily small axion masses.

\subsection{The case $N>1$}\label{ss:genericN}

In the previous Sections we have performed our analysis assuming that the U(1) breaking scale $v$ equals the axion decay constant $f_a$. Our results are however easily generalised when $N=v/f_a>1$. First we observe that the calculation of the axion and GW emission from the strings depends on $v$, which enters in eq.~\eqref{eq:LPhi} and therefore determines the string tension in eq.~\eqref{muth} and the emission rates $\Gamma_a$ and $\Gamma_g$. Thus, the GW spectrum in eq.~\eqref{eq:Omegaapprox} and Figure~\ref{fig:OmegaGW} is valid for a generic $v$ with the substitution $f_a\to v$. Similarly, the bounds from dark radiation and CMB anisotropies in Sections~\ref{ss:darkradiation} and~\ref{ss:cmb} and Figure~\ref{fig:param} apply to~$v$.

On the other hand, since the decay constant $f_a$ determines the axion potential $V\propto m_a^2f_a^2$, a $v\neq f_a$ affects the dark matter abundance and isocurvature perturbations discussed in Sections~\ref{ss:relic} and~\ref{ss:iso}. In all our calculations in those Sections the scale $v$ enters only through the inputted axion energy density spectrum from the scaling regime, and it appears together with $\xi_\star\log_\star$ (in particular, $\partial\rho_a/\partial k\propto v^2\xi_\star\log_\star$). The dark matter abundance in eq.~\eqref{eq:relic} is therefore easily generalised by substituting $\xi_\star\log_\star\to N^2\xi_\star\log_\star$. However, as discussed in Section~\ref{ss:relic}, if $N\gg1$ one needs to account for the number density non-conservation. This is done by multiplying eq.~\eqref{eq:relic} by the suppression factor in eq.~\eqref{eq:Q1}, which makes $\Omega_a^{\rm st}\propto (N^2\xi_\star\log_\star)^{3/4}$ (instead of a quadratic dependence on $N$). A similar change occurs in eq.~\eqref{eq:cmbiso} via the DM abundance (notice that $C$ depends on $\xi_\star\log_\star$, and needs to be recomputed).

The overall effect of $N>1$ can therefore be summarised by substituting $f_a\to v$ in Figures~\ref{fig:OmegaGW} and~\ref{fig:param}, but with a stronger constraint from dark matter overproduction (by a factor $\mathcal{O}(N^{3/2})$ on the vertical axis of Figure~\ref{fig:param}).

\section{Peccei-Quinn Restoration in the Early Universe} \label{sec:restore}

The post-inflationary scenario occurs if the U(1) symmetry has ever been restored after inflation. In this Section we first review two standard mechanisms that lead to such a restoration, as well as  another well known possibility (albeit more dependent on the details of the inflaton couplings and the axion sector). Then we discuss other ways that symmetry restoration can occur for axion decay constants $f_a\gtrsim 10^{14}\GeV$. 

\begin{itemize}
\item Quantum fluctuations during inflation induce perturbations in any effectively massless scalar field present at this time. These have amplitude of order $H_I/2\pi$, where $H_I$ is the Hubble parameter during inflation, so restore the PQ symmetry if $H_I/2\pi \gtrsim f_a$  \cite{Linde:1990yj,Lyth:1992tw}. However, the current bound on $H_I$ from the non-observation of tensor modes is \cite{planck_inflation}
\beq
\frac{H_I}{2\pi}<9.6\times 10^{12}~\GeV ~.
\label{eq:MaxHI}
\eeq
Consequently such fluctuations do not lead to the post-inflationary scenario for values of $f_a$ that give observable GWs.

\item The PQ symmetry  can be restored if the (maximum) temperature $T_{\rm max}$ during reheating  is greater than $f_a$. In particular, if the sector that gives rise to the axion is sufficiently close to thermal equilibrium at this time, finite temperature corrections drive the minimum of the scalar potential to the PQ-preserving vacuum for many choices of matter content and potentials (including  eq.~\eqref{eq:LPhi}) \cite{ Kirzhnits:1972ut,Kirzhnits:1974as,Weinberg:1974hy}.\footnote{High temperature does not restore symmetries in all theories, and the relevance of this to the formation of topological defects has been studied in \cite{Dvali:1995cc}.} In this case, the PQ symmetry spontaneously breaks when the temperature subsequently drops  \cite{Kibble:1976sj,Kibble:1980mv,Zurek:1985qw}.

If the inflaton decays relatively fast, $T_{\rm max}$ is parametrically larger than $H_I$. Consequently, the PQ symmetry is restored even for decay constants such that fluctuations during inflation are not sufficient. To illustrate this we parametrise the inflaton $\varphi$'s decay rate $\Gamma_{\varphi} = \epsilon H_I$. 
The Universe enters radiation domination at a temperature $T_{\rm RH}$ such that $\Gamma_{\varphi}= H(T_\textrm{RH})$, so
\begin{equation}
\begin{aligned}
T_\textrm{RH} &\simeq  \left(\frac{\epsilon}{0.001} \right)^{1/2} \left(\frac{H_{I}}{H_{\rm max}} \right)^{1/2} 3 \times 10^{14}~\GeV ~,
\label{reheating temperature}
\end{aligned}
\end{equation}
where $H_{\rm max}$ is the maximum value of Hubble during inflation allowed by eq.~\eqref{eq:MaxHI}.\footnote{In determining the numerical prefactors in these expressions we assume that the number of degrees of freedom that are thermalised is approximately similar to the standard model high temperature value $g_s \sim 100$.} For instance, suppose the inflaton's potential is  of the form  $V(\varphi)=\frac{1}{2}m_{\varphi}^2 \varphi^2$ in the part of field space where reheating occurs, and that  $\left<\varphi^2\right> \simeq M_{\rm P}^2$ at the start of reheating so  $m_\varphi\simeq H_I$. Then, if the inflaton decays to fermions $\psi$ though a coupling 
$\mathcal{L} \supset g \varphi \bar{\psi} \psi$, 
we have $\epsilon \simeq 0.03 g^2$ and $T_\textrm{RH}  \simeq g~  10^{15}\GeV$ for $H_I=H_{\rm max}$. 

Moreover, the energy density in radiation is even larger at the start of perturbative reheating, while the inflaton still dominates the energy density of the Universe. If the system is sufficiently close to thermal equilibrium that it can be assigned a temperature  at this point, the largest temperature it reaches is $T_{\rm max} \sim M_{\rm P}^{1/4} H_I^{1/4}T_{\rm RH}^{1/2}$  \cite{largest_temperature}.\footnote{The analysis of the inflaton's decay via perturbative processes is only consistent if $\Gamma_{\varphi} \ll m_\varphi \lesssim H_I$ so $\epsilon \ll 1$ and $T_\textrm{max} \gg T_\textrm{RH}$ as expected.} In particular, parametrising the inflaton decay rate as before,
\begin{equation}
\begin{aligned}
\label{eq:tmax}
T_\textrm{max} &=   \left(\frac{\epsilon}{0.001}\right)^{1/4}  \left(\frac{H_{I}}{H_{\rm max}} \right)^{1/2}  10^{15} ~ \GeV ~.
\end{aligned}
\end{equation}
Efficient thermalisation indeed occurs for some couplings and matter content (e.g. if this happens through interactions with massless gauge bosons, see \cite{thermalisation1,thermalisation2}). In this case temperatures close to $T_\textrm{max}$ are achieved and if $f_a \simeq 10^{14}$ $\GeV$ the PQ symmetry is restored even for $\epsilon\ll 1$, i.e. for inflaton decay rates $\Gamma_\varphi\ll H_I$. 

\item Additionally, for many types of interaction, the inflaton expectation value during inflaton restores the  PQ symmetry if it couples to the PQ sector, provided the coupling is large enough \cite{Kofman:1986wm,Vishniac:1986sk,Yokoyama:1989pa,Hodges:1991xs}. 
Depending on the details of the theory, the PQ symmetry might then break during inflation (as the inflaton evolves towards its final expectation value), or during reheating (once the inflaton is settling down e.g. to $\left<\varphi\right>=0$).  In many theories the inflaton expectation value is large compared to other scales, so this restores a PQ symmetry with large $f_a$ even if the inflaton only has relatively weak couplings. 

As an example, suppose the axion comes from a complex scalar $\phi$ with the potential of eq.~\eqref{eq:LPhi} that has an interaction with the inflaton
\beq
V_\textrm{int}=\frac{1}{2}g\varphi^2 |\phi|^2 ~,
\label{eq:interaction_potential}
\eeq
and the inflaton has a non-zero expectation value $\left<\varphi\right>$ during inflaton and evolves towards the minimum of its potential at $ \varphi =0$ (similar potentials and interactions have been widely considered, see e.g.  \cite{Shafi:1984tt,Kofman:1986wm,Kofman:1988xg,Hodges:1989dw}). The interaction eq.~\eqref{eq:interaction_potential}  brings the expectation value of $\phi$ to zero if, during inflaton, the inflaton satisfies
\beq 
g \left<\varphi^2 \right> >m_r^2 ~.
\label{eq:restore_inflation}
\eeq
Consequently, if $\left<\varphi\right> \sim M_{\rm P}$ the PQ symmetry is restored by couplings $g\simeq m_r^2/M_{\rm P}^2$, which is tiny, $\sim 10^{-6}$, even for $m_r \sim f_a \sim 10^{15}~\GeV$. If all of the inflaton's couplings are of this size the temperature of the Universe after inflation is never close to $f_a$ (as can be seen from eqs.~\eqref{reheating temperature} and \eqref{eq:tmax}, e.g. with $\epsilon \simeq 0.03 g^2$ in the previously discussed example  model).\footnote{Strings might also form in the case that interactions with the radial mode displaced $\phi$ to large field values $\gg f_a$ during inflation, as small initial fluctuations are amplified as the field subsequently oscillates around the $\phi=0$ \cite{Ballesteros:2016xej}, but we do not study this possibility in detail.}

\end{itemize}

Summarising, (perturbative) reheating restores the PQ symmetry by thermal corrections to the PQ potential even for large $f_a$ provided $H_I$ is fairly close to $H_{\rm max}$ and the inflaton decay rate and thermalisation are fast. In particular, from eq.~\eqref{eq:tmax}, inflaton decay rates corresponding to $\epsilon \sim 10^{-3}$ restore the PQ symmetry provided $f_a \lesssim 10^{15}~\GeV$ (this would require e.g. $g=\mathcal{O}(1)$ in the interaction with fermions).\footnote{We note that large inflaton couplings may lead to contributions to the inflaton potential, either directly or through radiative corrections, that necessitate fine tuning.} However, if the inflaton has interactions with the axion sector of this size then in many theories the PQ symmetry is restored directly during inflation regardless of reheating, and this also happens for much smaller couplings for which perturbative reheating is not enough to restore the symmetry. Moreover, as we will see in the following, if the inflaton has interactions with couplings $\gtrsim 10^{-5}$ non-perturbative processes are often important during reheating. 
This sometimes restores the PQ symmetry for couplings of a size that do not lead to restoration during the subsequent perturbative reheating.

In the remainder of this Section we discuss how couplings between the inflaton and the axion sector that restore the PQ symmetry during inflation can be accommodated in complete theories of inflation; and also how non-perturbative reheating, known as `preheating'  leads to symmetry restoration (both these possibilities have previously been considered in various contexts). We finally show that, in some axion models such as the KSVZ model, the PQ symmetry is restored at temperatures much less than $f_a$ if the mass of the radial mode is much smaller than $f_a$.

\subsection{Symmetry Restoration during Inflation} \label{ss:hybrid}

Interactions between the inflaton and the sector that gives rise to the axion might simply be present without playing a role in inflation. In Appendix~\ref{app:hybrid} we show that symmetry restoration via eq.~\eqref{eq:restore_inflation} indeed occurs without disrupting the slow roll conditions in an example model. Another possibility is that the axion sector plays a role in inflation. For instance, the complex scalar $\phi$ with the potential in eq.~\eqref{eq:LPhi} could be the additional field needed in the so-called `hybrid' inflation scenario. As reviewed in Appendix~\ref{app:hybrid}, in hybrid inflation the inflaton expectation value causes symmetry restoration in another sector, e.g. via the interaction in eq.~\eqref{eq:restore_inflation}. The parameters of the theory are fixed so that vacuum energy due to this symmetry restoration (e.g. $\sim f_a^4$)  dominates the energy density of the Universe. This relaxes the slow roll constraint on the inflaton potential, $V^\prime M_\textrm{Pl}/V\ll 1$,  
by giving a large contribution to $V$. Inflation is therefore possible with sub-Planckian field values (and small $H_I$) \cite{hybrid_initial,hybrid_Linde}.\footnote{This avoids the substantial fine tuning characteristic of many other small field models (see e.g. \cite{low_scale_fine_tuning, overshoot_fine_tuning}).} Inflation ends when the inflaton expectation value evolves enough that the (e.g. PQ) symmetry breaks, which naturally leads to topological defects. The fact that cosmic strings form in many models of hybrid inflation has previously been analysed  \cite{false_vacuum,cosmic_structure,gut_strings,Linde:2013aya,Fterm,Dterm, Dterm1}, focusing on gauged (i.e. local) strings, and 
 axion strings similarly form in some theories. In Appendix~\ref{app:hybrid} we analyse a hybrid model in which $\phi$ has the potential in eq.~\eqref{eq:LPhi} and the interaction in eq.~\eqref{eq:interaction_potential} and show that for any inflaton expectation value that satisfies $f_a< \left<\varphi \right> \lesssim M_{\rm P}$ there is an allowed range of $g\in [f_a^2/\left<\varphi \right>^2,1]$ such that hybrid inflation occurs.  Although the range of $g$ that leads to symmetry restoration is no larger than when $\phi$ plays no role in inflation, in these theories the restoration of PQ symmetry enables inflation in the first place rather than this being an ad hoc feature of the model.

\subsection{Parametric Resonances} \label{ss:preh}

For some inflaton couplings and potentials, non-perturbative processes transfer an order one fraction of the inflaton's energy to other fields within $\mathcal{O}(10)$ oscillations around the minimum of its potential. These effects are known as preheating, and they stem from a Bose enhancement of the inflaton decay rate due to the growing occupation number of the modes it decays into. A detailed analysis shows that  modes that have frequencies inside particular resonance bands are amplified exponentially fast (until either backreaction or scattering cuts off the growth or the expansion of the Universe moves the mode out of the resonance band), and the resulting energy densities  are far higher than are ever reached during perturbative reheating. 
Depending on the inflaton couplings, there are two different regimes, known as the broad and narrow resonance. Preheating and the resulting effective temperature have been considered analytically in the broad resonance \cite{Linde_broad,resonance_analytic} and narrow resonance cases \cite{shtanov_narrow}, and through lattice simulations in the broad resonance case \cite{khlebnikov_tkachev_1997,broad_simulation}.

We review preheating in more detail in Appendix~\ref{app:preheat}, and determine in which theories it restores a PQ symmetry with a large $f_a$.  One possibility is that  preheating occurs to the radial mode directly, e.g. through the interaction in eq.~\eqref{eq:interaction_potential} leading to fluctuations that are of order $f_a$. However, we show that for values of $f_a$ relevant for GW searches, preheating is only efficient enough for couplings such that the  inflaton expectation value restores the symmetry during inflation anyway. This is because the large radial mass reduces the width of the resonance band preventing efficient energy transfer unless $g$ is large. 
 On the other hand, we also show that preheating leads to the PQ symmetry being restored (in models in which this would otherwise not occur) if the preheating is to some other, effectively massless, scalar. 
  In particular, in some models a PQ symmetry with $f_a=10^{15}\GeV$ is restored for inflaton couplings that give a maximum temperature during perturbative reheating  $T_{\rm max} \lesssim 10^{12}~\GeV$.

\subsection{Symmetry Restoration with a Light Radial Mode} \label{sec:lightrad}

Finally, we show that if the radial mode (and additional PQ quarks) are light, the PQ phase transition happens at temperatures that are lower than $f_a$. Consequently, PQ symmetry restoration happens for reheating temperatures lower than if the mass of the radial mode is around $f_a$ (which as discussed, typically requires temperatures $\gtrsim f_a$).

As an example, we consider the Lagrangian in eq.~\eqref{eq:LPhi}, with a quartic $\lambda =  m_r^2/f_a^2 \ll 1$, with the radial mode coupled to fermions $\psi$ through a term of the form $V \supset g \phi \bar{\psi}   \psi $ (such interactions are present e.g. in the KSVZ QCD axion model) with $g\ll 1$. The thermal contribution to the finite temperature potential $V_T$ of $\phi$ from such fermions is $V_T \simeq T^2 m_\psi^2(\phi) = g^2 T^2 \phi^2$, and this produces a minimum at $\phi=0$ if $gT^2\phi^2\gtrsim m_r^2\phi^2$. Moreover, we  demand that $T \gtrsim g f_a$, which ensures that the fermions (which have mass $g\left< \phi \right>$) are present in the thermal bath at $\left<\phi\right>=f_a$. This guarantees that the finite temperature potential has no symmetry breaking minimum, so the PQ symmetry is restored regardless of the initial field values (which depend on e.g. the dynamics during inflation). In Appendix~\ref{app:lightR} we show that these parametric expectations are confirmed in a full analysis of $\phi$'s thermal potential (results are shown in Figure~\ref{fig:gspace}). Consequently the minimum temperature required to restore the PQ symmetry $T_{\rm  min}$ is approximately given by
\beq \label{eq:Tminmr}
T_{\rm  min} \simeq {\rm Max} \left( \frac{m_r}{g},g f_a  \right) ~.
\eeq
Depending on the value of $g$, this means that $T_{\rm min}$ can be as small as  $\sqrt{f_a m_r} $. For example, taking $f_a\sim 10^{14}~\GeV$ and $m_r \sim 10^8~\GeV$ the PQ symmetry could be restored at a temperature of $10^{11}~\GeV$, which is reached during perturbative reheating even for inflaton couplings that are much less than $1$.\footnote{The price of this is that small mass radial mode mass may necessitate fine tuning, although this could be evaded e.g. if this sector is supersymmetric with a breaking scale $\lesssim m_r$.} Importantly for this scenario, the energy that an axion string network emits to GWs is dominantly set by the value of $f_a$ rather than $m_r$ (having $m_r\ll f_a$ only affects the logarithm entering the string tension). Consequently, the GW predictions in Section~\ref{sec:GWs} are suppressed only by a factor of $\log^4(m_r/H(T_f))/\log^4(f_a/H(T_f))$, and remain observable for $f_a\gtrsim 10^{14}\GeV$ provided $m_r$ is not extremely small.

\section{Summary and Conclusions} \label{sec:Conclusion}

Gravitational wave observations can only be used to learn about  physics beyond the Standard Model if we understand the spectrum expected in motivated theories. 
In this paper we have made a step towards such a goal by studying the GWs emitted by the network of global strings that forms in the early Universe in a generic axion model if the U(1) symmetry has been ever restored after inflation.

During the subsequent scaling regime GWs are produced by the motion and interactions of strings. Calculating the resulting spectrum directly from numerical simulations is impossible, as these only have access to a small range of early times. However, in Section~\ref{ss:theory} we have shown that the Nambu--Goto effective theory with the Kalb--Ramond term -- which describes the parts of the string network for which the string thickness can be neglected  
-- gives us analytic control of the GW emission $\Gamma_g$ at all times, up to an (order one) constant coefficient $r$ related to shape of the string trajectories during scaling. In particular, $\Gamma_g/\Gamma_a=r G\mu^2/f_a^2$, where $\Gamma_a$ is the emission rate into axions and is fixed by energy conservation. 
As shown in Section~\ref{ss:gwsim}, this result is reproduced spectacularly well by first principles numerical simulations of the string network (at the accessible small values of $\log(m_r/H)$), from which we have extracted the value of $r$ together with the momentum distribution of $\Gamma_g$ (this is peaked at momenta of order Hubble and falls off at higher momenta as $\propto 1/k^2$). The existence of the attractor solution then allowed us to reconstruct the GW spectrum from the entire scaling regime, and our results are shown in Section~\ref{ss:presentday}. Due to the logarithmic increase of $\mu$ (and of $\xi$), the spectrum has substantial deviations from scale invariance. This enhances its amplitude at low frequencies, and makes it observable by multiple planned experiments for $f_a\gtrsim10^{14}$ GeV. Importantly, the detection prospects are not significantly altered by the remaining uncertainties. We note that similar scaling violations have previously been predicted in the GW spectrum from global strings, using models aiming to capture the emission from small loops, and we discuss the relation of our work to the existing literature in Appendix~\ref{app:compare}.
 
In passing, we noted the remarkably self-consistent picture of the scaling regime arising from numerical simulations. In particular, the logarithmic scaling violations of many of the observables seem to be a part of the scaling regime (for instance $\xi$ and the power law $q_a$ of the instantaneous emission spectrum of axions $\propto 1/k^{q_a}$ both increase logarithmically with time). This is particularly convincing because the energy emitted in axions and radial modes in the simulations is reproduced precisely by the energy emission rate in eq.~\eqref{eq:gammaemi} with the  $\xi$ in eq.~\eqref{eq:xivslog} and string tension in eq.~\eqref{muth}, see Appendix~\ref{ss:gwsim}. Moreover, the theoretical expectation that $\Gamma_g/\Gamma_a\propto G\mu^2/f_a^2$ (for trajectories with a fixed shape) is matched in the numerical simulations of the scaling regime, which suggests that the scaling solution in which $\xi$ increases logarithmically is self-similar (showing that this increase is not a transient). Additionally, the GW spectrum shows the similar features as the axion spectrum, but with $q>1$. This suggests that indeed our extrapolation of the axion spectrum is correct and that $q_a>1$ at large log (when the axion will be more weakly coupled to the string cores, like the GWs are).

In Section~\ref{sec:Constraints} we studied other general properties of axions in the post-inflationary scenario (including their dark matter abundance, contribution to dark radiation, and isocurvature perturbations), which led to complementary constraint on the axion mass and decay constant. This analysis shows that for all masses in the range $10^{-28}~\eV \lesssim m_a \lesssim 10^{-17}~\eV$ ultralight axions can have decay constants large enough ($f_a\gtrsim10^{14}$~GeV) to lead to observable GWs. The upper limit on $m_a$ comes from dark matter overproduction, while the lower limit comes from CMB observations (see Figure~\ref{fig:param}). As discussed in Section~\ref{ss:cmb}, this last limit is particularly uncertain and a revised analysis of $G\mu$ from CMB anisotropies could make it irrelevant by weakening the lower bound on $f_a$ for $m_a\lesssim 10^{-28}$~eV. In particular, our results exclude the possibility that GWs from QCD axion strings during scaling are observable, assuming a standard cosmological history, since the dark matter bound in this case requires $f_a\lesssim 10^{10}~\GeV$.\footnote{We cannot exclude signals from the subsequent dynamics of domain walls.}

 As well as constraints, the phenomenological features discussed in Section~\ref{sec:Constraints} also lead to  complementary observational signals of ultralight axions in the post-inflationary scenario, including $\Delta N_{\rm eff}$ and isocurvature perturbations. For values of $f_a$ that lead to observable GWs these are close to current bounds, and within reach of future improvements. Moreover, a PQ symmetry with large $f_a$ is most easily restored in the early Universe for large Hubble during inflation, so tensor modes with an amplitude close to the current observational upper bound might be present if GWs from strings are discovered. We also see that any interpretation of the recent possible GW signal by the pulsar timing experiment NANOgrav as being from global strings of an ultralight axion is in tension with cosmological observations. Although we do not attempt a detailed analysis of the possible signal, from Figure~\ref{fig:OmegaGW} a GW signal of the observed magnitude (with a standard cosmology) requires $f_a \gtrsim 10^{15}~\GeV$, which from Figure~\ref{fig:param} is in conflict with bounds on $\Delta N_{\rm eff}$ from BBN.\footnote{The power law of the spectrum extracted by NANOgrav might also suggest that the signal is not approximately scale invariant as is predicted by global strings. GWs from QCD axion strings in non-standard cosmologies has been recently considered in~\cite{Ramberg:2019dgi}, which modelled the GW emission from loops and  suggested that the NANOgrav result could be interpreted as the emission from QCD axion strings if the equation of state is $w<1/3$~\cite{Ramberg:2020oct}.} 
 
We also note that the small axion masses mentioned above are theoretically plausible, see e.g. \cite{Halverson:2017deq} and references therein. It appears inevitable that quantum gravity  will explicitly break all global symmetries \cite{Holman:1992us,Kamionkowski:1992mf,Barr:1992qq}, including the PQ one. The resulting breaking is often exponentially suppressed (e.g. in the ratio $f_a/M_{\rm P}$ \cite{Kallosh:1995hi,Alonso:2017avz}) and could therefore lead to the required ultra-light masses.

Finally we consider the remaining open questions and directions for future work. From a theoretical direction, it would be valuable to understand further which classes of axion models cosmic strings can form in. In particular, we have focused on axions that arise as the PNBG of a symmetry that is realised in four dimensional Lagrangian (e.g. from a scalar with a symmetry breaking potential or a new sector that runs into strong coupling, i.e. a composite axion model). Such axions can appear in string theory models from the closed string sector \cite{Ibanez:1999it}. However, in string theory compactifications axions often come from the open string sector. Although the conventional picture is that cosmic strings cannot form in this case, given the uncertainties surrounding early Universe cosmology and de Sitter vacua in string theory we believe this merits further study.

It would be also useful to explore the boundary between the pre- and post-inflationary scenarios (as a function of the masses of the radial mode and PQ fermions) in the QCD axion case in more detail. For instance, in the KSVZ QCD axion model \cite{Kim:1979if,Shifman:1979if} the extra fermions required to induce the PQ anomaly can lead to PQ symmetry restoration at temperatures parametrically below $f_a$ if the radial mode is light, in the same way as in the example theory we studied in Section~\ref{sec:lightrad} and Appendix~\ref{app:lightR}.

Another direction in which to extend our work is the study of local strings, which arise from a spontaneously broken gauged U(1) symmetry. In a system of local stings all the degrees of freedom are massive, and, when the Hubble parameter is much smaller than the UV physics scale, the energy might be radiated only in GWs. However, the emission of heavy modes is efficient at small values of $\log(m_r/H)$ (where $m_r$ is the mass of the heavy modes). Consequently, extrapolation will be essential to determine the GW spectrum at observable frequencies.\footnote{Indeed, at small log the system of local strings seems to resemble the global string one in many ways.}

We also note that the GW signals from axion strings are fairly close to the maximum reach of proposed detectors. Consequently, continued detailed analysis of astrophysical foregrounds will be essential if such signals  are to be identified.\footnote{Note also that, if an ultralight axion makes up an order one percentage of the observed DM, the oscillations in its gravitational field  lead to signals at pulsar timing array frequencies that are of the same order as the stochastic GW background from black hole binaries~\cite{Khmelnitsky:2013lxt}. Axion miniclusters that appear in the postinflationary scenario could also leave imprints in pulsar timing arrays through their gravitational interactions~\cite{Lee:2020wfn}, though there are still large uncertainties on the properties of these objects.} Although challenging, subtraction of e.g. the foreground from neutron star and black hole binaries in the BBO frequency range appears feasible~\cite{Cutler:2005qq}. A careful study of the impact of the foregrounds on the prospects of the detecting  GW signals from strings in different frequency ranges would be worthwhile in the the future.

\section*{Acknowledgements}

We are very grateful to Giovanni Villadoro for countless invaluable discussions and collaboration on related work. We thank Anson Hook, Junwu Huang, Robert Lasenby and Giovanni Villadoro for helpful comments on a draft. We thank Joan Elias-Miró for useful discussions on the effective theory of strings. We acknowledge SISSA and ICTP for granting access at the Ulysses HPC Linux Cluster, and the HPC Collaboration Agreement between both SISSA and CINECA, and ICTP and CINECA, for granting access to the Marconi Skylake partition. We also acknowledge use of the University of Liverpool Barkla HPC cluster.

\appendix

\section{More Details on GWs from Strings} \label{app:analytics}

In this Appendix we give more details on the analytic derivations of Section~\ref{sec:GWs}. In particular, we will discuss the GW emission from Nambu--Goto strings in Appendix~\ref{app:moretheory} and from the whole scaling regime in Appendix~\ref{app:strings}.

\subsection{Axions and Gravitational Waves from Nambu--Goto Strings} \label{app:moretheory}

In this Appendix we derive eq.~\eqref{EaEg}  in detail and compute the coefficients $r_a[X]$ and $r_g[X]$, showing that they are invariant under rescaling of the string trajectory.

Eq.~\eqref{EaEg} can be rewritten in terms of the transverse metric fluctuation $H^{\mu\nu}\equiv h^{\mu\nu}-\frac12\eta^{\mu\nu}h$ and reads $\partial_\alpha\partial^\alpha H^{\mu\nu}=16\pi G T_s^{\mu\nu}$. At large distance $r\equiv\abs{\vec{x}}$ from the region where the trajectory $\vec{X}$ is localised, eq.~\eqref{NGeqs2} and this last equation provide the axion and gravitational wave field
\begin{align}\label{solA}
A^{\mu\nu}=&\frac{f_a}{2\sqrt{2}r}\int{d\sigma}\left(\dot{X}^{\mu}{X'}^{\nu}-{X'}^{\mu}\dot{X}^{\nu}\right),\\
H^{\mu\nu}=&\frac{4G\mu}{r}\int{d\sigma}\left(\dot{X}^{\mu}\dot{X}^{\nu}-{X'}^{\mu}{X'}^{\nu}\right) ,
\label{solH}
\end{align}
where we 
neglected terms that decay faster than $1/r$. The right hand sides of eqs.~\eqref{solA} and~\eqref{solH} are evaluated at the retarded time $t'=t-|\vec{x}-\vec{X}|$, which at this order in $1/r$ reads $t'=t-r+\vec{X}\cdot \vec{n}$ with $\vec{n}\equiv\vec{x}/r$. Clearly eqs.~\eqref{solA} and~\eqref{solH} verify the wave relations $\partial_\alpha H^{\mu\nu}=n_\alpha \dot{H}^{\mu\nu}$ with $n^0\equiv 1$ (so that $n^2=0$) up to $1/r^2$ terms. The harmonic gauge condition is rewritten as $\partial_\mu H^{\mu\nu}=0$ which implies the relation $n_\mu\dot{H}^{\mu\nu}=0$ for the solutions of eqs.~\eqref{solA} and~\eqref{solH}.

The axion and gravitational wave energy radiated per unit time at infinity is given by $dE/dt\equiv -\int d^3x\dot{T}^{00}$, where $T^{\mu\nu}$ is the energy momentum tensor of the axion (from the second term in eq.~\eqref{NGaction}) or the gravitational waves, which are respectively
\begin{align}\label{Ta}
T_a^{\mu\nu}=& \, F^{\mu\alpha\beta}F^{\nu}_{\ \,\alpha\beta}-\frac16\eta^{\mu\nu}F^{\alpha\beta\gamma}F_{\alpha\beta\gamma} , \\
T_g^{\mu\nu}=& \, \frac{1}{32\pi G}\left(\partial_\mu H^{\alpha\beta}\partial_\nu H_{\alpha\beta}-\frac12\eta^{\mu\nu}\partial_\alpha H\partial^\alpha H \right) \ ,\label{Tg}
\end{align}
with $H=H^\alpha_\alpha$. We can use the conservation of $T^{\mu\nu}$ and Gauss theorem to rewrite $dE/dt=\int d^3x\partial_j T^{j0}=\int d{\Sigma^j}T^{j0}$ where the last integral is done on the sphere at spatial infinity. So,
\begin{equation}\label{dEdt}
\frac{dE}{dt}=\lim_{r\to\infty}r^2\int {d\Omega}\, n^jT^{j0} \ . 
\end{equation}
Plugging eqs.~\eqref{solA} and~\eqref{solH} into eqs.~\eqref{Ta} and~\eqref{Tg}, and then plugging these in eq.~\eqref{dEdt}, the whole (neglected) subleading dependence on $1/r$ vanishes in the limit and we obtain eq.~\eqref{EaEg}. For instance, the coefficient for the gravitational wave emission is 
\begin{align}\label{rg}
r_g[X]=\int \frac{d\Omega}{2\pi} \left\{\left[\int d\sigma \partial_t(\dot{X}^\mu\dot{X}^\nu-{X'}^\mu{X'}^\nu) \right]^2-\left[\frac12\int d\sigma \partial_t(\dot{X}^2-{X'}^2)\right]^2\right\} \ ,
\end{align}
where again the right hand side is evaluated in $t'=t-r+\vec{X}\cdot\vec{n}$. A similar equation holds for the axion emission. The parameter $\sigma$ can be taken to parametrise the string trajectory at a fixed time, and in this case it runs in the interval $[0,L]$ where $L$ is the length of the string trajectory. As initially claimed, eq.~\eqref{rg} is invariant under the rescaling $t\to \alpha t$ and $L\to \alpha L$ (and $\vec{x}\to\vec{x}$). This can be easily seen by noticing that\footnote{This is because, by dimensional analysis, $X^{\mu}(t,\sigma)$ must be proportional to $\sigma$ times a dimensionless function of all the other parameters on which $X^{\mu}$ depends, i.e. $t/\sigma$, $L/\sigma$ and $t/L$, etc.\,. Such a function is therefore left invariant by the mentioned rescaling.} $X^{\mu}(\alpha t,\sigma)=\alpha X^{\mu}(t,\sigma/\alpha)$ and making the change of variable $\sigma\to\alpha \sigma$.

Note that in the nonrelativistic limit the dependence on $\vec{n}$ in $t'$ is subleading and one can perform the angular integral in eq.~\eqref{rg} exactly. The result is the well known quadrupole approximation
\begin{align}
r_a[X]=&\frac{2\pi}{3}\left[\int d\sigma\partial_t(\dot{X}^i {X'}^j-\dot{X}^j {X'}^i)\right]^2 ~, \\
\quad r_g[X]=&\frac15\left\{\left[\int d\sigma\partial_t^3(\dot{X}^i  \dot{X}^j)\right]^2-\frac13\left[\int d\sigma\partial_t^3(\dot{X}^i \dot{X}^i)\right]^2\right\}.
\end{align}

\subsection{GW Spectrum during the Scaling Regime} \label{app:strings}

In this Appendix we give more details on the derivation of the GW spectrum in eqs.~\eqref{totalspectrum} and~\eqref{eq:Omegaapprox}. We perform in eq.~\eqref{eq:drhogdk} the change of variable $x=k'/H(t')$ with $k'\equiv kR/R'$, and use eqs.~\eqref{gammagw} and~\eqref{eq:xivslog}. As a result, for $R\propto t^{1/2}$ and in the large log limit
\begin{equation}\label{eq:drhodlogkint}
\frac{\partial \rho_g}{\partial \log k}\left[k,t\right]=8\pi^3 c_1 r Gf_a^4 H^2\int^{k/H}_{\max\left[x_0,\frac{k}{\sqrt{HH_1}}\right]} dx\, F_g\left [x,y\right ] \log^4\left(x^2\frac{m_rH}{k^2}\right)\,,
\end{equation}
where $y\equiv m_rk^2/(Hx^2)$ and, as mentioned in Section~\ref{ss:setup}, we approximated the momentum distribution to be $F_g[x,y]=0$ for $x<x_0$. For $k<x_0\sqrt{HH_1}$, the lower extreme of the integral in eq.~\eqref{eq:drhodlogkint} is just $x_0$. Approximating the momentum distribution with a single power law, i.e.
\begin{equation}{}\label{eq:Fsimple}
F_g[x,y]= \left \{
\begin{array}{l c} 
\frac{(q-1) x_0^{q-1}}{x^q} & x\in [x_0,y] \\
0 & x\notin [x_0,y]
\end{array} \right. \,,
\end{equation}
eq.~\eqref{eq:drhodlogkint} for $k<x_0\sqrt{HH_1}$ leads to
\small
\begin{equation}\label{eq:drhodlogkfull}
\begin{split}
&\frac{\partial \rho_g}{\partial \log k}\left[k,t\right]=8\pi^3 c_1 r Gf_a^4 H^2\log^4\left(\frac{m_r}{H}\right)\Bigg\{\left(1-2\frac{\log(k/k_0)}{\log}\right)^4-\left(\frac{k_0}{k}\right)^{q-1}+\\
&+\frac{8}{(q-1)\log}\left[\left(1-2\frac{\log(k/k_0)}{\log}\right)^3-\left(\frac{k_0}{k}\right)^{q-1}\right]+\frac{48}{(q-1)^2\log^2}\left[\left(1-2\frac{\log(k/k_0)}{\log}\right)^2-\left(\frac{k_0}{k}\right)^{q-1}\right]+\cdots\Bigg\},
\end{split}
\end{equation}
\normalsize
where $k_0\equiv x_0H$ and the dots stand for terms proportional to further inverse powers of $(q-1)\log$ (up to $(q-1)^{-4}\log^{-4})$. In the large log limit and as long as $q-1$ remains definitely larger than $1/\log$, only the first two terms in eq.~\eqref{eq:drhodlogkfull} survive (see also the discussion in Section~\ref{ss:setup}). The second term in eq.~\eqref{eq:drhodlogkfull} is negligible for $k\gtrsim x_0H$,\footnote{This term encodes the IR profile of $\partial \rho_g/\partial \log k$, and as visible in Figure~\ref{fig:Fg} is not well resembled by the simple assumption of a hard IR cutoff for $F_g[x,y]$.} and the first terms indeed corresponds to the full eq.~\eqref{totalspectrum}. For higher momenta, $k>x_0\sqrt{HH_1}$, the lower extreme in eq.~\eqref{eq:drhodlogkint} is $k/\sqrt{HH_1}$ and the integration gives
\begin{equation}\label{eq:drhodlogkfullUV}
\begin{split}
&\frac{\partial \rho_g}{\partial \log k}\left[k,t\right]=8\pi^3 c_1 r Gf_a^4 H^2\log^4\left(\frac{m_r}{H}\right)\left[\frac{x_0\sqrt{HH_1}}{k}\right]^{q-1}\left[1-\left(\frac{H}{H_1}\right)^{\frac{q-1}{2}}+\cdots\right],
\end{split}
\end{equation}
which gives the dependence on $1/k^{q-1}$ mentioned in Section~\ref{ss:setup} (the dots stand again for subdominant $(q-1)^{-1}\log^{-1}$ corrections).

Note that if the effective number of degrees of freedom in thermal equilibrium $g$ is not constant, entropy conservation ($g R^3T^3=\rm{const}$) and the Friedmann equations ($H^2=1/(2t)^2\propto g T^4$, valid away from particle thresholds) imply $R\propto g^{-1/12}t^{1/2}$. In this case, neglecting time derivatives of $g$, the same change of variable as before provides the overall factor $(g(t_k)/g(t))^{1/3}$ in eqs.~\eqref{eq:drhodlogkint},~\eqref{eq:drhodlogkfull} and~\eqref{eq:drhodlogkfullUV}, plus a change in the argument of the logarithms of a factor $(g(t_k)/g(t))^{1/6}$, which is extremely negligible. Finally, we observe that the spectrum in eq.~\eqref{eq:drhodlogkfull} is has a very similar form as the axion spectrum in Appendix~E.1 of~\cite{moreaxions}, with however two more powers of $\log$ since $\Gamma_g/\Gamma_a\propto\log^2$ at large log.

From the end of the scaling regime (when $H=H_\star$) to today the GWs redshifts freely, i.e. $\frac{d\rho_g}{d\log k}[t_0,k]=(R_\star/R_0)^4\frac{d\rho_g}{d\log k}[t_\star,k R_\star/R_0]$. From eq.~\eqref{totalspectrum} and conservation of entropy density, we get the spectrum today in the $\log\gg1$ limit
\begin{equation}\label{eq:Omegaapproxanalytic}
\frac{d \Omega_{\rm gw}}{d \log k}=\frac{c_1\pi^4 r f^4_a}{90M^4_{\rm P}}\frac{g_0T_0^4}{\rho_c}\left(\frac{g_0}{g_k}\right)^{1/3}\log^4\left[{\frac{3\sqrt{10}g_\star^{1/6}M_{\rm P}k^2}{\pi x_0^2g_0^{2/3}m_r T_0^2}}\right],
\end{equation}
where $T_0$ and $g_0$ are the temperature of photons and the effective number of relativistic degrees of freedom today. Eq.~\eqref{eq:Omegaapprox} is the numerical evaluation of eq.~\eqref{eq:Omegaapproxanalytic} and is valid in the momentum range $x_0 H_\star(R_\star/R_0)<k<x_0 \sqrt{H_\star H_1}(R_\star/R_0)$. Although eq.~\eqref{eq:Omegaapproxanalytic} is a very good approximation of the spectrum at large log, as mentioned in Section~\ref{ss:presentday} to obtain the lines in Figure~\ref{fig:OmegaGW} we integrated numerically eq.~\eqref{eq:drhogdk}, accounting for the smooth change in $g$ and using a double power-law form for $F_g[x]$ (this resembles well the simulation results of Figure~\ref{fig:Fg}). In particular we used $F_g[x]\propto x^3$ for $x\lesssim x_0$ and $F_g[x]\propto 1/x^2$ for $x\gtrsim x_0$, as in eq.~(38) of~\cite{Gorghetto:2018myk} with $x_2\to\infty$.

\section{Details of Simulations} \label{app:sim}

In our numerical simulations of the scaling regime we evolve the equations of motion of eq.~\eqref{eq:LPhi},
\begin{equation} \label{eq:eomscaling}
\ddot{\phi} +3 H\dot{\phi}- \frac{\nabla^2\phi}{R^2}+\phi \frac{m_r^2}{f_a^2}\left(|\phi|^2 -\frac{f_a^2}{2}\right)=0  \ , 
\end{equation}
on a discrete lattice. The details of our implementation are as in~\cite{Gorghetto:2018myk,moreaxions} where extensive further discussion may be found. In this Appendix we simply summarise some key aspects, and describe the extra features in our present work.

As mentioned in the main text, we study both the fat and physical string systems. The latter has two advantages: 1) it takes a longer physical time for the maximum log to be reached, and 2) the string core scale changes with time at the same rate as the momentum redshifts, which means that axion waves emitted at the string core scale remain in the UV part of the spectrum (this leads to a cleaner spectrum in the physically relevant region $k<m_r/2$, as discussed in \cite{moreaxions}). In all cases we start from initial conditions that are close to the attractor solution identified in \cite{Gorghetto:2018myk}, with corresponding number of strings per Hubble volume $\xi$ represented in Figure~\ref{fig:xi}. This maximises the interval in $\log$ over which the properties and changes in the attractor solution can be extracted. Such initial conditions are obtained as in \cite{moreaxions} by a procedure that gives strings with the correct core size, which avoids a large amount of energy being emitted as axions and radial modes as the strings relax (and would contaminate the subsequent axion spectrum, especially for the physical string system).

We calculate the GWs emission following the method developed in~\cite{GarciaBellido:2007af}, to which we refer for the details. In this approach, six independent fields $u_{ij}$, with $u_{ij}=u_{ji}$, are evolved according to eq.~\eqref{eq:gweom}, but sourced by the total energy momentum tensor (rather than its TT part). The Fourier Transform (FT) of the GW field, $h_{ij}(t,\vec{k})$, 
 is then obtained by projecting the FT of $u_{ij}$ as
\beq\label{eq:hk}
h_{ij}(t,\vec{k})= \left(P_{il}(\hat{k})P_{jm}(\hat{k})-\frac{1}{2} P_{ij}(\hat{k})P_{lm}(\hat{k})  \right) u_{lm}(t,\vec{k}) ~,
\eeq
where $P_{lm}(\hat{k})\equiv \delta_{ij}- k_i k_j/|\vec{k}|^2$ (it is indeed easy to see that $h_{ij}$ solves eq.~\eqref{eq:gweom} if and only if $u_{ij}$ solves the same equation sourced by the total energy momentum tensor).\footnote{See~\cite{Figueroa:2011ye} for an analysis demonstrating that the lattice version of the projector does not introduce systematic errors.} This procedure avoids the need to obtain the TT part of the energy momentum tensor at every simulation timestep, which greatly reduces the computational cost. Instead FTs are only required at particular timeshots when the $h_{ij}$ are needed to evaluate the GW energy and spectrum. We use the same approach when evaluating the GW emission at the time of the axion mass turn on in axion only simulations, as studied in Appendix~\ref{app:massGW}, simply by substituting the appropriate energy momentum tensor.

As mentioned in Section~\ref{ss:gwsim}, numerical simulations including the GW backreaction (analysed  in Appendix~\ref{app:back}) require that eq.~\eqref{eq:eomscaling} is evolved with the additional term $R^{-2}h_{ij}\partial_i\partial_j\phi$ on the left hand side, and thus the expression of $h_{ij}$ needs to be known at every timestep. For such simulations we calculate $h_{ij}$ from $u_{ij}$ via eq.~\eqref{eq:hk} at all time steps (by carrying out all of the required FTs and anti-FTs). Given the computational cost, we limit ourselves to grids of size $800^3$ for this analysis, which allows $\log \simeq 6$ to be reached. Such a log is sufficient for the system to be in scaling and emitting axions with a momentum distribution with a gap between the IR peak and UV scale for an interval $\Delta \log \simeq 2$, enough to extract the time-dependence of the relevant physical observables. For our main simulations, which are much less computationally demanding, we use larger grids, with $2500^3\div3000^3$ lattice points.

\section{Further Results from Simulations} \label{app:simR}
In this Appendix we provide further results from the numerical simulations of the string system. In particular, in Appendix~\ref{app:simscal} we give more details on the scaling regime and the emission of radiation from long stings
. In Appendix~\ref{app:back} we discuss the effect of the backreaction of the GWs on the string network. Finally, in Appendix~\ref{app:destroy} and~\ref{app:density} we describe the end of the scaling regime for a temperature-independent mass and the power spectrum of axion overdensities after the string network is destroyed.

\subsection{The Scaling Regime} \label{app:simscal}

\begin{figure}[t]
	\begin{center}
		\includegraphics[width=0.46\textwidth]{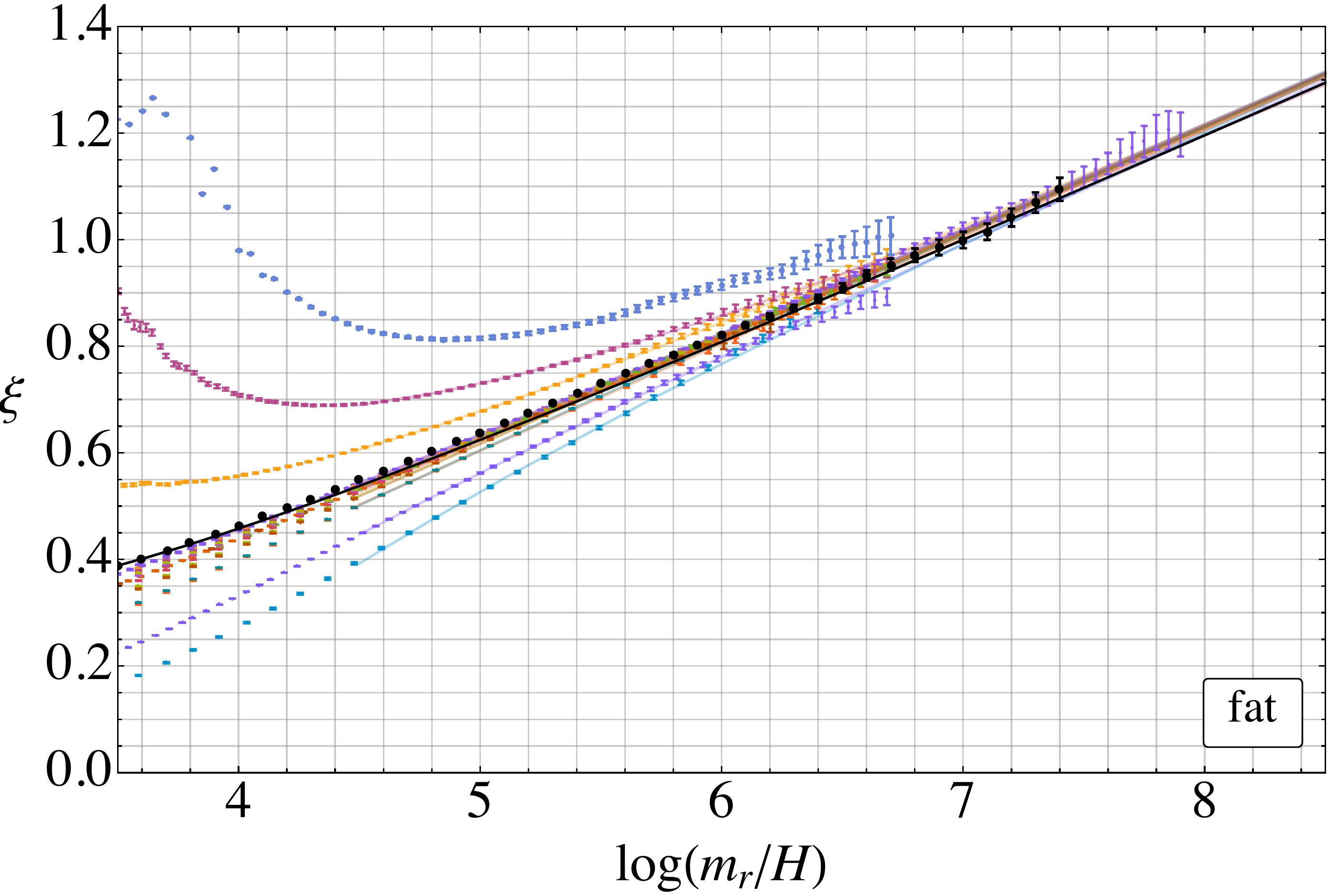}
		\qquad 		\includegraphics[width=0.46\textwidth]{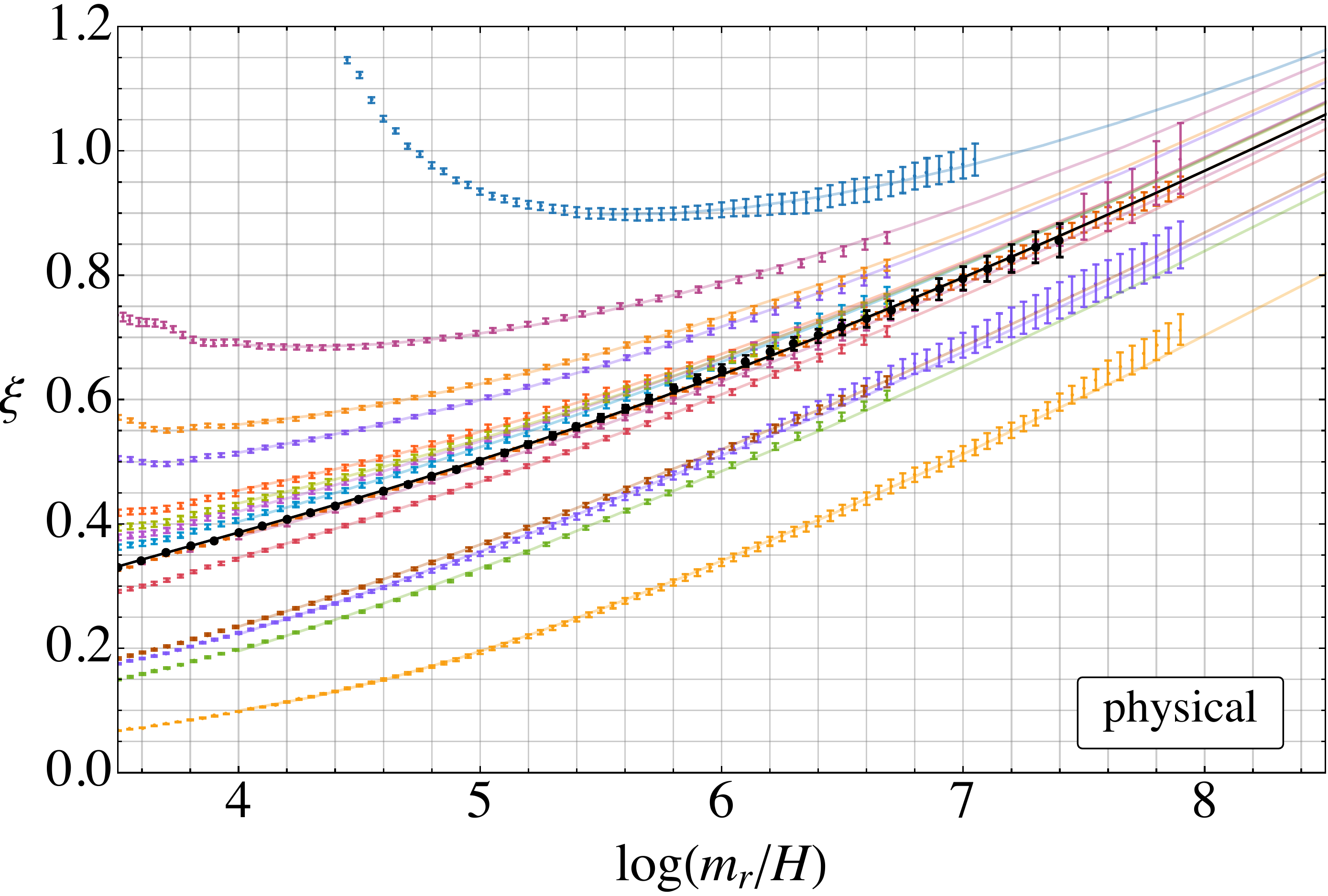}
	\end{center}
	\caption{The number of strings per Hubble volume $\xi$ for the fat (left) and physical (right) string systems. The initial conditions that we use for the GW analysis, starting close to the attractor, are plotted in black. The evolution of the system starting from other initial conditions (studied in \cite{Gorghetto:2018myk,moreaxions}) is shown for comparison.	 \label{fig:xi}} 
\end{figure} 

In Figure~\ref{fig:xi} we show the evolution of the number of strings per Hubble volume $\xi$, where the logarithmic increase mentioned in Section~\ref{ss:Review} is evident. The black points correspond to the simulations used in this work, while for comparison we also show data from~\cite{Gorghetto:2018myk,moreaxions} starting from different initial conditions and reaching larger log.\footnote{As mentioned, in our present simulations, evolving the fields $u_{ij}$ adds to the computational cost, so to gain sufficient statistics we use slightly smaller grids.} As discussed in~\cite{Gorghetto:2018myk,moreaxions}, the data clearly rules out any behaviour that saturates soon after $\log=8$. Moreover, for the most overdense initial conditions, $\xi$ first drops and then starts rising again, suggesting that the logarithmic growth is not a transient but a part of the attractor solution.

As shown in~\cite{Gorghetto:2018myk}, the logarithmic growth affects both long strings (defined to be the strings with length much larger than $H^{-1}$) and loops (which are all the other strings). In particular, $\xi$ restricted only to long strings -- which we call $\xi_L$ -- is at any time during the scaling regime a fixed fraction of $\xi$, i.e. $\xi_L=f_L \xi$ with $f_L\simeq 0.814$ for the fat string system (this can be extracted from the jump of the cumulative distribution of $\xi$ in Figure~3 of~\cite{moreaxions}; this jump makes the distinction between long strings and loops sharp). As a result, the fraction of strings in loops is also constant. The fact that the proportion of length in long strings and loops remains the same  provides another convincing evidence that the logarithmic growth is a property of the scaling solution.

\subsubsection*{\emph{Energy Emission Rate and Long Strings}}

As mentioned in Section~\ref{ss:Review}, in order to maintain the scaling regime the energy that needs to be released from the string system is $\Gamma=\dot{\rho}_s^{\rm free}-\dot{\rho}_s$. We derived such a quantity in eq.~\eqref{eq:gammaemi} assuming that $\rho_s^{\rm free}\propto R^{-2}$ (see~\cite{Gorghetto:2018myk} for the details). However, this derivation, strictly speaking, applies only to the part of the string network whose energy decreases proportionally to $R^{-2}$ in the free limit, which are only the long strings.\footnote{Sub-horizon loops redshift as nonrelativistic matter in the free limit.}  For the sake of a completely consistent treatment, we therefore split the energy density in strings as $\rho_s=\rho_s^L+\rho_s^{\rm loops}$, where $\rho_s^{L}=\xi_L\mu_L/t^2$ is the energy in long strings and $\rho_s^{\rm loops}$ the energy in loops. The tension of long strings $\mu_L$ is expected to take the form $\mu_{\rm th}$ in eq.~\eqref{muth} during the scaling regime. The total energy lost by long strings, $\Gamma=\dot{\rho}_{\rm free}-\dot{\rho}_s^L$, is then correctly given by eq.~\eqref{eq:gammaemi} but with $\xi$ substituted with $\xi_L$, i.e. 
\begin{equation} \label{eq:gammaemiLong}
\Gamma=\frac{\xi_L\mu_{\rm th}}{t^2} \left[2H-\frac{\dot \xi}{\xi}-\frac{\pi f_a^2}{\mu_{\rm th}} \left(H+\frac{\dot\eta}{\eta} 
-\frac12 \frac{\dot\xi}{\xi} \right)\right] \,,
\end{equation}
where for convention we evaluate $\mu_{\rm th}$ with $\xi$ (rather than $\xi_L$: the difference is a constant reabsorbed in $\eta$) and in the square bracket we can use $\xi$ instead of $\xi_L$ given that they are proportional.

The energy $\Gamma$ lost by long strings is either directly converted into axion and radial mode radiation ($\Gamma^{\rm rad}$), or lost by the formation of string loops ($\Gamma^{\rm loops}$), which continually arise from the intersection of long strings and then decay into radiation. Therefore $\Gamma=\Gamma^{\rm rad}+\Gamma^{\rm loops}$. The total emission rate into axions and radial modes $\Gamma_a+\Gamma_r$ from the network is the sum of the energy emitted into radiation directly from long strings ($\Gamma^{\rm rad}$) and the one from loops. If such loops decay into radiation efficiently, this last quantity also equals the energy lost by long strings by the formation of loops, $\Gamma^{\rm loops}$. As a result,  we expect $\Gamma=\Gamma_a+\Gamma_r$. In the following we will show that numerical simulations reproduce this expectation  remarkably well.

We stress that in this picture the energy emitted into radiation $\Gamma_a+\Gamma_r$ originally comes from long strings. However a part of it comes \emph{directly} from long strings ($\Gamma^{\rm rad}$), and the other \emph{via} loops that decay ($\Gamma^{\rm loops}$), which thus act as an efficient mean dissipation of the energy that is originally in long strings. So, although fixed by $\Gamma$ in eq.~\eqref{eq:gammaemiLong}, the eventual emission into axions happens through both long strings and loops, which is why in Section~\ref{ss:theory} we studied a \emph{generic} trajectory, which includes long strings and loops. Similarly, the total emission into GWs, which for a trajectory with a fixed shape is proportional to the one into axions (see Section~\ref{ss:theory}), depends on both long strings and loops, but we will refer to $\Gamma_g$ as the sum of the two ($\Gamma_g$ will have a contribution from long strings and another from loops, which we will not distinguish, as we cannot distinguish them in simulations). In the following, by proving that $\Gamma=\Gamma_a+\Gamma_r$, we will see that this picture agrees with the evolution of the physical system at least at small log. We will comment on the possible changes of this picture if the energy is not emitted efficiently from the decay of loops at large log.

Similarly to~\cite{Gorghetto:2018myk,moreaxions}, during the evolution of the field we extract the energy densities in axions $\rho_a$ and radial modes $\rho_r$ (that are present at a generic time) from the kinetic energies $\langle\dot{a}^2\rangle$ and $\langle\dot{r}^2\rangle$, where the averages are done over spatial points away from the string cores (to avoid the contribution from the strings). The emission rates are then calculated as the time variation of these energies, namely $\Gamma_a=R^{-4}\frac{d}{dt}(R^4\rho_a)$ and $\Gamma_r=R^{-z}\frac{d}{dt}(R^z\rho_r)$ where $z=4$ for the fat strings.\footnote{Irrespectively of whether they are relativistic or not, free radial modes in the fat string system have energy $\sqrt{k^2+m^2}$ that redshifts as $R^{-1}$, given that $m_r\propto R^{-1}$ (so the energy density redshift as $R^{-4}$ assuming comoving number density conservation).} For the physical system the redshift factor is $3<z<4$, and can be calculated as $\int dk z[k/m_r] \partial\rho_r/\partial k  $ where $z[k/m_r]\equiv3+(k/m_r)^2/((k/m_r)^2+1)$ is the redshift factor of one mode with momentum $k$ (the spectrum of radial modes $\partial\rho_r/\partial k$ can be found in~\cite{moreaxions}).\footnote{This is easily seen by calculating $d\log E/d\log R$ where the energy $E$ is $E=n_{\rm com}(R_0/R)^3\sqrt{m_r^2+k_{\rm com}^2(R_0/R)^2}$ and assuming comoving number density conservation $dn_{\rm com}/dR=0$ and using $dk_{\rm com}/dR=0.$}

\begin{figure}[t]
	\begin{center}
		\includegraphics[width=0.46\textwidth]{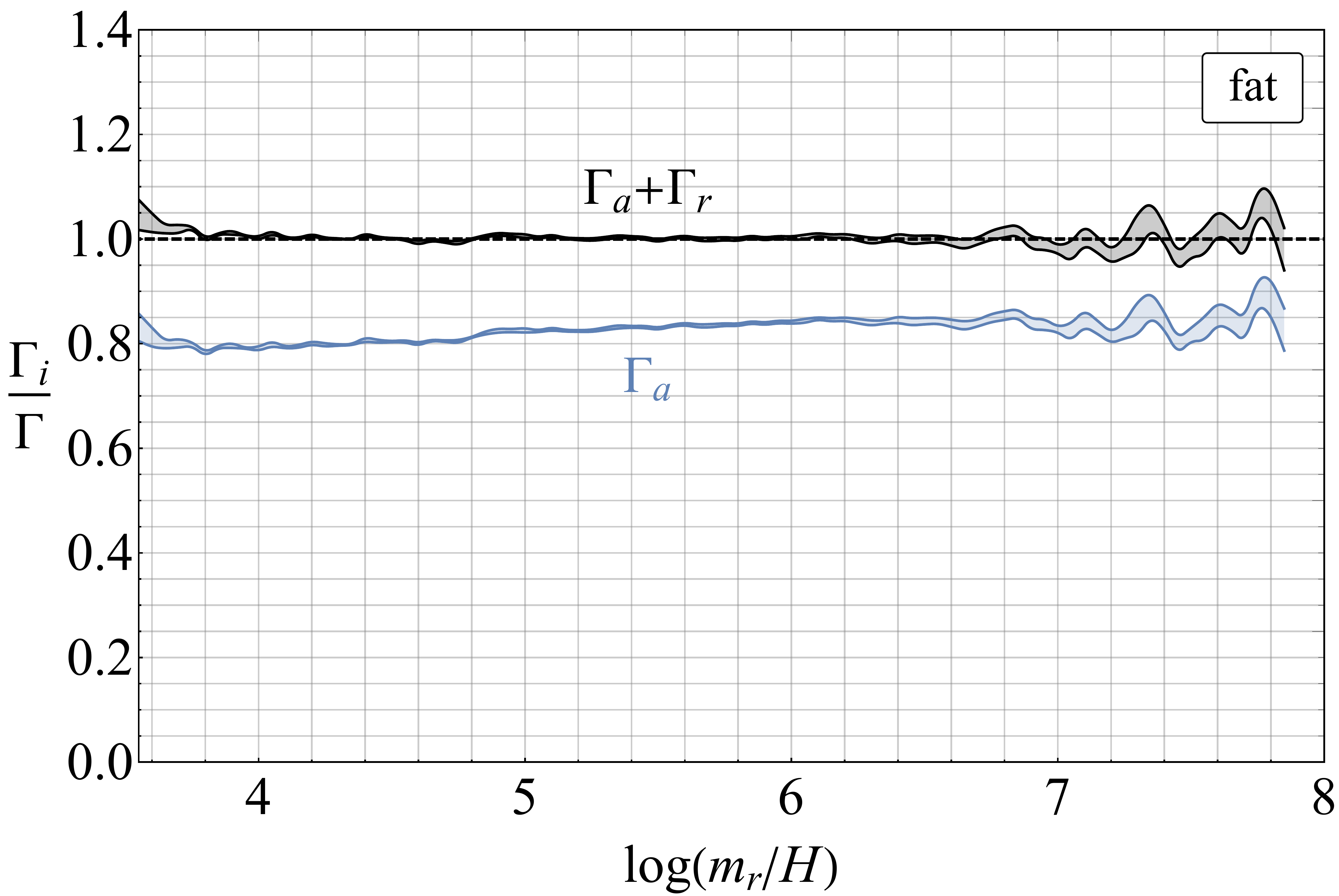}
		\qquad 		\includegraphics[width=0.46\textwidth]{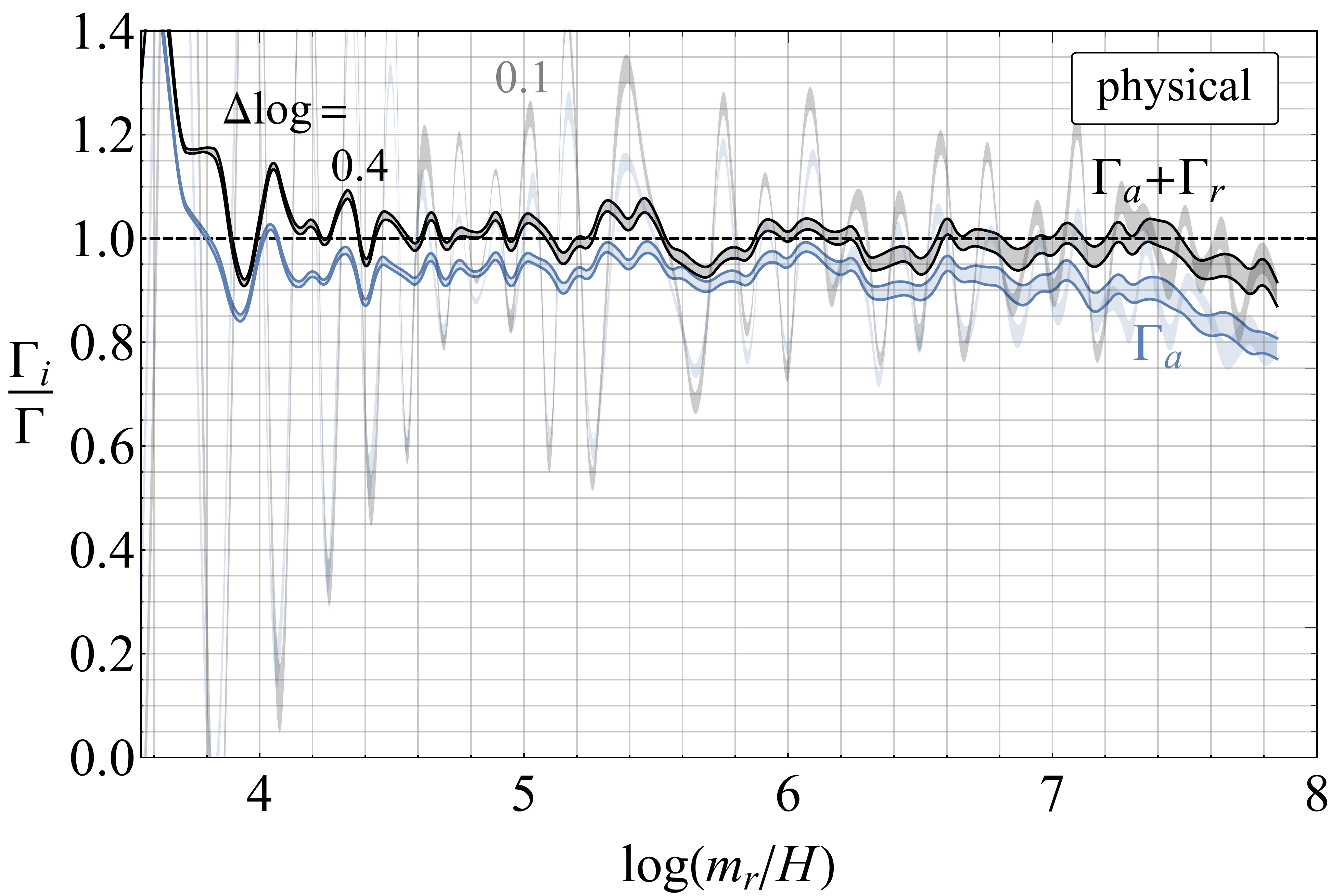} 
	\end{center}
	\caption{The energy density emitted in axions and radial modes, $\Gamma_a$ and $\Gamma_r$, during the scaling regime calculated in the simulation, normalised to the theoretical expectation of the total emission rate from (long) strings $\Gamma$ in eq.~\eqref{eq:gammaemi} after fitting the (constant) value $\eta$, defined in eq.~\eqref{muth}. The fact that $\Gamma_a+\Gamma_r$ coincides with the theoretical expectation $\Gamma$ is a confirmation that the emission rate is described by eq.~\eqref{eq:gammaemi} and that the string tension is reproduced by $\mu_{\rm th}$ in eq.~\eqref{muth} for a fixed value of $\eta$. 
	  \label{fig:Gamma_ratio}} 
\end{figure}

\begin{figure}[t]
	\begin{center}
		\includegraphics[width=0.46\textwidth]{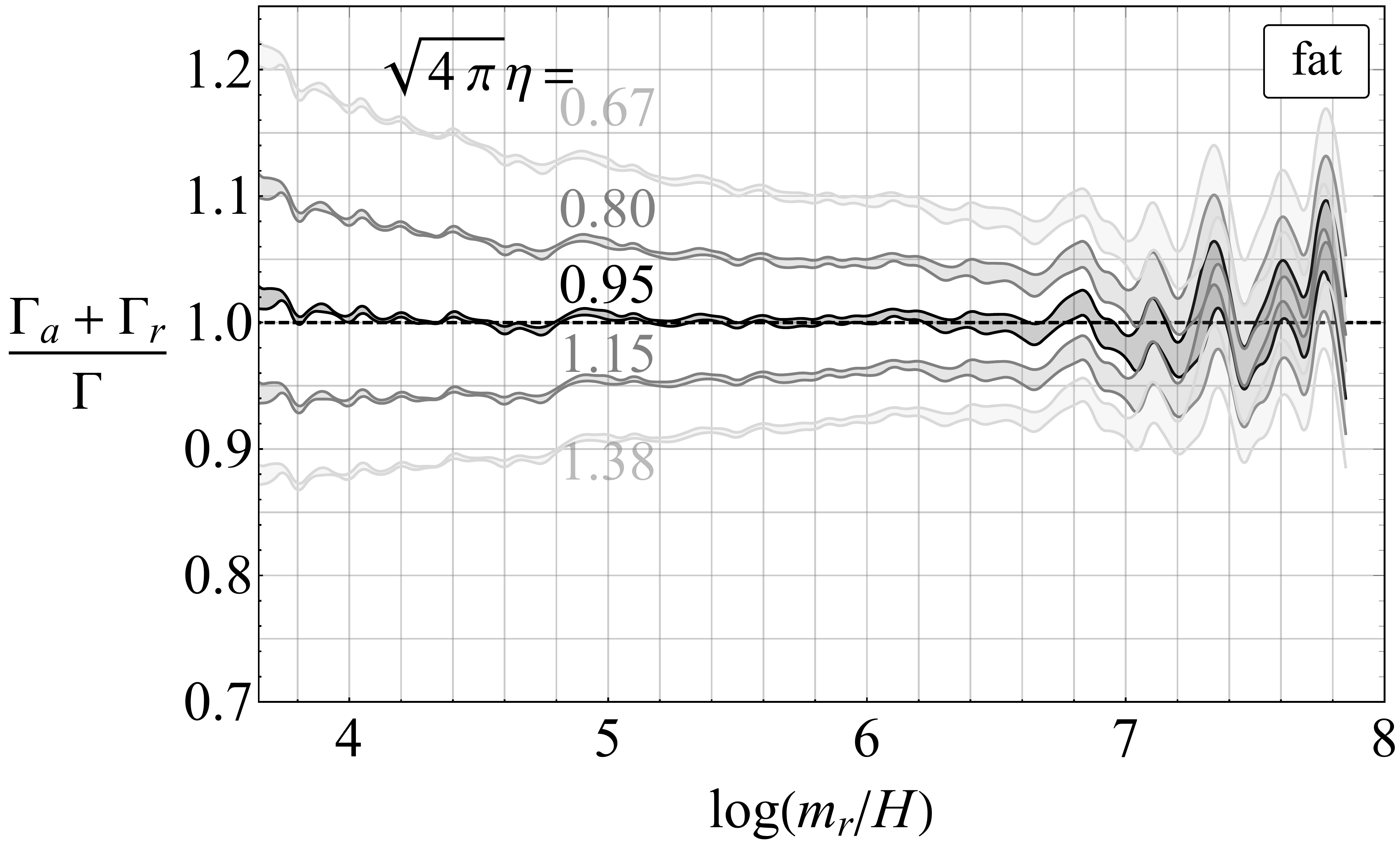}
		\qquad 		\includegraphics[width=0.46\textwidth]{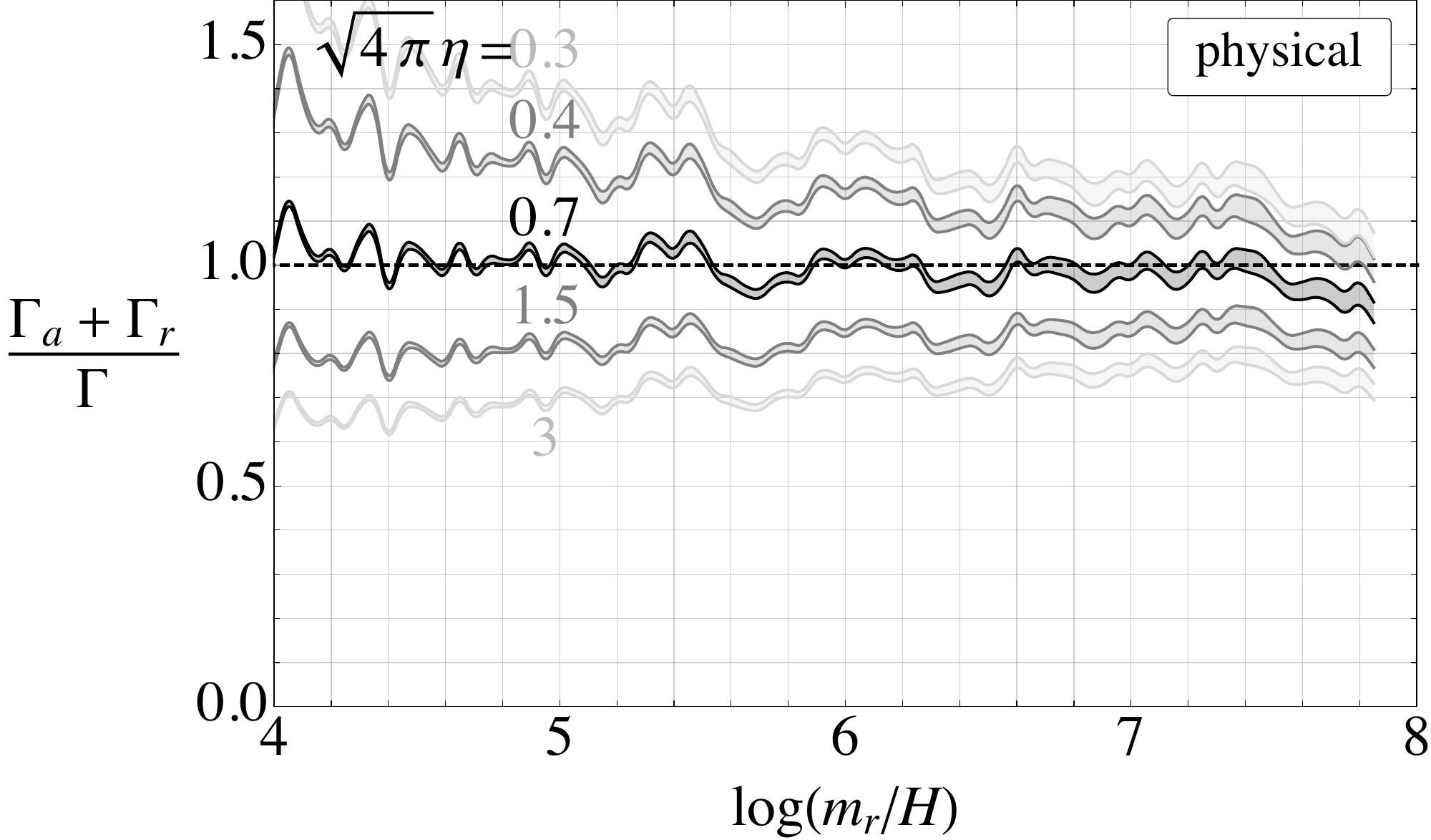} 
	\end{center}
	\caption{ The dependence of the ratio $(\Gamma_a+\Gamma_r)/\Gamma$ on the choice of the parameter $\eta$. Values of $\eta$ outside the intervals $0.8\div1.1$ and $0.4\div1.5$ for the fat and physical strings respectively (shown in light gray) do not lead to a constant ratio, signalling that they are not able to reproduce the emission and therefore the sting tension.  
		\label{fig:Gammaeta}} 
\end{figure}

In Figure~\ref{fig:Gamma_ratio} we show the numerical results for $\Gamma_a$ and $\Gamma_a+\Gamma_r$ for the fat and physical systems (we use the data from the $4500^3$ grids of~\cite{moreaxions}, reaching $\log=7.9$). To allow the direct comparison with the previous discussion, the results are normalised to the theoretical expectation in eq.~\eqref{eq:gammaemiLong} for the total emission rate $\Gamma$. In this last equation the (only) free constant parameter $\eta$ has been fitted in such a way that $\left(\Gamma_a+\Gamma_r\right)/\Gamma$ equals unity. The fact that $\Gamma$ reproduces $\Gamma_a+\Gamma_r$ over more than four $e$-foldings in time (especially clearly for the fat system) is a remarkable confirmation of the theoretical picture described above. In particular this shows that the form eq.~\eqref{eq:gammaemi} captures the total emission rate during scaling when $\xi$ is restricted to the only long strings, that such emission happens either directly or via loops, and that eq.~\eqref{muth} reproduces the effective string tension $\mu_L$ even at small $\log$ for a fixed choice of $\eta$.\footnote{Another approach to measure the string tension would be by subtracting the energy in waves, as was done in~\cite{moreaxions}, which leads to compatible results.}

We now discuss the details for fat and physical strings separately. For the fat string system, shown in Figure~\ref{fig:Gamma_ratio} (left), the agreement between $\Gamma$ and $\Gamma_a+\Gamma_r$ is excellent and $\eta$ is fixed precisely. The value of $\eta$ selected by the fit (in which we consider only $\log>4$ data and, as mentioned, we evaluate $\Gamma$ using $f_L=0.84$), is $\sqrt{4\pi}\eta \simeq 0.95$. This is close to $1/\sqrt{4\pi}$, which is the value this parameter would get if all the long strings were straight and parallel to each other.\footnote{This is seen requiring that the argument of the log in eq.~\eqref{muth} equals the inter-string distance in units of $m_r^{-1}$.} Although such $\eta$ reproduces the emission, we assign to it a conservative $15\%$ uncertainty, i.e $\sqrt{4\pi}\eta_{\rm fat}=0.95(15)$. This uncertainty is estimated by looking at how much the ratio $(\Gamma_a+\Gamma_r)/\Gamma$ varies with log for different choices of $\eta$: as shown in Figure~\ref{fig:Gammaeta}, if $\sqrt{4\pi}\eta$ is outside the range $0.8\div1.1$ the ratio is not constant, suggesting that the choice of this parameter outside this range is not appropriate to describe the emission.\footnote{Note that indeed the definition of long strings is not completely fixed (and so $f_L$), so in principle one can just require the ratio to be a constant close to one rather than exactly one.}\footnote{For the fat string time derivatives are done averaging over $\Delta\log=0.2$.} Notice that the agreement between $\Gamma$ and $\Gamma_a+\Gamma_r$ for the value of $f_L=0.84$ extracted from the jump of loop distribution is remarkable, and is a particularly convincing confirmation of the theoretical discussion above. Even more remarkably, leaving both $\eta$ and $f_L$ as free parameters in the fit $(\Gamma_a+\Gamma_r)/\Gamma=1$ provides a similar value of $\eta$ and a value of $f_L$ that differs by less than $1\%$ from the one extracted from the loop distribution. Notice also that, as mentioned in Section~\ref{ss:Review}, and shown more in detail in~\cite{moreaxions}, radial modes are still produced (i.e. $\Gamma_r\neq0$), though increasingly less with respect to axions (i.e. $\Gamma_r/\Gamma_a$ diminishes), and it is indeed the sum $\Gamma_a+\Gamma_r$ that reproduces~$\Gamma$.

The results for the physical system are shown in Figure~\ref{fig:Gamma_ratio} (right), where we calculate the derivative averaging over $\Delta\log=0.1$ (shaded lines) and over $\Delta\log=0.4$ (solid lines). For such a system some fluctuations are visible, and are particularly evident at early times and with the smaller time averaging. As already noted in~\cite{moreaxions}, these are related to parametric resonance effects between the axion and radial modes, and possibly to the emission from excited string cores due to imperfect initial conditions.\footnote{These effects are clearly visible in the axion instantaneous emission spectrum as studied in \cite{moreaxions}, which has large oscillations at around the string core scale.} As the value of $f_L$ is not known for physical strings, in the fit $(\Gamma_a+\Gamma_r)/\Gamma=1$ (done for $\log>4.5$) both the parameters $\eta$ and $f_L$ are allowed to vary. The best fit values correspond to $f_L\simeq0.9$ and $\sqrt{4\pi}\eta\simeq0.7$, which are those for which we evaluate $\Gamma$ in Figure~\ref{fig:Gamma_ratio} (right). As expected, the fit selects $f_L<1$, and the parameters $f_L$ and $\eta$ turn out to be remarkably very close to the fat string values. As shown in Figure~\ref{fig:Gamma_ratio} (right), despite the fluctuations, for such values of $\eta$ and $f_L$ the ratio $(\Gamma_a+\Gamma_r)/\Gamma$ is in average close to unity at all times. The large fluctuations however lead to a greater uncertainty on the actual value of $\eta$ that reproduces the emission. Similarly to the fat string system, we estimate this uncertainty by varying $\eta$ and looking at the impact on $\Gamma$. As is clear from Figure~\ref{fig:Gammaeta} (right), all the $\eta$'s in the range $0.4\div1.5$ give an approximately constant ratio, while if they are outside this range the ratio starts tilting. We can take this range as a conservative estimate for $\eta_{\rm phys}$. 

As in the fat system, the production of radial modes is non-negligible. However the fluctuations make the time evolution of $\Gamma_r/\Gamma_a$ more unclear than in the fat string system. Notice that such radial modes are mildly relativistic (in particular, $z\simeq3.3$ in the whole time range; this turns out to be an essential information for getting a $\Gamma_r\neq0$ from $R^{-z}\frac{d}{dt}(R^z\rho_r)$).

\begin{figure}[t]
	\begin{center}
		\includegraphics[width=0.46\textwidth]{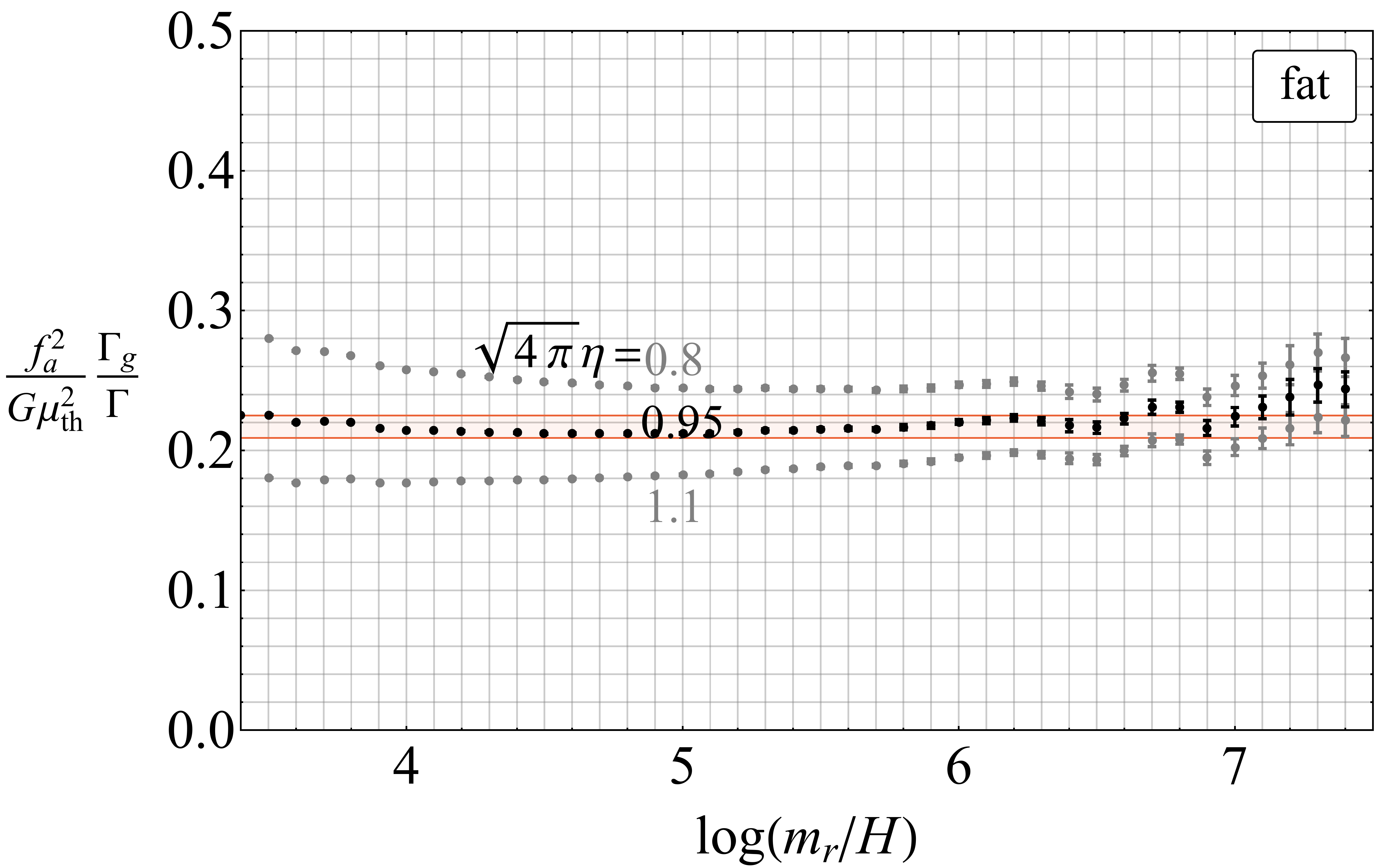}
		\qquad 		\includegraphics[width=0.46\textwidth]{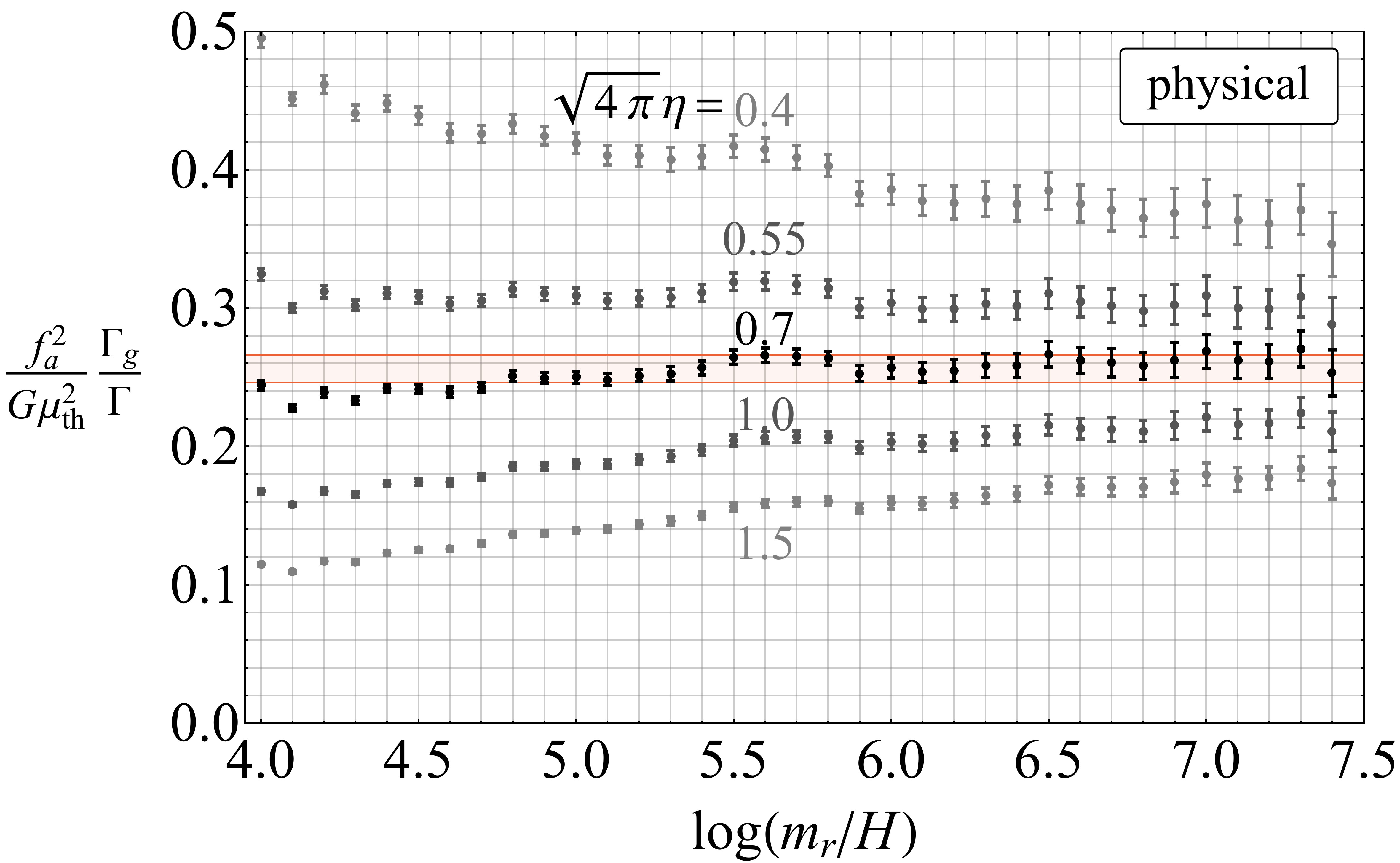}
	\end{center}
	\caption{ The dependence of $f_a^2\Gamma_g/(G\mu^2_{\rm th}\Gamma)$ on the parameter $\eta$, which enters the string tension $\mu_{\rm th}$ for the fat (left) and physical (right) string systems. We plot results for the range of $\eta$ compatible with the measured instantaneous energy emitted to axions and radial modes of Figure~\ref{fig:Gammaeta}. For the value of $\eta$ Across this range, the plotted ratio approaches a constant value, as is expected from the theoretical analysis in Section~\ref{ss:theory}.}
		\label{fig:reta}
\end{figure} 

Let us now discuss the GW emission $\Gamma_g$ and $r$, defined in eq.~\eqref{gammagw}. As mentioned in Section~\ref{ss:gwsim}, the uncertainty on $\eta$ translates into an uncertainty on the string tension $\mu_{\rm eff}$ (more precisely on $\mu_L$; we discuss later the contribution of $\rho_s^{\rm loops}$ to $\rho_s$). This uncertainty also affects the value of $\mu_{\rm th}$ to be used in the calculation of $r_{\rm sim}\equiv f_a^2 \Gamma_g/(G\mu_{\rm th}^2 \Gamma)$ defined in Section~\ref{ss:gwsim}.\footnote{As in the main text, $\Gamma_g$ is calculated by averaging the time derivative of $\rho_g$ over $\Delta\log=0.2$, which is already consistent with the continuum limit.} In Figure~\ref{fig:Gammaeta} we show the value of $r_{\rm sim }$ from different $\eta$'s chosen in the ranges mentioned above (note that $\eta$ enters both in $\mu_{\rm th}$ and in $\Gamma$). Remarkably, the best fit value of $\eta$ selected by the above discussion leads a constant $r_{\rm sim}$. As described in the main text, this matches the expectation from the Nambu-Goto effective theory, further confirms that the value of $\eta$ reproduces the string tension and ensures the energy emitted into GWs can safely be extrapolated to large $\log$. As expected, larger (smaller) values of $\eta$ lead to an increasing (decreasing) $r_{\rm sim}$, which however tends asymptotically to the same constant. Although the best fit values of $\eta$ lead to a constant $r_{\rm sim}$, we conservatively estimate the uncertainty on $r$ from the one on $\eta$ by varying $\eta_{\rm fat}$ and $\eta_{\rm phys}$ in the intervals mentioned before for which $(\Gamma_a+\Gamma_r)/\Gamma$ is constant, and considering the range in which $r_{\rm sim}$ varies at the largest available $\log=7.4$ for such $\eta$'s. For instance, $r_{\rm phys}$ ranges in the interval $0.17\div0.34$.\footnote{Notice that strictly speaking  $\Gamma_a=\xi_L\mu_{\rm th}/t^3$ at large log (rather than $\Gamma_a=\xi \mu_{\rm th}/t^3$). This makes the extraction of the numerical value of $r$ from $r_{\rm sim}$ a factor of $f_L^{-1}$ larger and $\Gamma_a$ a factor of $f_L$ smaller than what mentioned in the main text, leaving in any case the GW emission rate $\Gamma_g$ in eq.~\eqref{gammagw} invariant.}

The discussion above confirms that the tension of long strings $\mu_L$ is reproduced by $\mu_{\rm th}$, i.e. $\rho_s^L=\xi_L\mu_{\rm th}/t^2$, with the $\eta$ fixed as before. As mentioned at the beginning, the energy density in strings $\rho_s=\rho_s^L+\rho_s^{\rm loops}$ however contains also the contribution from loops $\rho_s^{\rm loops}$. This in principle changes $\rho_s$ and therefore could make the effective string tension $\mu_{\rm eff}$ not match $\mu_{\rm th}$ anymore for the same value of $\eta$ (note that $\mu_{\rm eff}$ -- and not $\mu_L$ -- is the one determining the GW emission, as GWs are emitted both from long strings and from loops, and therefore should be the one that leads to a constant $r=f_a^2\Gamma_g/(G\mu^2\Gamma_a)$, since $\Gamma_g$ is the \emph{total} GW emission rate). 

First, we expect a possible overall (constant, see~\cite{moreaxions}) factor on the total string tension $\mu_{\rm eff}$ due to the non-trivial boost factor of loops but, as mentioned in footnote~\ref{ft1}~in Section~\ref{ss:gwsim}, this gives an overall correction to $\mu_{\rm th}$ and could be simply reabsorbed in the definition of $r$. In particular, $\Gamma_g$ defined in eq.~\eqref{gammagw} with $r$ extracted from Figure~\ref{fig:rsim} still provides the correct large log behaviour of the emission.

Moreover, the fact that $\eta$ predicted by long strings works well in providing a constant $r$ suggests that even the time dependence of $\mu_{\rm eff}$ is close to that of $\mu_{\rm th}$ with the same value of $\eta$ (or that, alternatively, most of the GW emission is from long strings).\footnote{The fact that the effective tension of the strings is close to $\mu_{\rm th}$ with $\eta\approx1/\sqrt{4\pi}$ is also suggested by the direct measurement of the string tension in~\cite{moreaxions}.}\footnote{During the scaling regime, the loop distribution is scale invariant, i.e. contains a fixed number of loops per decade of loop length, for lengths between the IR and UV cutoffs $\simeq H^{-1}$ and $\simeq m_r^{-1}$ (see~\cite{moreaxions}  for more details). For such a distribution the energy takes the form $(\xi-\xi_L)\pi f_a^2\log(m_r/H\eta')/t^2$, where $\eta'$ is a (possibly time-dependent) parameter that depends on the precise location of the cut-offs in units of $H^{-1}$ and $m_r^{-1}$. [To show this, notice that for such a distribution the number of loops per unit length and unit volume is $dn_\ell/d\ell=4H^3(\xi-\xi_L)/\ell$ and the energy of a loop of length $\ell$ is $E_\ell=\pi f_a^2\log(m_r \ell)$. The mentioned energy density follows from $\int_{m_r^{-1}}^{H^{-1}} d\ell \, E_\ell\, dn_\ell/d\ell$ in the limit $m_r\gg H$.] This formula implies $\rho_s^{\rm loops}\approx (\xi-\xi_L)\mu_{\rm th}/t^2$ (except for the value of $\eta'$), which justifies why $\mu_{\rm eff}$ in eq.~\eqref{eq:rhoscal} is approximated by $\mu_{\rm th}$, at least at large log. We stress in any case that the precise form of $\rho_s^{\rm loops}$ is still uncertain as we do not have complete control of the cutoffs of the loop distribution and of the boost factors of the loops, which could change the formula above.} 

\begin{figure}[t]
	\begin{center}
		\includegraphics[width=0.445\textwidth]{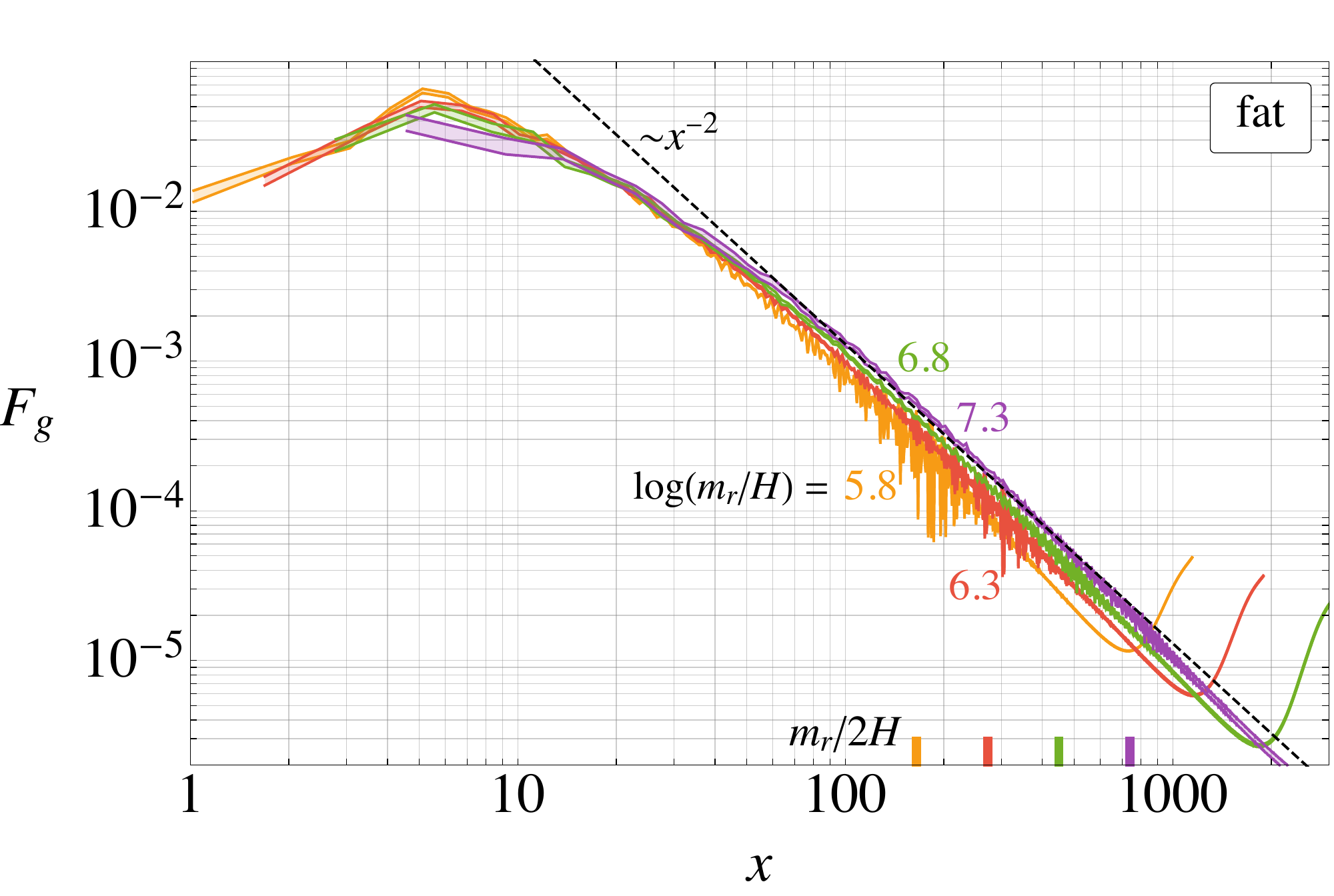}
		\qquad					\includegraphics[width=0.495\textwidth]{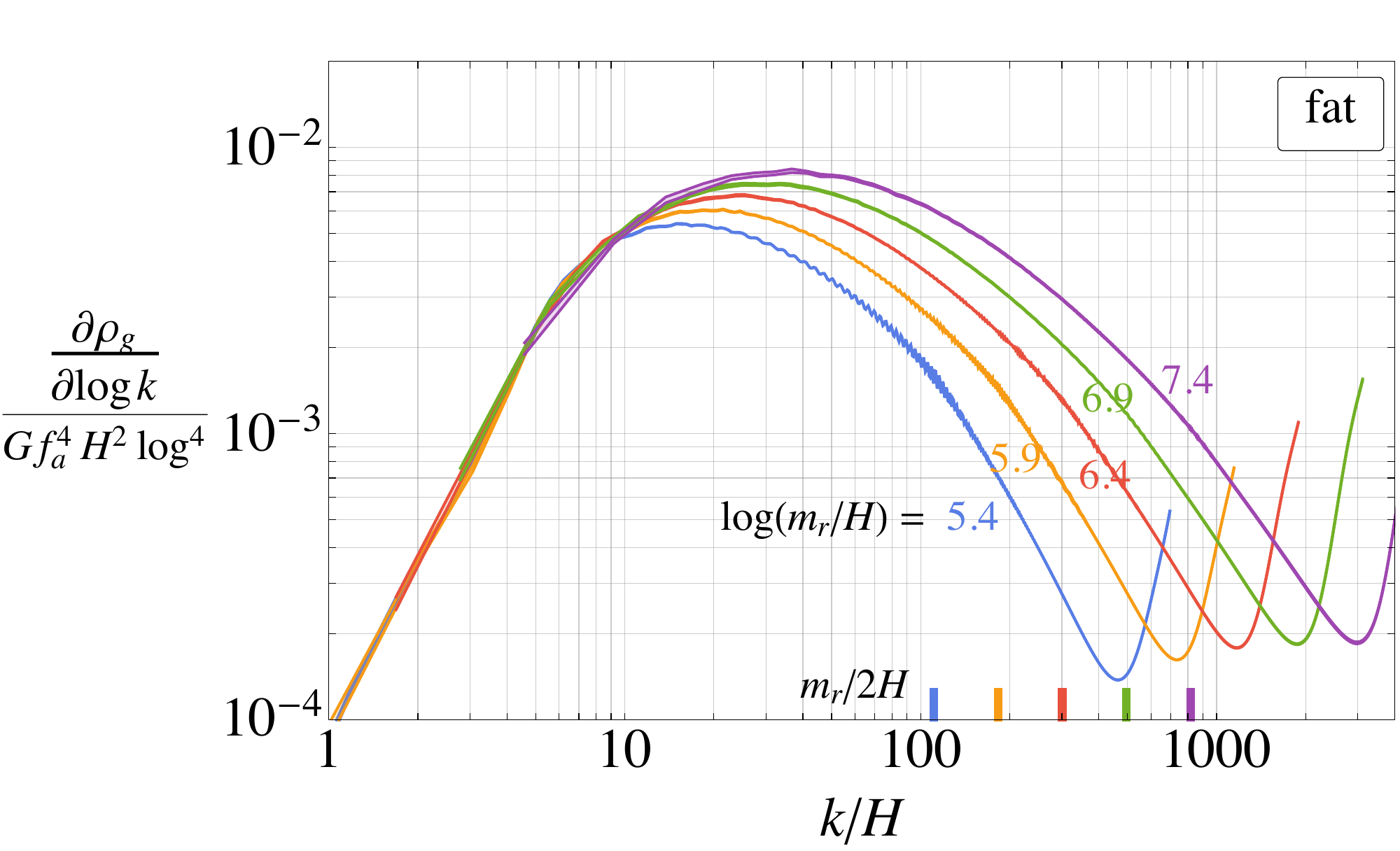}
	\end{center}
	\caption{The normalised instantaneous emission spectrum (left) and the total spectrum of GWs in simulations (right) for the fat string system. The key features match those of the physical system plotted in Figure~\ref{fig:Fg}.  \label{fig:GWfat}} 
\end{figure} 

We finally comment on the validity of this picture at larger values of log. If the loops become very long lived, the energy could be radiated less efficiently from loops into radiation, therefore the total emission rate will be enhanced with respect to the energy lost by long strings in eq.~\eqref{eq:gammaemiLong} (i.e.~$\Gamma_a+\Gamma_r>\Gamma$). Indeed, in the limit of infinite oscillations the loop energy density will redshift as $R^{-3}$, which diminishes slower than $R^{-4}$, which is the rate at which the energy would decrease if it were immediately radiated into axions. Although we do not see a sign of this in the simulation between $\log=4$ and $\log=8$, we cannot exclude this possibility at very large log. If this is the case, this would enhance most likely the axion and GW emission, and our predictions would still be reliable lower bounds.

\subsubsection*{\emph{The GW Spectrum}}

\begin{figure}[t]
	\begin{center}
		\includegraphics[width=0.46\textwidth]{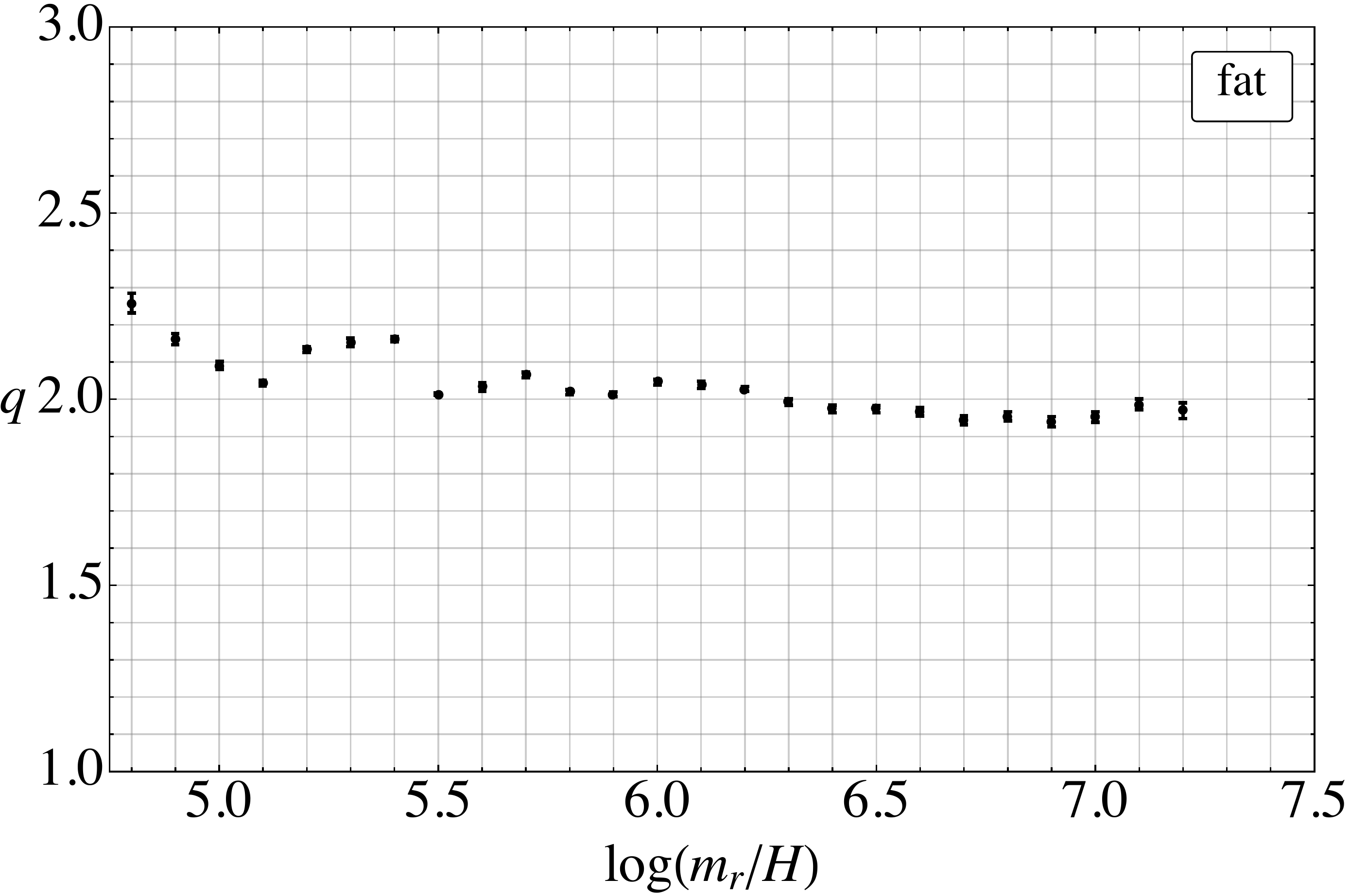}
		\qquad 		\includegraphics[width=0.46\textwidth]{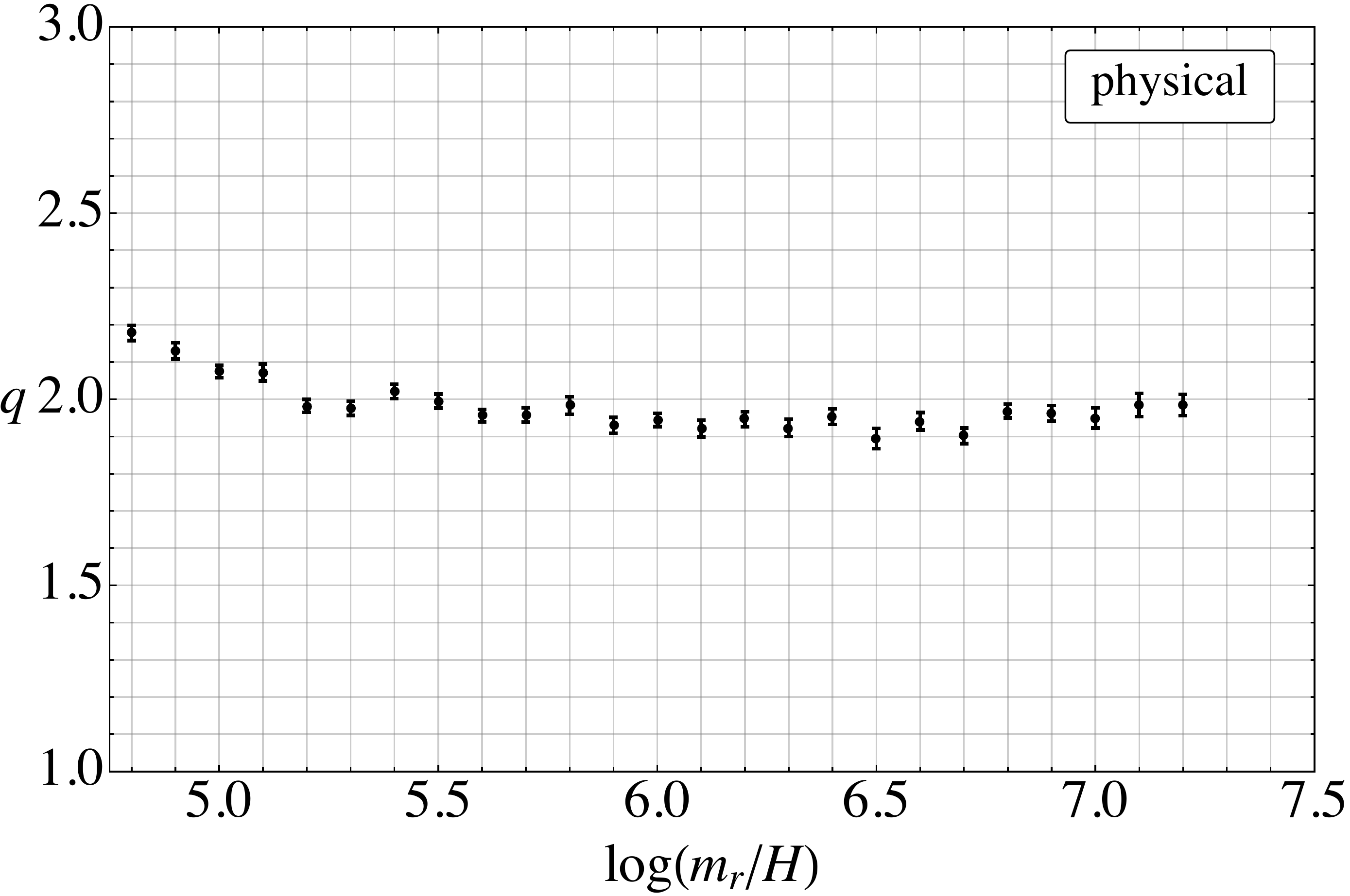}
	\end{center}
	\caption{The time-evolution of the power-law $q$ of the instantaneous GW spectrum $F_g[x,y]\propto 1/x^{q}$ for the fat (left) and physical (right) string systems. In both cases after $\log=5$ the values are approximately constant (and compatible with 2). For smaller times the GW spectrum fluctuates and has no definite power law. \label{fig:q_gw_t}} 
\end{figure} 

Finally, we give more details on the GW spectrum $\partial\rho_g/\partial k$. To extract $\partial\rho_g/\partial k$ in the simulation we used its explicit expression in terms of $\dot{u}_{ij}(\vec{k})$ defined in eq.~\eqref{eq:hk}, easily derived from its definition $\int dk\partial\rho_g/\partial k\equiv T_g^{00}=(32\pi G)^{-1}\langle\dot{h}_{ij}\dot{h}_{ij}\rangle$, see e.g. eq.~(29) in~\cite{GarciaBellido:2007af}. From $\partial\rho_g/\partial k$ we calculate the instantaneous emission spectrum $F_g$ defined in eq.~\eqref{Fdef} as (we calculate the time derivatives numerically considering $\Delta\log=0.2$)
\begin{equation}\label{eq:Fnum}
F_g\left [\frac{k}{H},\frac{m_r}{H}\right ] = \frac{H/\Gamma_g}{R^3}
\frac{\partial}{\partial t} \left ( R^3 \frac{\partial \rho_g}{\partial k}\right) ~.
\end{equation}

\begin{figure}[t]
	\begin{center}
		\includegraphics[width=0.46\textwidth]{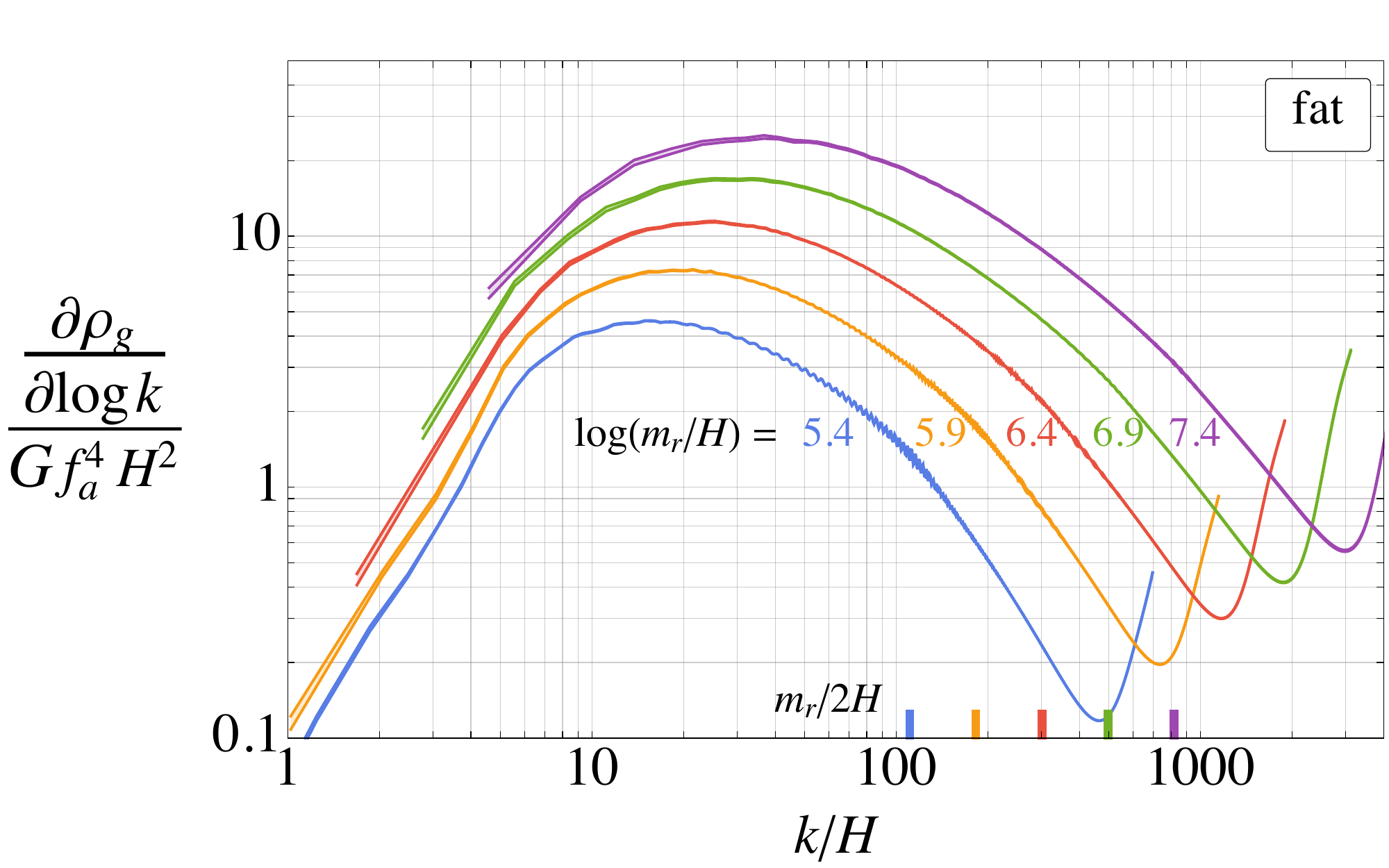}
		\qquad				\includegraphics[width=0.46\textwidth]{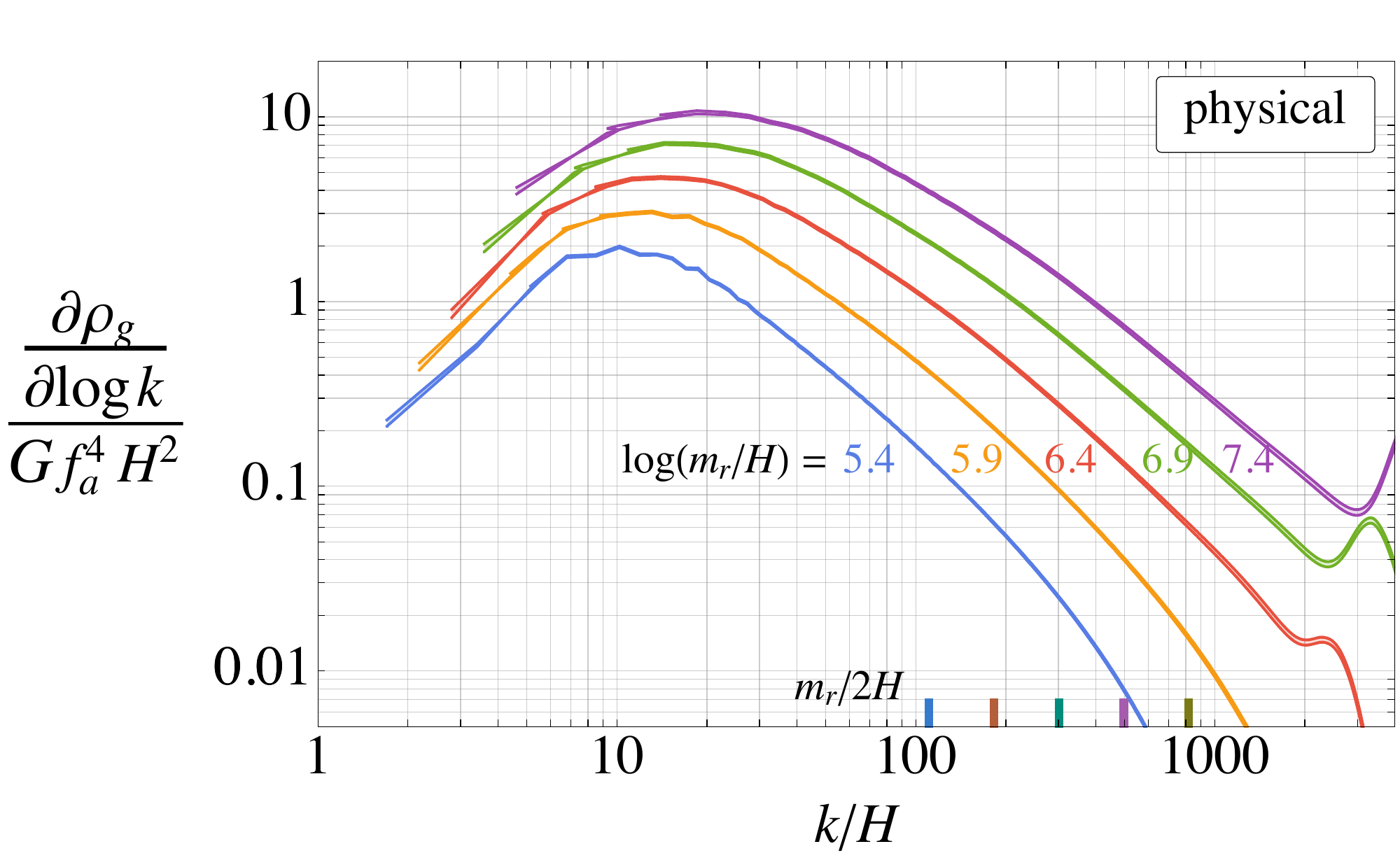}
	\end{center}
	\caption{The total GW spectrum for the fat (left) and physical (right) string networks. The results are identical to those in Figures~\ref{fig:Fg} and~\ref{fig:GWfat} except without the $\log^4$ normalisation. The growth of the spectrum is clear in both cases.  \label{fig:GWnolog}} 
\end{figure}

In Figure~\ref{fig:GWfat} we plot $\partial\rho_g/\partial k$ and $F_g[x,y]$ for the fat string system. Both these observables have similar features to the physical system, shown in Figure~\ref{fig:Fg}. In particular, the approximate power law $q\simeq 2$ is reproduced and is time-independent. As a further study, in Figure~\ref{fig:q_gw_t} we show the best best fit value for the slope of $F_g$ in the momentum range $30H<k<m_r/4$. As mentioned in Section~\ref{ss:gwsim}, given that $q$ is safely above 1 (and appears constant), we do not analyse further its time-dependence and the possible dependence of the fitted value of the slope on the momentum range.\footnote{Such detail is more important for the power law of the axion instantaneous spectrum since it changes with the log.} Although already clear from the previous plots, to show explicitly the time-dependence of the spectrum we also plot the total GW spectrum without the $\log^4$ normalisation in Figure~\ref{fig:GWnolog} for the fat and physical system.

\subsection{GW Backreaction on the Strings} \label{app:back}

In this Appendix we discuss the GW backreaction on the string network in the physical system of eq.~\eqref{eq:LPhi}  by solving the coupled eqs.~\eqref{eq:gweom} and~\eqref{eq:eomscaling} (including the backreaction term $R^{-2}h_{ij}\partial_i\partial_j\phi$ in the left hand side of  this last equation). As mentioned in Appendix~\ref{app:sim}, this is numerically expensive as requires to perform FT and anti-FT every time step, and we therefore limit ourselves to small $800^3$ grids that can explore values of $\log<6$.\footnote{Similarly to the simulations in the main text, this corresponds at the time when $HL=1.5$ for $ m_r\Delta=1$.}

The theoretical discussion of Section~\ref{sec:GWs} suggests that the (evolving) effective parameter controlling the relevance of the GW backreaction on the string network during scaling is $G\mu^2/f_a^2=\pi/8 (f_a\log/M_{\rm P})^2$ (as in the main text, $M_{\rm P}=1/\sqrt{8\pi G}$). Numerical simulations will confirm this expectation. In particular we will see that, as long as $G\mu^2/f_a^2\lesssim 0.5$ (corresponding to $f_a\lesssim M_{\rm P}/\log$), (a) gravity is always in the perturbative regime and (b) the effects of the GWs on the properties of the string network (e.g. $\xi$ and $\rho_a$) is smaller than few percent. Therefore the backreaction is relevant only for $f_a\gtrsim M_{\rm P}/\log$ which is well beyond the region allowed by the bounds in Figure~\ref{fig:param}. A detailed analysis of the impact of the backreaction is therefore not necessary (and, as mentioned, we did not include the backreaction in the simulations presented the main text).

\begin{figure}[t]
	\begin{center}
		\includegraphics[width=0.5\textwidth]{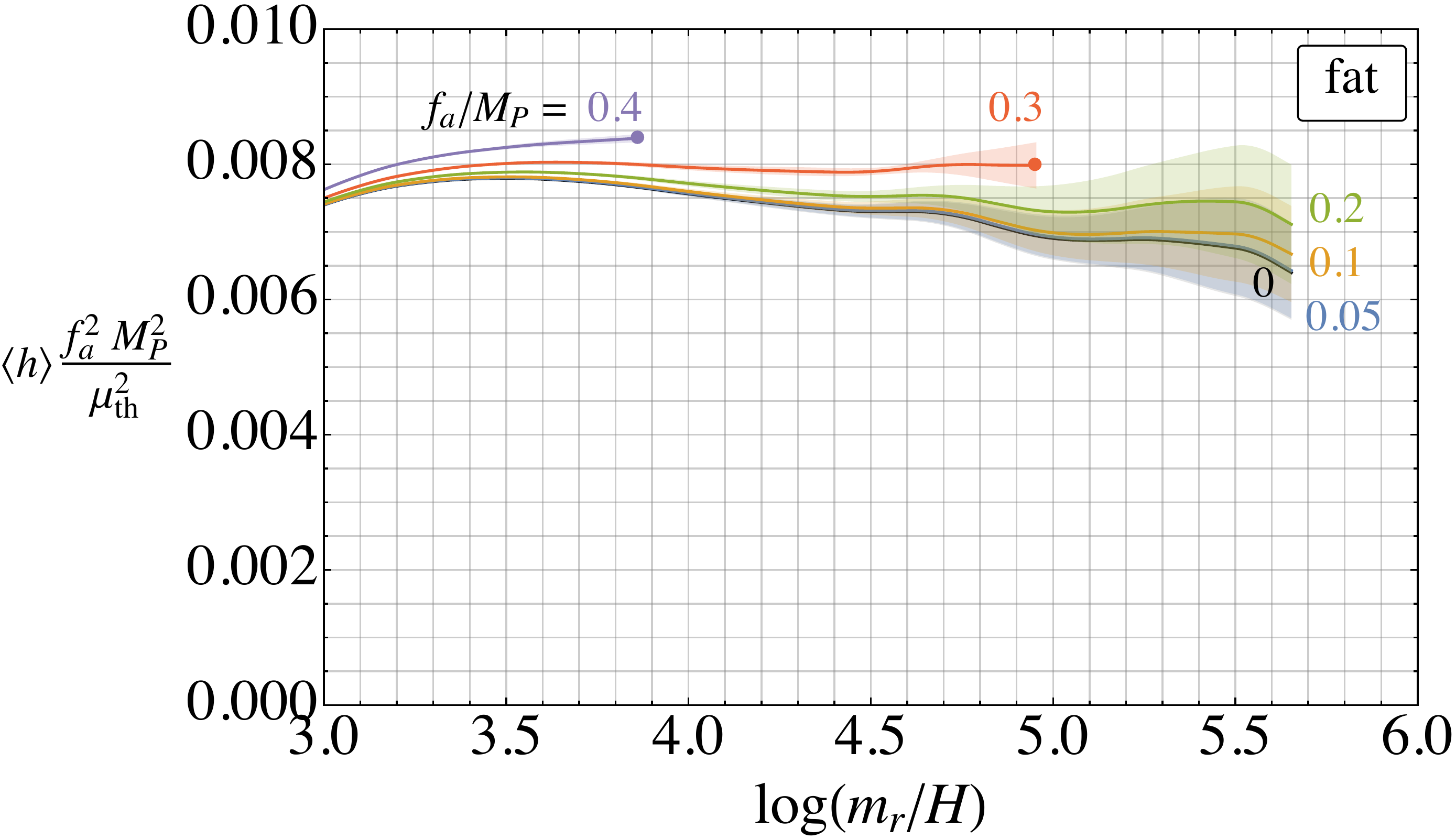}
		\qquad 		\includegraphics[width=0.44\textwidth]{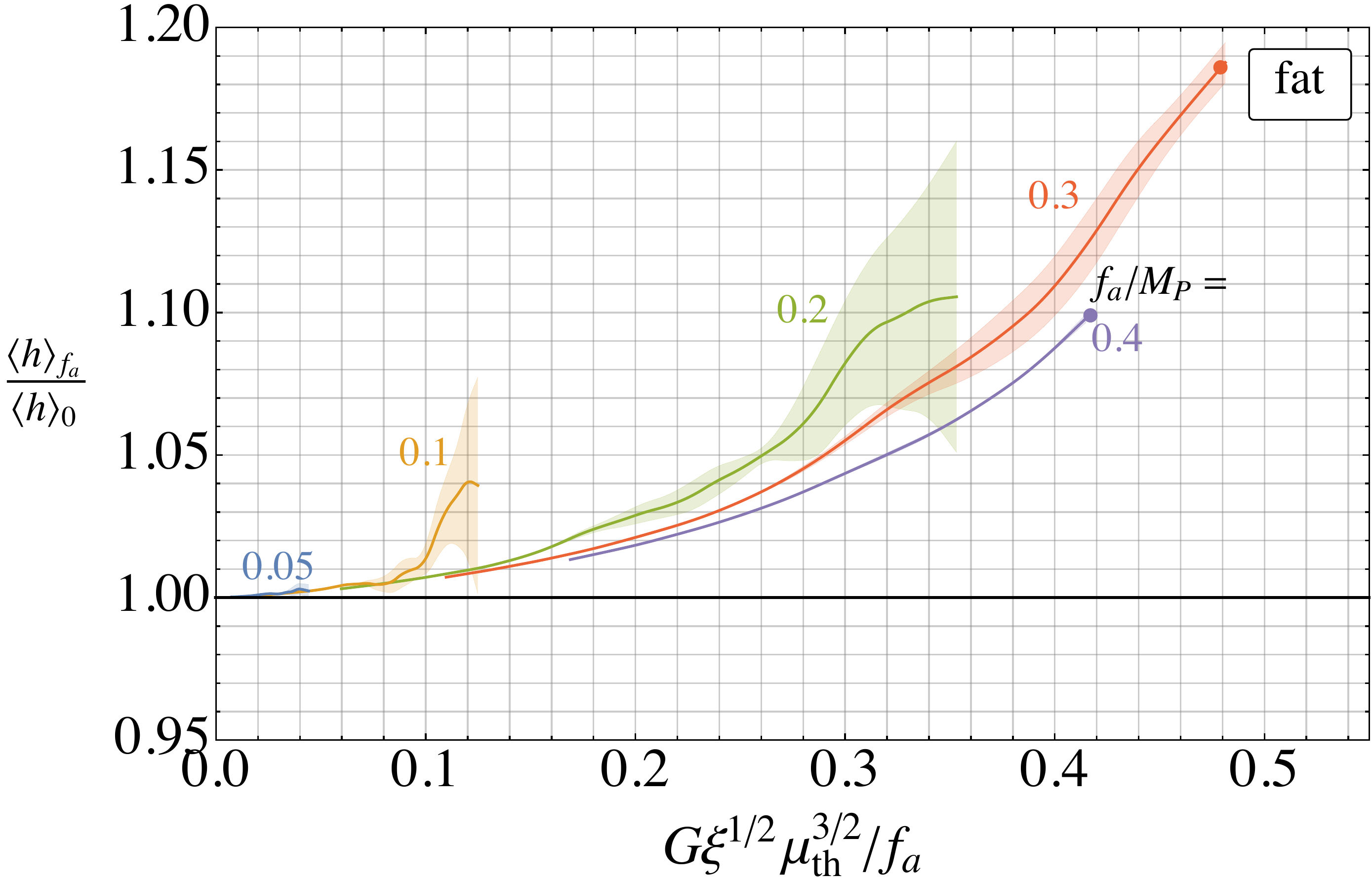}
	\end{center}
	\caption{The evolution of the average square value of the GW field $\langle h\rangle\equiv\langle h_{ij} h_{ij}\rangle^{1/2}$ for different values of $f_a/M_{\rm P}$, and in the limit $f_a/M_P\to0$ normalised to the theoretical expectation for the string tension squared $\mu_{\rm th}^2$ (left). We also show the relative deviation of $\langle h\rangle$ for different non-zero $f_a/M_{\rm P}$ from its value for $f_a/M_P\to0$ (right). On the $x$-axis we plot $G\xi^{1/2} \mu^{3/2}_{\rm th}/f_a$, which is the parameter expected to control the deviation of $\left<h\right>$.
  \label{fig:h_back}} 
\end{figure}

First, notice that the equations of motion~\eqref{eq:gweom} and~\eqref{eq:eomscaling} depend only on the dimensionless ratio $f_a/M_{\rm P}$.\footnote{This can be seen by redefining $\phi\to\phi f_a$ and $h_{ij}\to h_{ij} f^2_a/M^2_{\rm P}$ in such equations.} We evolve these equations for fat strings and different values of $f_a/M_{\rm P}=0.05,0.1,0.2,0.3,0.4$, for the same set of initial conditions (similar to those in Figure~\ref{fig:xi}), as well as in the limit $f_a/M_{\rm P}\to 0$, i.e. not taking into account the backreaction term in eq.~\eqref{eq:eomscaling}. 
In Figure~\ref{fig:h_back} (left) we show the evolution of the average square value of the GW field $\langle h\rangle\equiv\langle h_{ij} h_{ij}\rangle^{1/2}$ in the limit $f_a/M_{\rm P}\to 0$. As expected from the form of the instantaneous GW emission in eq.~\eqref{gammagw} and of the energy density in GWs, $\langle h\rangle$ is of order $f_a^2/M_{\rm P}^2$ for $\log=\mathcal{O}(1)$ and increases proportionally to $\log^2$ (up to subleading corrections).

\begin{figure}[t]
	\begin{center}
		\includegraphics[width=0.46\textwidth]{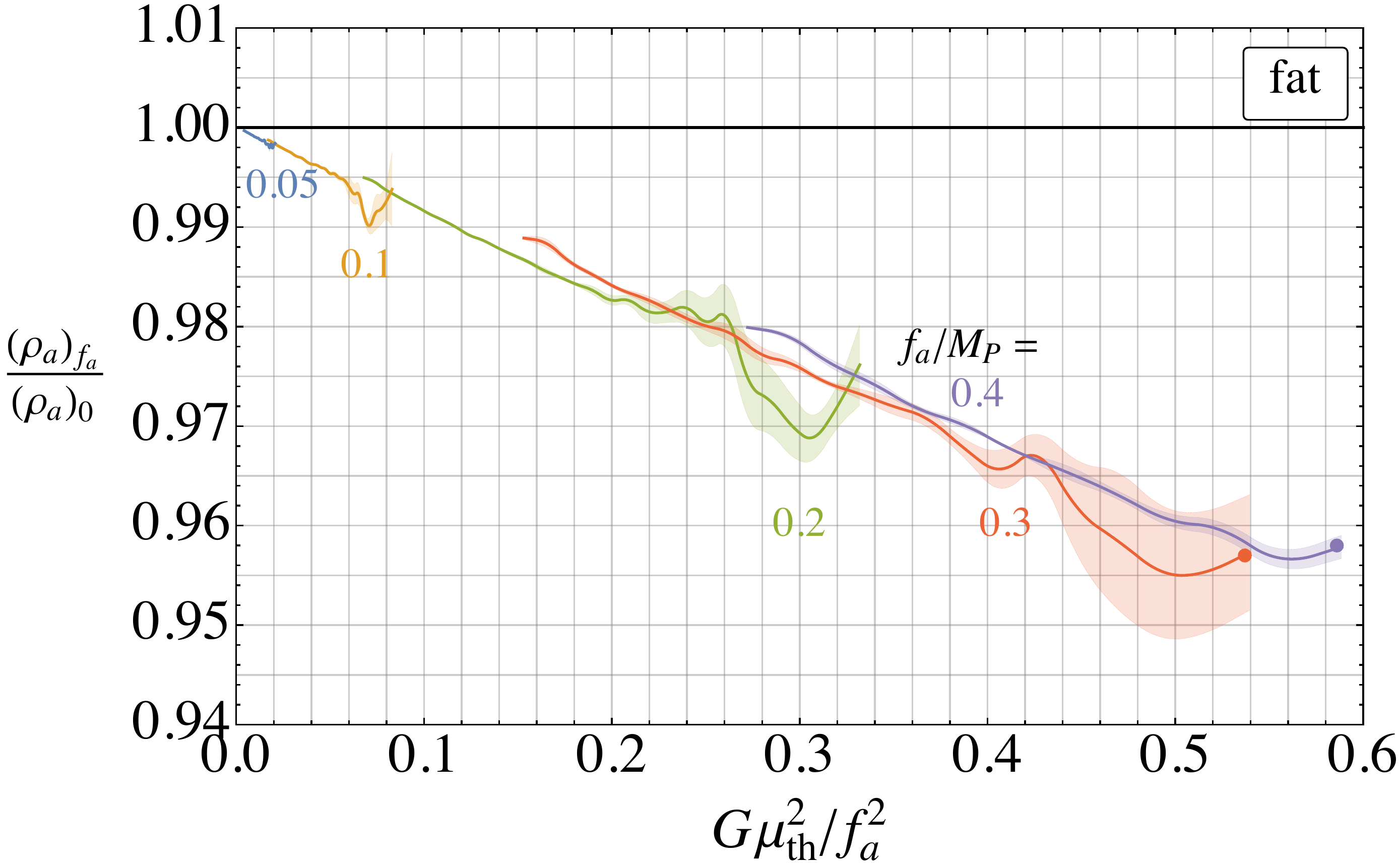}
		\qquad 		\includegraphics[width=0.46\textwidth]{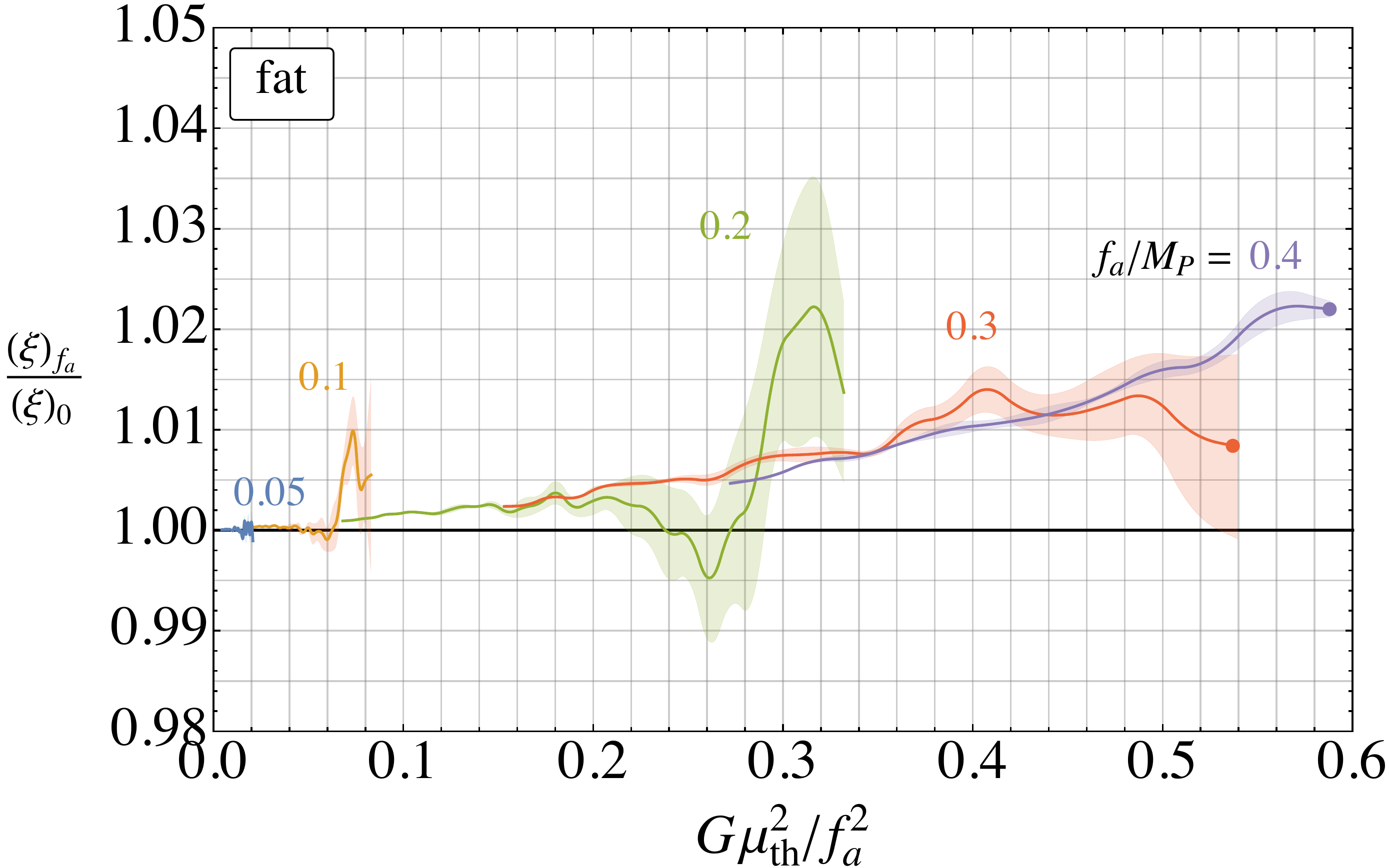}
	\end{center}
	\caption{The evolution of $\xi$ and $\rho_a$ for different values of $f_a/M_{\rm P}$, normalised to their value in the limit $f_a/M_{\rm P}\to0$ (i.e. without GW backreaction, as in the main Sections). We plot  $G\mu^2_{\rm th}/f_a^2$ rather than $\log(m_r/H)$ on the $x$-axis to highlight that the deviations of these observables from  GW backreaction depend on the effective (evolving) parameter $G\mu_{\rm eff}^2/f_a^2$. The filled circles correspond to the time when the numerical evolution of the linear approximation of the Einstein equations breaks down, and at this point $G\mu_{\rm eff}^2/f_a^2\simeq0.5$. For smaller values of this parameter, the linear approximation is valid and the observables deviate by less than few percent from their value in the $f_a/M_{\rm P}\to0$ limit.\label{fig:xirhoback}} 
\end{figure}

Figure~\ref{fig:h_back} (right) and Figure~\ref{fig:xirhoback} show the effect of a finite value of $f_a/M_{\rm P}$ on $\langle h\rangle$, $\xi$ and $\rho_a$ by plotting the time evolution of these observables for different values of $f_a/M_{\rm P}$, normalised to their value in the absence of backreaction. To make the role of the effective parameter $G\mu^2/f_a^2$ manifest, we trade $\log(m_r/H)$ with $G\mu^2_{\rm th}/f_a^2=\pi/8 f_a^2/M_{\rm P}^2\log(m_r/H\eta/\sqrt{\xi})$ in the $x$-axis.\footnote{We fix the same value of $\eta$ as in the main text.} For a non-zero value of $f_a/M_{\rm P}$, the quantities $\langle h\rangle /(f_a^2/M^2_{\rm P})$ and $\xi$ increase with respect to their value in the limit $f_a/M_{\rm P}\to0$, while $\rho_a$ decreases. Indeed, as expected, more energy is transferred to GWs rather than to axions, making $h_{ij}$ larger and $\rho_a$ smaller. Crucially, the deviation of these observables from their value at $f_a/M_{\rm P}\to0$ is controlled by the combination $G\mu_{\rm th}^2/f_a^2$, rather than by $\log$ and $f_a/M_{\rm P}$ separately. Indeed, simulations with different value of $f_a/M_{\rm P}$ present the same deviation at a different value of $\log$ but at the same $G\mu^2_{\rm th}/f_a^2$, as suggested by the fact that the lines overlap in Figure~\ref{fig:xirhoback}. 

When $G \mu^2_{\rm th}/f_a^2\simeq 0.5$ the numerical evolution of the equations of motion breaks down for all $f_a/M_{\rm P}$,\footnote{This is seen by the non-convergence of the numerical algorithm that integrates eqs.~\eqref{eq:gweom} and~\eqref{eq:eomscaling}.} signalling that the linear approximation of the Einstein equations is not valid, and the full general relativity description should be used. We therefore show the results for $\langle h\rangle,~\xi$ and $\rho_a$ until the evolution makes sense, with a filled circle at the time when the numerical evolution breaks. From Figure~\ref{fig:h_back}, it is easy to see that this value of $G\mu_{\rm th}^2/f_a^2$ corresponds to $\langle h\rangle \simeq 0.1$ (the local value of $h_{ij}$ will be larger than this). Moreover, at this value of $G \mu^2_{\rm th}/f_a^2$, the observables $\xi$ and $\rho_a$ have changed only of a few percent with respect to their value in the absence of backreaction. This suggests that a dramatic change of the evolution of the network can be only captured by the full general relativity description of the system.

The decrease in energy in axions seen for non-zero $f_a/M_{\rm P}$ in Figure~\ref{fig:xirhoback} left is quantitatively consistent with the energy that is found to be in gravitational waves, and therefore the value of $r$ obtained in Section~\ref{ss:gwsim} (there is a small increase in the sum of the energy in all components as $f_a$ increases, which is expected since the effect of backreaction is to slightly increase $\xi$, as seen in Figure~\ref{fig:xirhoback}, right).

We observe that, while the dependence on the parameter $G\mu_{\rm th}^2/f_a^2$ has only been tested for small logs and relatively large values of $f_a/M_{\rm P}$, the theoretical discussion in Section~\ref{sec:GWs} allows to extrapolate the results of Figure~\ref{fig:h_back} and~\ref{fig:xirhoback} also at large logs (and smaller $f_a/M_{\rm P}$), in particular ensuring that the GW backreaction is negligible for all $f_a$ in Figure~\ref{fig:param}, despite the large log. Finally notice that, as the evolution of the system is not known for $f_a\gtrsim M_{\rm P}/\log$, there are in principle two orders of magnitude of $f_a$ below $M_{\rm P}$ for which the bounds in Figure~\ref{fig:param} do not apply, and a complete general relativity reanalysis would be needed.

\subsection{Analysis of Systematics} \label{app:simUn}

It is essential that systematic uncertainties from simulations are under control if their results are to be reliable. The most important sources of systematic errors 
 come from the lattice spacing $\Delta$ and the number of Hubble lengths in a box $HL$, where $L$ is the physical box length. In~\cite{Gorghetto:2018myk,moreaxions} it was shown that $HL \geq 1.5$ and $m_r \Delta \geq 1$ are accurate for $\xi$ and most observables relevant to the axion emission. Here we will show that the same values of $HL$ and $\Delta$ are accurate for the GW observables of interest as well, and we therefore use these for our main simulations. We fix the time step to be $a_\tau = a_c /3 $ where $a_\tau$, $a_c$ are the comoving time and space steps respectively, which is small enough to introduce a negligible error in all quantities \cite{Gorghetto:2018myk}.

In particular, we analyse the effect of $m_r \Delta$ and $HL$ on two of the most important observables: the fraction of energy going into GWs, defined by $r_{\rm sim}$, and the instantaneous GW emission spectrum, $F_g$. The latter is more sensitive to finite box size systematics than the total GW spectrum, which, even at the end of simulations when $HL$ is small, includes emission from earlier times when $HL$ was larger. In combination with the analysis of the effects on the axion observables described in~\cite{moreaxions}, this assures us that the effects of systematics on the results in the main text are negligible.

\begin{figure}[t]
	\begin{center}
		\includegraphics[width=0.485\textwidth]{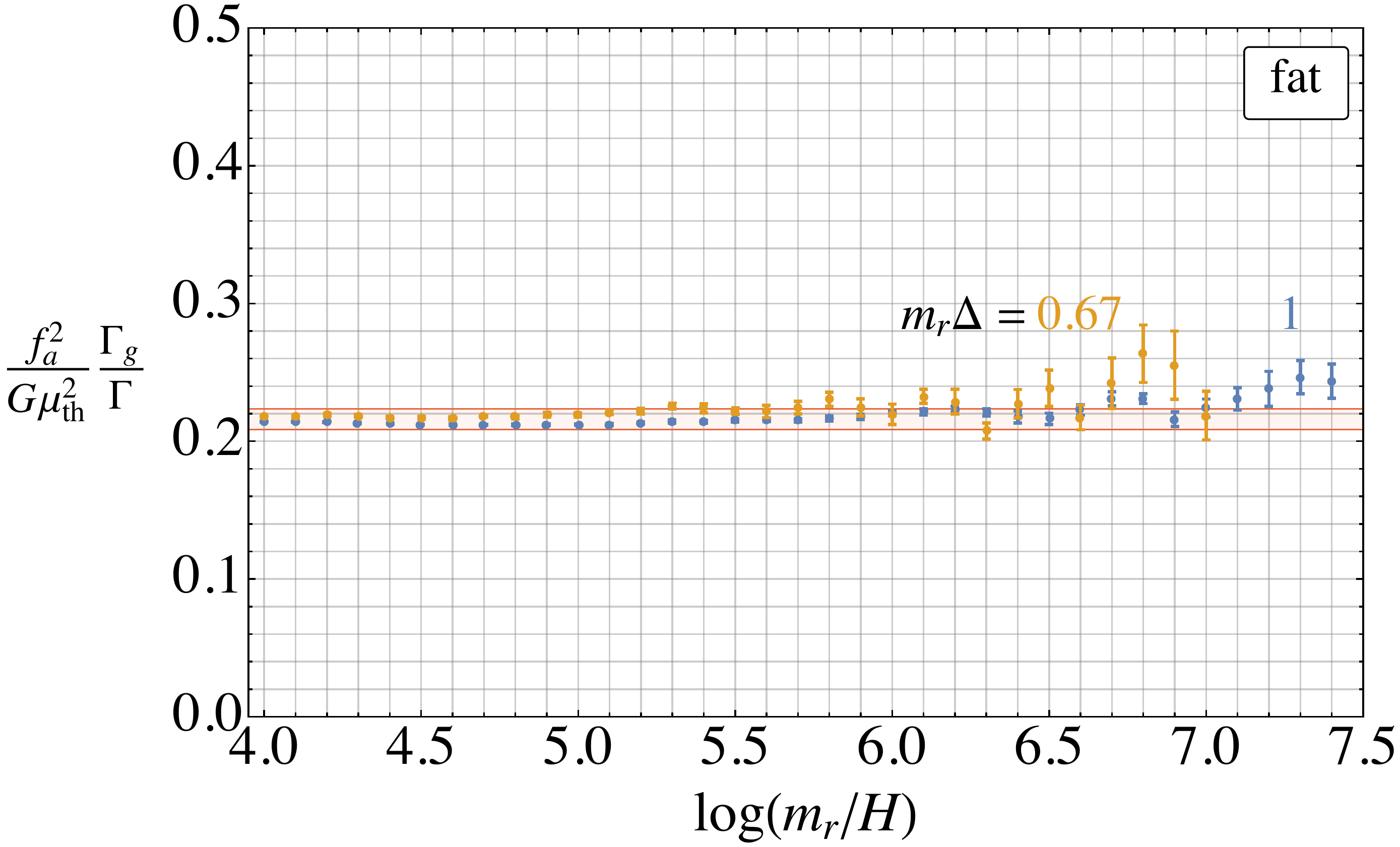}
		\qquad				\includegraphics[width=0.45\textwidth]{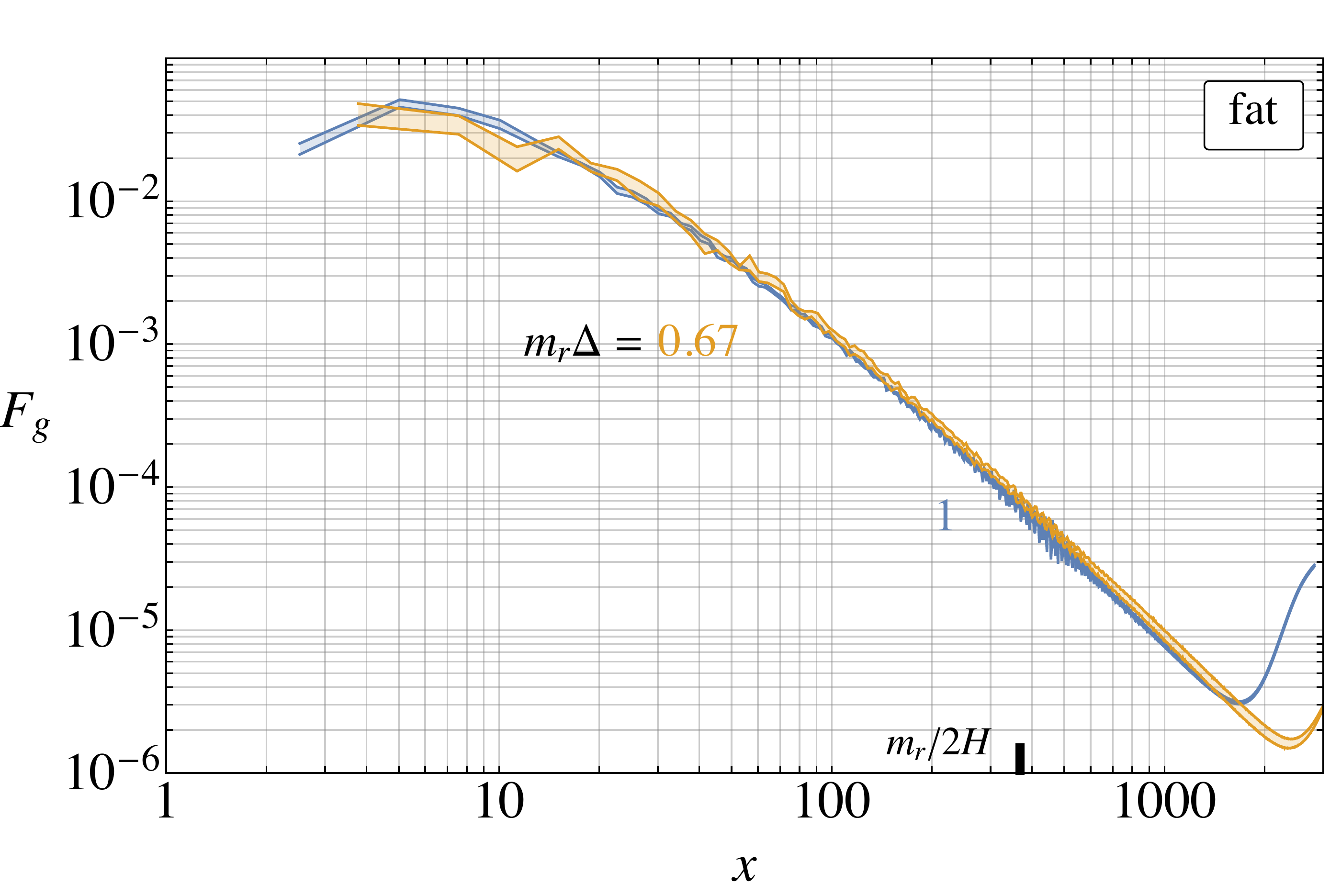}
	\end{center}
	\caption{ The dependence on the lattice spacing $m_r \Delta$ of the energy emission rate into into GWs (left), and of the GW instantaneous emission spectrum, plotted at $\log=6.6$  (right). Data is shown for $m_r \Delta= 1$ and also for a finer lattice with $m_r \Delta=0.67$. The agreement between the two data sets indicates that $m_r \Delta=1$ does not introduce significant systematic uncertainties (in particular, they are smaller than the statistical fluctuations in our main data set).}   \label{fig:mra_sys} 
\end{figure}

In Figure~\ref{fig:mra_sys} we plot $r_{\rm sim}\equiv f_a^2 \Gamma_g/(G\mu_{\rm th}^2 \Gamma)$ and $F_g$ for $m_r \Delta=1$ and for the finer lattice spacing $m_r \Delta=2/3$, both for the fat string system (the results is an average over 10 simulations; the grid sizes were taken equal in both cases so the finer lattice spacing simulations finish sooner). The only deviations between $r_{\rm sim}$ the two data sets are small statistical fluctuations at late times (when the number of independent Hubble patches is indeed smallest), and the form of $F_g$ is consistent in both cases (as for the axion instantaneous spectrum $F_a$, a finer lattice spacing reduces the energy going into modes with momentum $k > m_r$, however it has no effect on the part of $F_g$ from which we extract the spectral index $q$).

\begin{figure}[t]
	\begin{center}
		\includegraphics[width=0.485\textwidth]{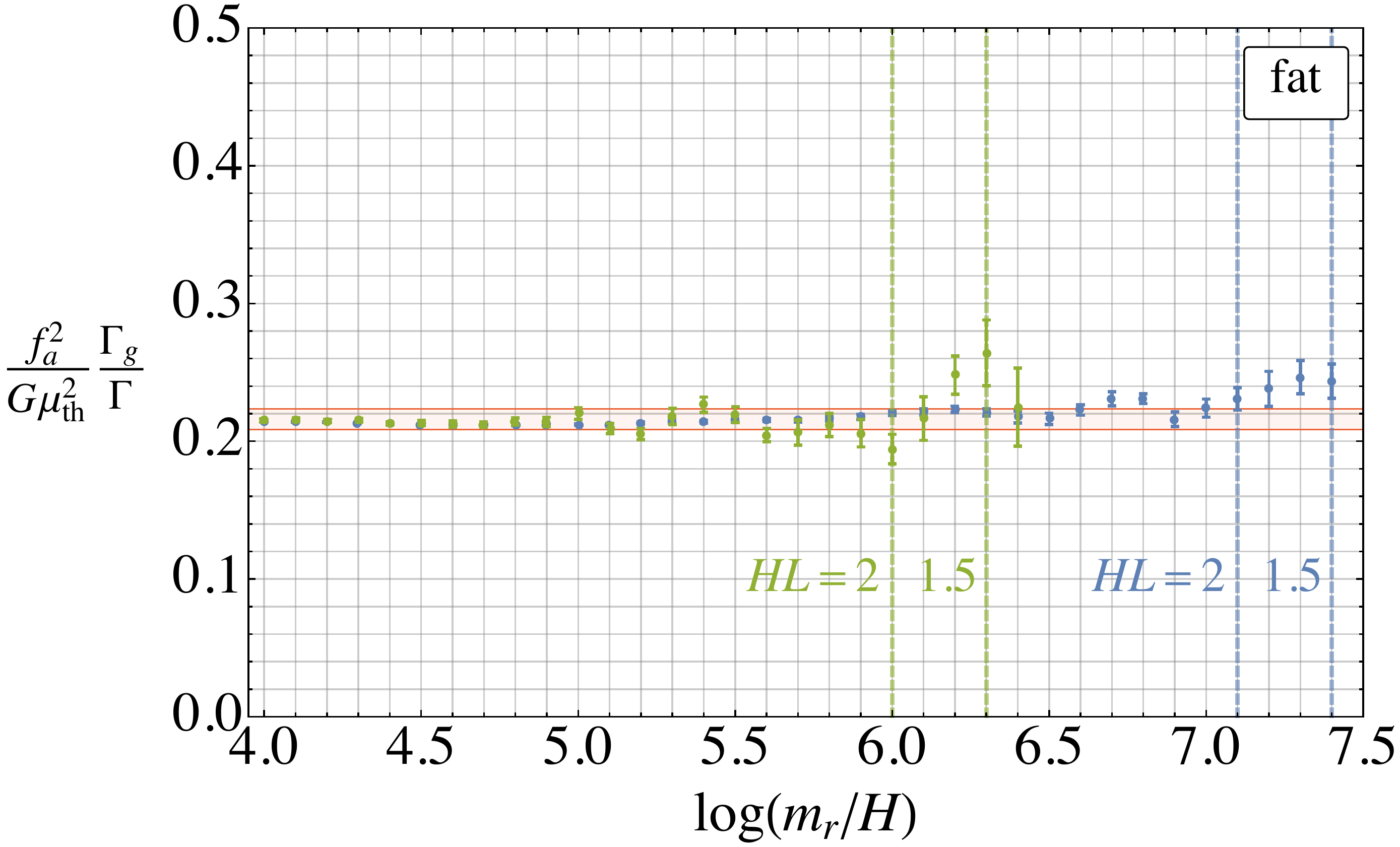}
				\qquad							\includegraphics[width=0.45\textwidth]{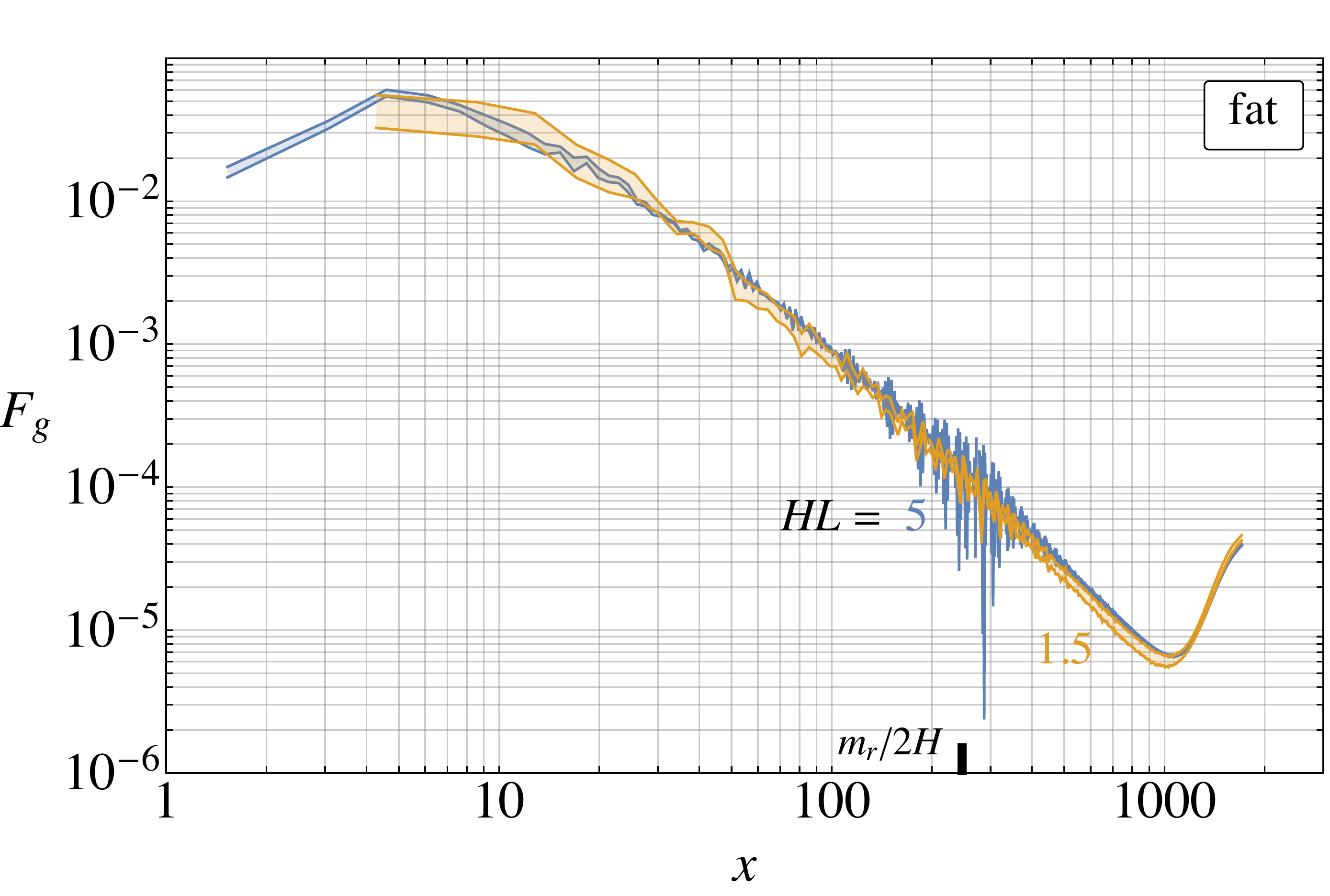}
	\end{center}
	\caption{The effect of the finite box size $L$ on the fraction of energy going into GWs, left, and the normalised GW instantaneous emission spectrum (plotted at $\log=6.2$), right. The two data sets are from simulations on different sized grids. The smaller grids reach $HL\simeq 1$ earlier, when there are many Hubble patches left in the larger simulations. The agreement between these two data sets up to when the smaller grid has $HL=1.5$ indicates that $HL\geq 1.5$ does not lead to large systematic uncertainties in these quantities. \label{fig:HL_sys}} 
\end{figure}

Similarly, in Figure~\ref{fig:HL_sys} we study the $HL$ systematics. To do so we carried out a set of simulations on small grids, of size $800^3$, with smaller value of $HL$ (but with otherwise identical properties as the main simulations, e.g. identical initial string density and also with $m_r \Delta =1$). At a log such that $HL\simeq 1$ on the small grids, $HL$ is safely $\gg 1$ on the large grid, enabling the finite volume effects to be easily identified. From Figure~\ref{fig:HL_sys} it is clear that $HL\gtrsim 1.5$ has no visible effect on the result for $r_{\rm sim}$. For smaller values of $HL$, the IR part of $F_g$ starts being distorted, but the momentum range of interest for extracting the spectral index $q$ (i.e. $k/H\gtrsim10$) is unchanged.

\subsection{The End of the Scaling Regime and Nonlinear Transient} \label{app:destroy}

In this Appendix we give more details on the end of the scaling regime for a temperature-independent axion mass. We will also study in more depth the non-conservation of the comoving number density of axons at $H\simeq H_\star$. Most of the discussion of this Appendix builds on the material of Section~3 of~\cite{moreaxions} and Appendices~D and E of the same reference, to which we refer for a more pedagogical presentation.

\subsubsection*{\emph{End of the scaling regime}}
As mentioned in Section~\ref{ss:Review}, the axion potential becomes relevant in the evolution of the string system only for $H\lesssim H_\star\equiv m_a$, at which time a network of domain walls forms and destroys the string system. To determine precisely the critical value of $H$ (which we call $H_{\rm crit}$) when the scaling regime starts getting affected by the axion potential, we evolve eq.~\eqref{eq:eomscaling} with the same initial conditions as in the main text for the fat string system, but with the additional term $-m_a^2f_a^2/\sqrt{2}$, which corresponds to including the axion potential $V=m_a^2f_a^2(1-\cos(a/f_a))$ in the Lagrangian of eq.~\eqref{eq:LPhi} (see Appendix~D of~\cite{moreaxions}). Notice that the dependence on $m_a$ of the equations of motion enters only through the ratio $m_a/m_r$, and we therefore refer to different axion masses via the value of $\log(m_r/m_a)=\log(m_r/H_\star)\equiv\log_\star$.

\begin{figure}[t]
	\begin{center}
		\includegraphics[width=0.4\textwidth]{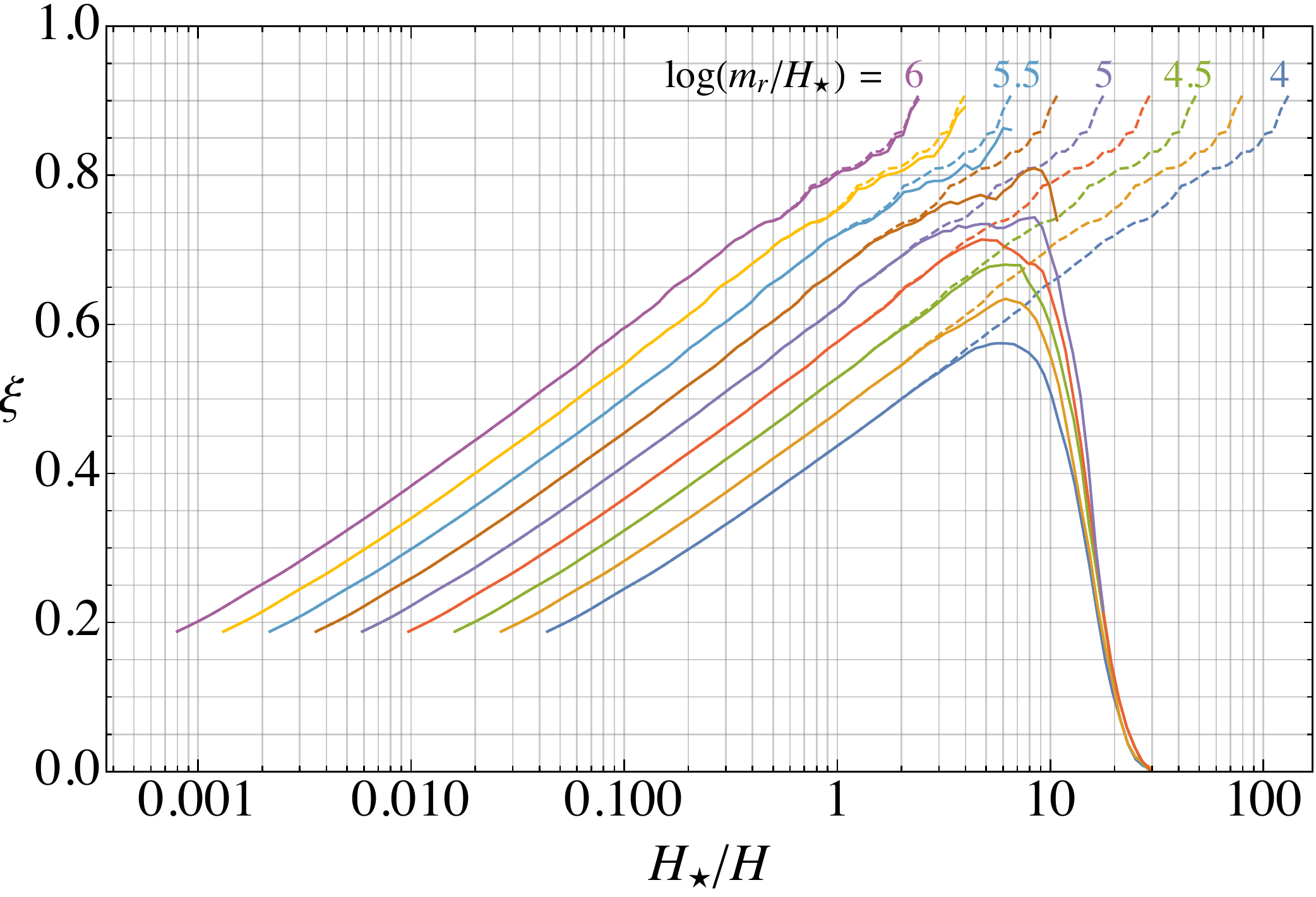}
		\qquad 		\includegraphics[width=0.5\textwidth]{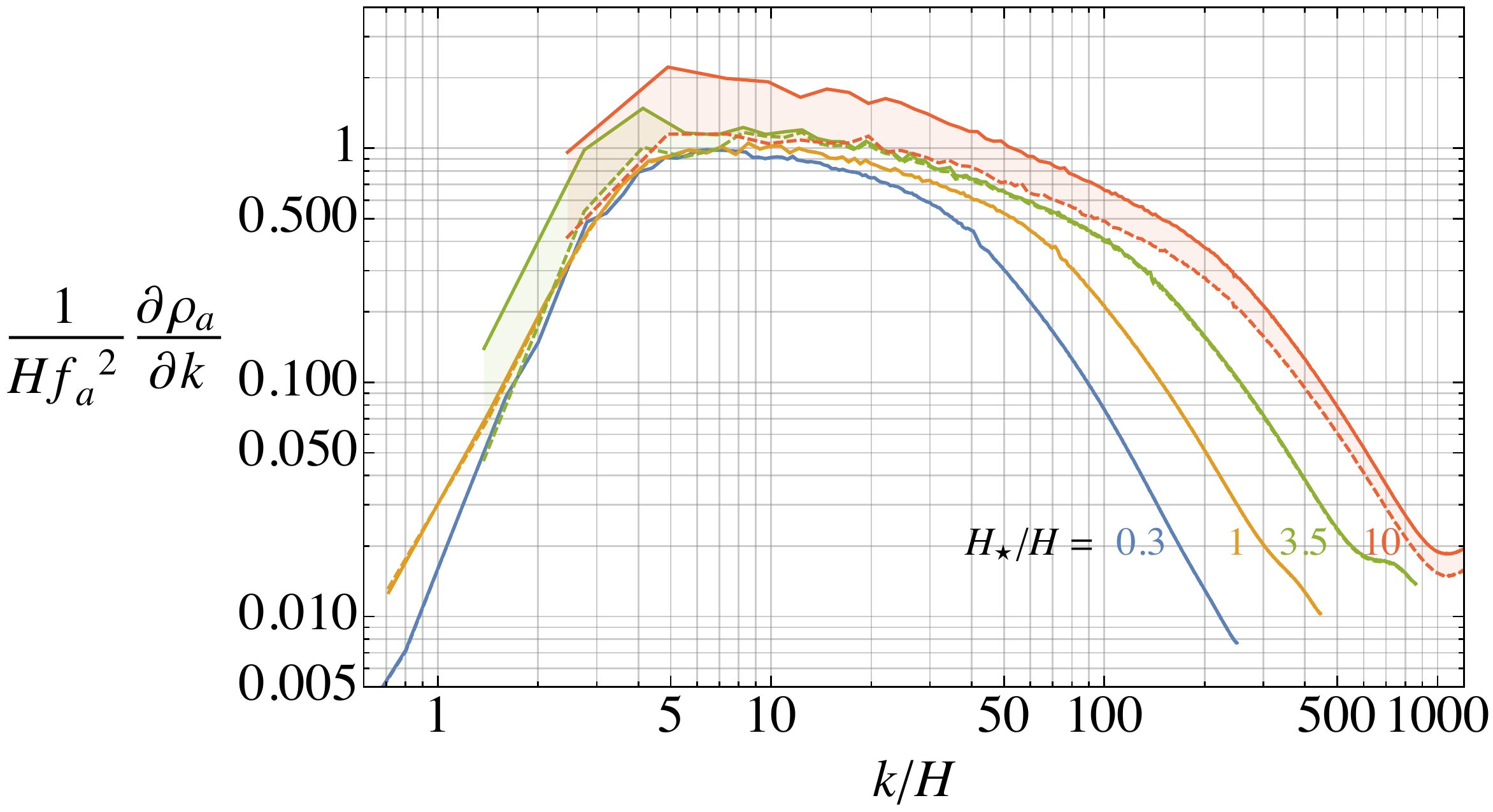}
	\end{center}
\caption{\label{fig:destroy}  Left: The evolution of $\xi$ for a non-zero constant axion mass at different $\log_\star\equiv\log{(m_r/m_a)}$ (solid lines), and for $m_a=0$ throughout with the same initial condition (dashed lines). Independently of $\log_\star$, $\xi$ is unaffected by the mass before $H=H_{\rm crit}\equiv H_\star/3$.  Right: The axion spectrum in the presence of the axion mass with $\log_\star=5$ (upper lines) at different times labelled by $H_\star/H$, and for vanishing axion mass (lower lines, dashed). Before $H=H_{\rm crit}$ the effect of the non-zero axion mass has on the spectrum is negligible.}\end{figure}

In Figure~\ref{fig:destroy} (left) we show the time-evolution of $\xi$ for different values of $\log_\star$, plotted as a function of $H_\star/H$, together with the evolution of the equations for $m_a=0$ (dashed lines). In Figure~\ref{fig:destroy} (right) we also show the time-evolution of the axion spectrum $\partial\rho_a/\partial k$ for $\log_\star=5$. It can be easily seen that for values of $H$ larger than $H_{\rm crit}\equiv H_\star/3$ both $\xi$ and the axion spectrum are not significantly affected by the axion potential, as they closely follow the evolution for $m_a=0$. For smaller values of $H$, $\xi$ diminishes (as the network starts being destroyed) and the spectrum gets affected, starting from its IR part. 
Although the value of $H_{\rm crit}$ can be numerically studied only at small values of $\log_\star$, it is reasonable to expect that it will not change at larger $\log_\star$.

\subsubsection*{\emph{Nonlinear evolution of the axions waves}}

As mentioned in Section~\ref{ss:relic}, and explained in detail Section~3 of~\cite{moreaxions}, the axion waves produced up until $H_\star$ have kinetic energy much larger than the potential energy at $H=H_\star$ (we momentarily assume that $H_{\rm crit}=H_\star$, and discuss to the modification to a generic $H_{\rm crit}$ later).\footnote{Recently it has been claimed that this cannot be true because the compactness of the axion field bounds the energy that can be stored in low momentum modes~\cite{Dine:2020pds}.  In fact, the periodicity of the axion only affects the zero-mode: all the other modes can be populated by arbitrarily large amplitudes.}  
This results in a period of relativistic redshift and a nonlinear transient, which implies a partial non-conservation of the comoving number density. However we now show that the number density non-conservation is small for the value of $\xi_\star\log_\star$ discussed in Section~\ref{ss:relic} for a temperature-independent mass.

The number density after the nonlinear transient follows the analytic description and eq.~(36) of~\cite{moreaxions} (evaluated for a constant axion mass, i.e. $\alpha=0$). The corresponding number density non-conservation reads
\begin{align}\label{eq:Q1}
\frac{n_a^{\rm st}}{n_a^{\rm st}|_{\rm linear}} & =\frac{c_n c_V}{\frac{8\pi\xi_\star\log_\star}{x_{0,a}}}\left[ \frac{W_{-1}\left(-\frac{c_V}{2\pi \xi_\star \log_\star}
\left(\frac{x_{0,a}}{c_m}\right)^4 \right)}{-\frac{c_V}{2\pi \xi_\star \log_\star}}
\right]^{\frac34}\\
&=\frac{c_n c_V}{\frac{8\pi\xi_\star\log_\star}{x_{0,a}}}\left[2\pi\xi_\star\log_\star\log\left(\frac{2\pi\xi_\star\log_\star}{c_V}\left(\frac{c_m}{x_{0,a}}\right)^4\right)\right]^{\frac34} ~,
\end{align}
where $W_k$ is the Lambert $W$-function evaluated on the $k$-th Riemann sheet and in the second equality we expanded $W_{-1}$ for large negative values of its argument. The coefficients $c_m,~c_V,~c_n$ have been extracted in~\cite{moreaxions} by fitting eq.~\eqref{eq:Q1} with the number density obtained from the numerical evolution of
\begin{equation}\label{eq:axeom}
\ddot{a}+3H\dot{a}-R^{-2}\nabla^2a+m_a^2 f_a\sin(a/f_a)=0 ~,
\end{equation}
 with $m_a=H_\star(H_\star/H)^{\alpha/4}$ and $\alpha=4,6,8$, with initial conditions (at $H=H_\star$) given by a superposition of axion waves with energy density spectrum $\partial\rho_a/\partial k$ from the (reconstructed) scaling regime at $H=H_\star$ (see~\cite{moreaxions} and the following eq.~\eqref{drhodk} for the explicit expression of the initial conditions). Note that such simulations (that include only the axion field) can study directly the physical point (i.e. without extrapolation, unlike simulations of the physical system in~\eqref{eq:LPhi}, which must include modes with $k \simeq m_r$).
\footnote{The coefficients read $c_m=2.40,~c_V=0.12,~c_n=1.20$ and $c_m=2.08,~c_V=0.13,~c_n=1.35$, respectively for an initial axion energy density spectrum given by the convolution of $F_a$ sharp IR cutoff at $x=x_{0,a}$ and a more physical form as described in~\cite{moreaxions} (in both cases with $q_a=5$).}

Let us now discuss what changes if $H_{\rm crit}< H_\star$. In this case the number density at $H=H_{\rm crit}$ is approximately $n_a=8\pi  \xi_\star \log_\star H_{\rm crit}f_a^2 /x_{0,a}$.\footnote{As defined in Section~\ref{ss:relic}, the number density is $n_a^{\rm st}\equiv\int dk\partial\rho_a/\partial k $, and can be approximated with $\rho_{\rm IR}/(x_{0,a}H)$ during the scaling regime (up until $H_{\rm crit}$), where $\rho_{\rm IR}= 8 \pi f_a^2\xi\log H^2$ is the energy density in IR modes. The difference $\xi_\star\log_\star$ from $\xi_{\rm crit}\log_{\rm crit}$ is insignificant with respect to the change in $H$.}  If this number density is thought as coming from an axion field at $H=H_\star$ (which redshifts relativistically from $H_\star$ to $H_{\rm crit}$), such a field has the same energy density spectrum $\partial\rho_a/\partial k|_\star$ but with $x_{0,a}\to x_{0,a}(H_{\rm crit}/H_\star)^{1/2}=x_{0,a}/\sqrt{3}$ (i.e. an IR cutoff smaller by $\sqrt{3}$).\footnote{In particular, at $H=H_\star$ its (relativistic) number density will be enhanced by a factor $\sqrt{3}$ with respect to $8\pi  \xi_\star \log_\star H_{\star}f_a^2 /x_{0,a}$.} To evaluate the suppression of the number density for $H_{\rm crit}< H_\star$ we can therefore use eq.~\eqref{eq:Q1} with $x_{0,a}\to x_{0,a}/\sqrt{3}$, which gives a $20\%$ non-conservation of the comoving number density for $\xi_\star\log_\star=3000$ (and $x_{0,a}=10$).\footnote{In doing this estimate we used a $c_V$ that is $(1-1/q_a)^{-1}= 5/4$ larger than what mentioned before, in order to account for the fact that the coefficients have been extracted for $q_a=5$ instead of $q\to\infty$ (which is the limit in which $n_a=8\pi\xi\log/x_{0,a}f_a^2H$ is valid.)}

\begin{figure}[t]
	\begin{center}
		\includegraphics[width=0.6\textwidth]{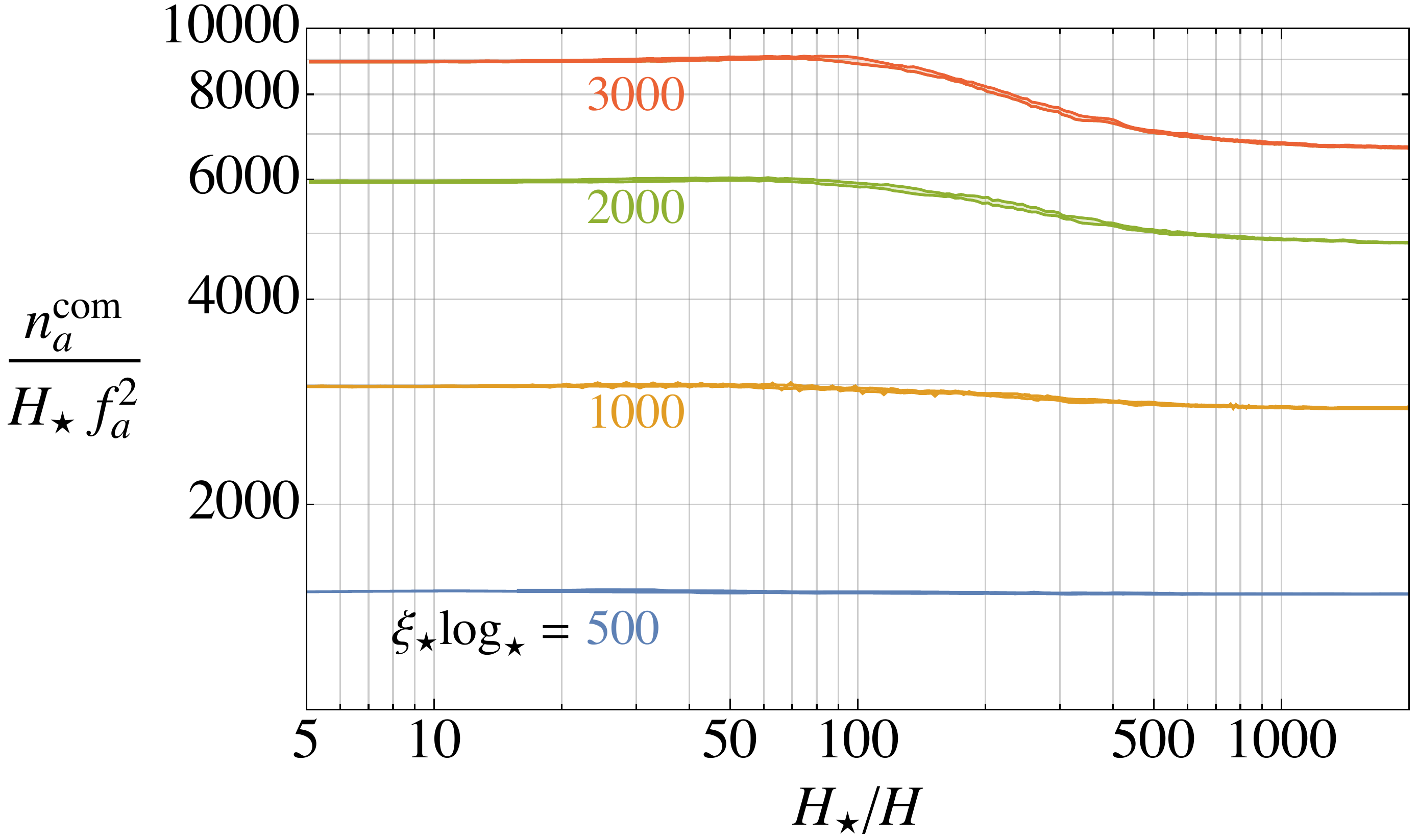}
	\end{center}
	\caption{ The  evolution of the comoving number density of the axions (produced during the scaling regime) through the nonlinear transient that occurs when the axion potential becomes cosmologically relevant. The initial conditions are set at $H=H_{\rm crit}=H_\star/3$ to be a superposition of axion waves with the energy density spectrum $\partial\rho_a/\partial k$ from the (reconstructed) scaling regime at $H=H_{\rm crit}$ (see~\cite{moreaxions}). For $\xi_\star\log_\star\lesssim3000$ (which are the relevant values for ultralight axions) the number density non-conservation is at most $20\%$, and was therefore neglected in the derivation of $\Omega_a^{\rm st}$ of Section~\ref{ss:relic}. \label{fig:num_den}	} 
\end{figure}

In support of this, in Figure~\ref{fig:num_den} we also show the evolution of the comoving number density for different values of $\xi_\star\log_\star$ from the numerical evolution of the equation of motion~\eqref{eq:axeom}.
The set up of the simulations is explained in~\cite{moreaxions}, to which we refer for the details. As in~\cite{moreaxions}, we start from $H=H_{\rm crit}=H_\star/3$ with initial conditions given by an axion field made of a random superposition of waves with the energy density spectrum  of the scaling regime at $H=H_{\rm crit}$, i.e. (for $k>k_0\equiv x_{0,a}H_{\rm crit}$)
\begin{equation}
\begin{split}\label{drhodk}
 \frac{\partial \rho_{a}}{\partial k} (t_{\rm crit},k) =&
\frac{8\xi_\star \mu_\star H_{\rm crit}^2}{k} \left[
\left(1-2\frac{\log(k/k_0)}{\log_\star} \right)^2-\left(\frac{k_0}{k}\right)^{q_a-1} \right. \\
& \left. +4 \frac{1-2\frac{\log(k/k_0)}{\log_\star}-\left(\frac{k_0}{k}\right)^{q_a-1}}{(q_a-1)\log_\star
}+8 \frac{1-\left(\frac{k_0}{k}\right)^{q_a-1}}{(q_a-1)^2\log_\star^2} \right] , \
\end{split}
\end{equation}
which follows from eq.~\eqref{eq:drhogdk} with $\Gamma_g$ replaced with $\Gamma_a$ and $F_g$ with $F_a$, and we assume $F_a\propto1/x^{q_a}$ for $x>x_{0,a}$ and $q_a>1$ (and $F_a=0$ for $x<x_{0,a}$).\footnote{We used $q_a=5$ and as mentioned $x_{0,a}=10$.} Notice from Figure~\ref{fig:num_den} that for values of $\xi_\star\log_\star=\mathcal{O}(10^3)$ the suppression is the predicted one, and for smaller values the conservation of the number density is even more accurate (and eq.~\eqref{eq:Q1} breaks). Finally, for larger (not relevant, unless $N>1$, see Section~\ref{ss:genericN}) values of $\xi_\star\log_\star$ the non-conservation of the number density becomes substantial, and the relic density in eq.~\eqref{eq:relic} must be multiplied by the suppression factor in eq.~\eqref{eq:Q1}. In particular, as mentioned in the main text, $\Omega_a^{\rm st}\propto(\xi_\star\log_\star)^{3/4}$.

\subsection{The Density Power Spectrum} \label{app:density}

In this Appendix we give more details on the determination of the power spectrum of axion overdensities $\Delta_a^2(k)$ defined in eq.~\eqref{eq:delta2k}, and discuss the uncertainties on this.

As described in Section~\ref{ss:iso} and Appendix~\ref{app:destroy}, we consider only the axion radiation emitted during the scaling regime up to $H=H_{\rm crit}=H_\star/3$. 
In particular, we neglect the strings and the domain walls that are present in the field at this time. The evolution of such radiation follows the axion equations of motion~\eqref{eq:axeom}. 
Similarly to Appendix~\ref{app:destroy}, we start at $H=H_{\rm crit}$ with initial conditions given by a superposition of waves with the axion spectrum $\partial\rho_a/\partial k$ from the extrapolated scaling solution (with $\xi_\star \log_\star = 2000$) in eq.~\eqref{drhodk}%
.\footnote{Although this spectrum derives from a simplified form of $f_a$, we have confirmed that starting with a more realistic $F_a$ (discussed in~\cite{moreaxions}) with the same $x_{0,a}$ but with an IR tail $F(x)\sim x^3$ for $x<x_{0,a}$ changes the fit of the constant $C$ to the IR of $\Delta_k^2$ by less than 20\%, which is much smaller than the uncertainties we subsequently discuss.} As discussed in Appendix~\ref{app:destroy}, the axion number density (and in general the dynamics of the IR part of this radiation) can be directly studied at the physical value of $\log_\star$ in these simulations without the need of extrapolations (thanks to the absence of strings).

Such simulations capture the dynamics of axion field up the momentum mode $k_{\rm UV}\simeq N H_\star / (H_\star L_\star)$  at $H_\star$, where $N$ is the number of lattice points  and $H_\star L_\star$ is the number of Hubble lengths in the box at this time. While most of the energy density of the field is not included in the simulations due to the almost scale invariant form of the energy density spectrum $\partial\rho_a/\partial k$ (see eq.~\eqref{drhodk}), the axion number density $n_a=\int{dk/\omega_k\partial\rho_a/\partial k}$ dominantly comes from IR modes, which are included (at $k>m_a$ the contribution to the number density from the mode $k$ is proportional to $k^{-2}$). Consequently, the IR modes of the field also dominantly determine the axion dark matter power spectrum $\Delta_a^2(k)$ at least at IR momenta.\footnote{In particular, the UV modes evolve freely without affecting the dynamics of the IR modes at any point including during the previously discussed non-linear transient \cite{moreaxions}. Subsequently, the energy in the UV modes simply redshifts away leaving a negligible contribution to the DM abundance. The fact that simulations do not include the (large fraction of the total axion) energy that is such UV modes therefore does not introduce uncertainty in the power spectrum that we extract.}

Since we are primarily interested in the coefficient of the $k^3$ IR part of the density power spectrum, the simulations are carried out starting with $HL=20$ at $H_{\rm crit}$ so that modes with momentum down to $k/H_\star\simeq 0.2$ are included, giving a large enough momentum range for the $k^3$ slope to be present and $C$ to be fit. Moreover, we set $N=1300$, which is large enough that modes with $k/H_\star\simeq 200$ are included, corresponding to $95\%$ of the number density. During the evolution of the field, we calculate the (discretized version of the) power spectrum as $\Delta_a^2(k)=k^3/(2\pi^2L^3)\langle\tilde{\delta}^2(\vec{k})\rangle|_{|\vec{k}|=k}$ where $\tilde{\delta}(k)$ is the Fourier Transform of $\delta(x)$ and $\langle\cdot\rangle|_{|\vec{k}|=k}$ stands for the average over the momenta with modulus $\vec{k}$.

\begin{figure}[t]
	\begin{center}
		\includegraphics[width=0.45\textwidth]{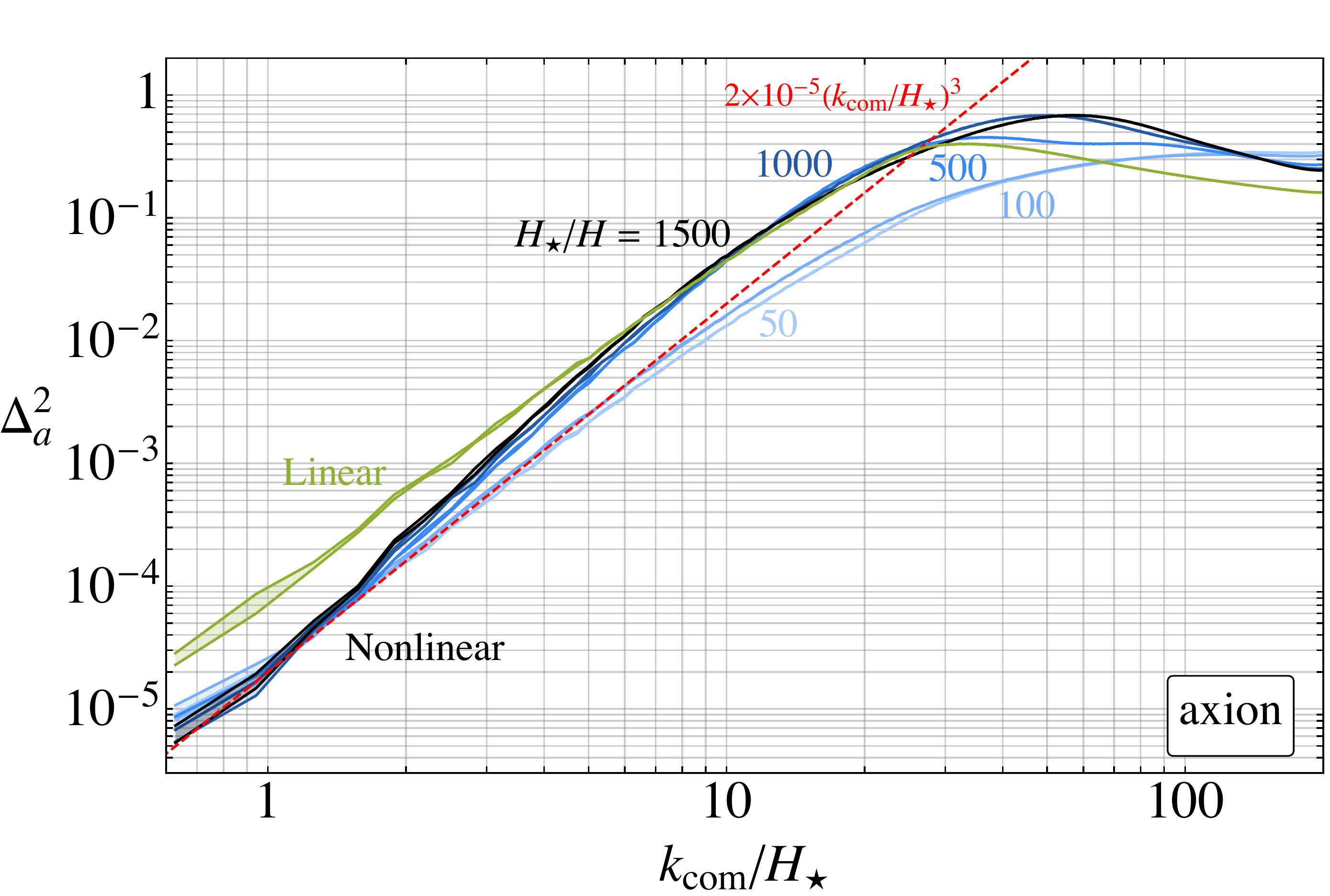}
		\qquad	 	\includegraphics[width=0.45\textwidth]{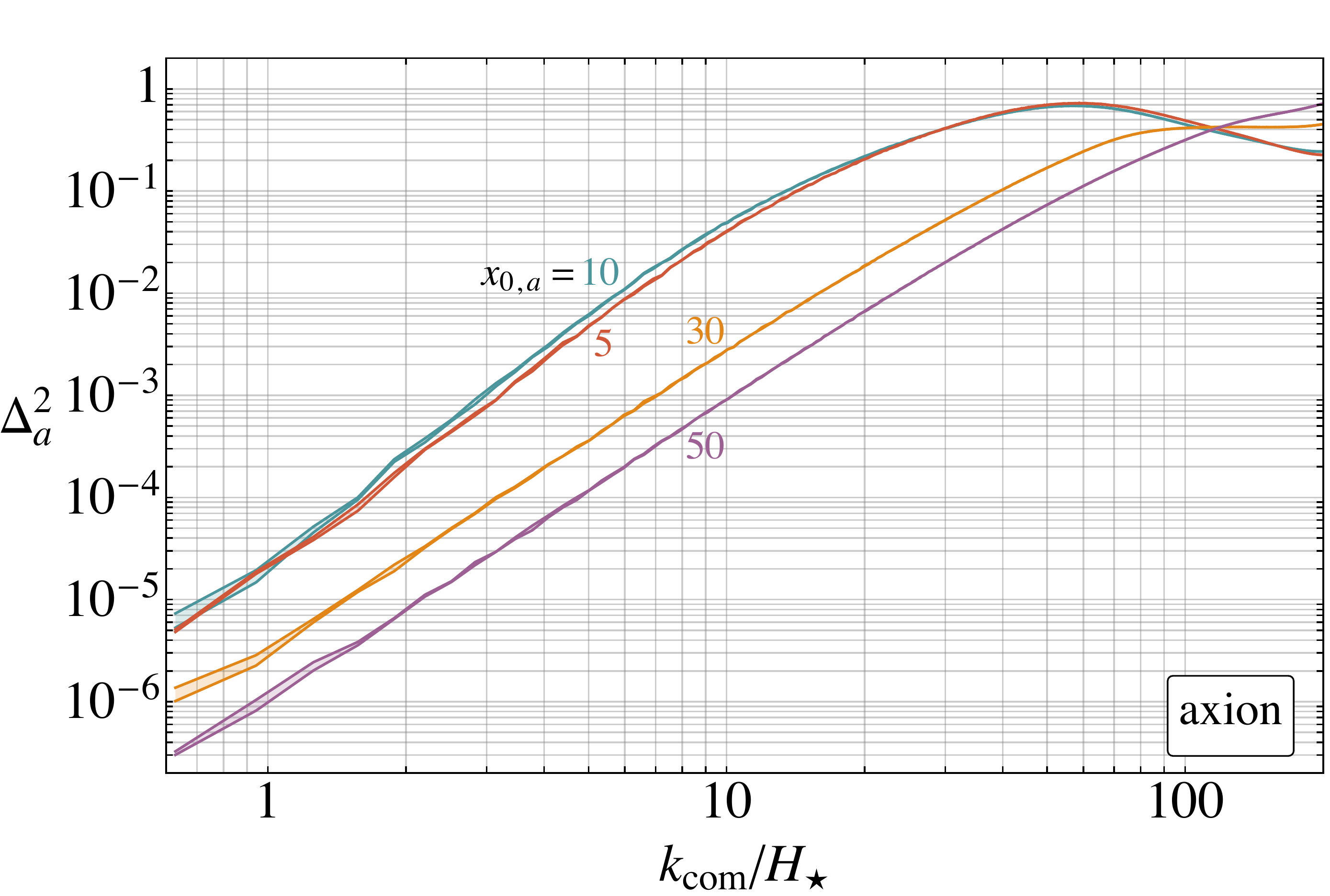}		
	\end{center}
	\caption { \emph{Left:} The power spectrum $\Delta_a^2$, as a function of comoving momentum $k_{\rm com}\equiv k (R/R_\star)$, during the evolution of the axion waves (produced during the scaling regime) when the axion mass becomes cosmologically relevant (blue), and at the final simulation time (black), after the the nonlinear transient and once it has reached an approximately constant form. The simulation starts at $H=H_{\rm crit}=H_\star/3$ with waves with the energy density spectrum $\partial\rho_a/\partial k$ predicted from the scaling regime in eq.~\eqref{drhodk} at $\xi_\star\log_\star=2000$ and $x_{0,a}=10$. The green line is the result at the final time for a purely linear evolution. 
	\emph{Right:} The result of $\Delta_a^2$ at the final time for different values of the IR cutoff of $F_a$, i.e. the parameter $x_{0,a}$, in the initial energy density spectrum.
		\label{fig:denpow}} 
\end{figure}

\begin{figure}[t]
	\begin{center}
		\qquad	 	\includegraphics[width=0.55\textwidth]{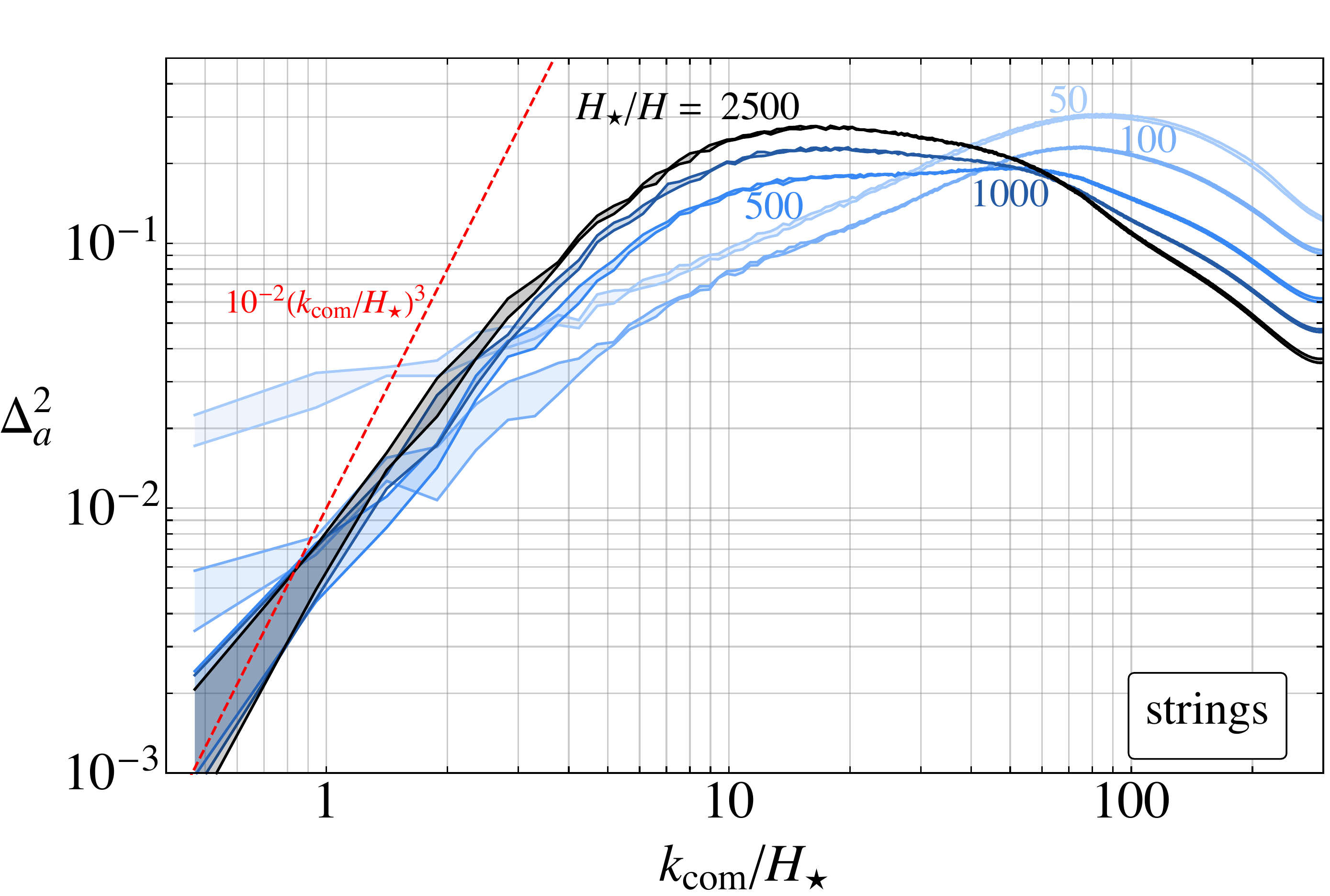}		
	\end{center}
	\caption {The evolution of the power spectrum $\Delta_a^2$ during the destruction of the string network at the unphysical value of $\log(m_r/H_\star)=5$. See Figure~\ref{fig:destroy} for the corresponding evolution of $\xi$ and the axion energy density spectrum at these times. 
		\label{fig:denst}} 
\end{figure} 

In Figure~\ref{fig:denpow} left we show the results for $\Delta_a^2(k)$ at increasing times during the evolution of the system at $H<H_{\rm crit}$ (as the mass becomes relevant), for the input spectrum in eq.~\eqref{drhodk} with $x_{0,a}=10$. As expected, $\Delta_a^2(k)$ changes, in particular growing at scales $k_{\rm com}/H_\star \simeq 10$. Thanks to the relatively large $N$ used, the transient has finished and the axion energy in the simulation is redshifting non-relativistically by the end of the simulation.\footnote{As discussed, in reality soon after $H_\star$ a large fraction of the axion energy is in UV modes that are not captured by simulations. However, these continue to redshift until the energy they contain is negligible so the  true power spectrum will eventually reach the form measured in simulations.} As it is clear in Figure~\ref{fig:denpow}, at this time the IR part of $\Delta_a^2(k)$ has reached a time-independent form. The UV part of $\Delta_a^2(k)$ is not fully constant, due to the presence of high momentum modes and oscillons (these contain only a small fraction of the total energy and will eventually decay into high momentum modes so will not alter the IR of the density power spectrum), and as mentioned in Section~\ref{ss:iso} its understanding is beyond the scope of this paper.

The IR part of $\Delta_a^2(k)$ approaches the expected $k^3$ dependence, with coefficient $C\simeq 2\times 10^5$ (defined in Section~\ref{ss:iso}). 
Note that even with a linear evolution (i.e. with potential $V=1/2 m_a^2 a^2$), $\Delta_a^2(k)$ would evolve simply due to modes turning non-relativistic. However the results we obtain differ from those with a linear potential due to the, previously discussed, relativistic redshift and the small non-linear transient. To understand the importance of these two effects, in Figure~\ref{fig:denpow}  we also plot the $\Delta_a^2$  that is obtained at the final simulation time evolving with a linear potential. From this is can be seen that the non-linear effects decrease $\Delta_a^2(k)$ by a factor of $4$ in the IR. 
This is reasonable since the transient moves energy to higher momentum modes, and it is also not surprising that the effect is relatively small given that the transient only has a minor effect on the axion number density. Although we have fixed a potential of the form $V(a)=m_a^2 f_a^2 (1-\cos(a/f_a))$ we expect any other bounded potential to lead to a similar $\Delta_a^2(k)$ (since the main effect of the non-linear potential on $\Delta_a^2(k)$ comes from the extra era of relativistic redshifting).

Let us discuss the possible uncertainties in $C$. One uncertainty comes from the value of the IR cutoff $x_{0,a}$ of $F_a$. As mentioned, at $\log\simeq7\div8$ the value of $x_{0,a}=10$ (as we have used) fits well the simulation results\cite{moreaxions}, with no evidence for a strong log dependence. Nevertheless, we cannot exclude that $x_{0,a}$ has a log dependence that is smaller than would be visible with current simulations (for instance the growth of $\xi$ might increase $x_{0,a}$). In Figure~\ref{fig:denpow} (right) we plot $\Delta_a^2(k)$ with $x_{0,a}$ between $5$ and $50$, from which it can be seen that $x_{0,a}>10$ decreases $\Delta_a^2(k)$ in the IR leading to a smaller $C$ and a weaker limit via eq.~\eqref{eq:cmbiso}. Decreasing $x_{0,a}$ below $10$ actually barely affected the fitted $C$, since the non-linear transient has a greater effect in this case, removing energy from IR modes. For $x_{0,a}$ between $5$ and $30$, which we take as a plausible range, $C$ varies by a factor of $20$. For a fixed $f_{\rm iso}$, the bound on $f_a \sim C^{1/4}$, so this leads to a $100\%$ uncertainty on the limit on $f_a$. In Figure~\ref{fig:param}, we correspondingly blur the limit above a lower edge corresponding to the constraint for $x_{0,a}=10$ to reflect this uncertainty. Together these uncertainties mean that the isocurvature bounds that we plot should be treated with substantial caution as discussed in the main text.

Finally, as discussed in Section~\ref{ss:iso}, our analysis of density perturbations in DM axions from the scaling regime gives a conservative isocurvature bound, since it misses the DM axions produced by the network of strings and domain walls formed when the mass becomes cosmologically relevant. To get an idea of how much these could strengthen the bound, we  also calculate $\Delta_a^2$ from simulations of the string network through the mass turn on until its destruction at a small (unphysical) value of $\log_\star$, by numerically solving eq.~\eqref{eq:LPhi} with an additional mass term, as in Appendix~\ref{app:destroy}. Of course, such simulations are at small scale separations (i.e. small tension) and have no hope of accurately reproducing the dynamics of the system at the physical point, and our results are solely to give an indication of the possible magnitudes of effects. The various competing requirements in such simulations discussed in Appendix~F of\cite{moreaxions} dramatically limit the value of $\log_\star$, and the results we show are for $\log_\star = 5$.\footnote{These requirements include, in particular, the lattice spacing $\lesssim m_r^{-1}$, a large enough hierarchy between axion and radial model mass, and that $HL$ is sufficiently large that the $k^3$ IR part of the density power spectrum can be fit once $\Delta_a^2$ has reached a constant form in the IR part.}

The results for $\Delta_a^2$ are shown in Figure~\ref{fig:denst} at increasing times until the network is destroyed (see Figure~\ref{fig:destroy} for the behaviour of $\xi$ and the axion energy density spectrum at the corresponding times). Given the more challenging simulations, the minimum values of $k_{\rm com}/H_\star$ (where $k_{\rm com}=k (R/R_\star)=k (H_\star/H)^{1/2}$ is the comoving momentum) are larger than in simulations in which only the axion field is evolved. Nevertheless,
the expected $k^3$ IR power law is reproduced  and in this momentum region $\Delta_a^2$ is time independent after the network disappears. The coefficient  $C \simeq 5\times 10^{-2}$ is much larger than in the axion only simulations (this is perhaps not surprising, since at $H=H_{\rm crit}$ most of the string length is in long strings, which are expected to lead to fluctuations on Hubble scales).\footnote{The UV part of $\Delta_a^2$ is still evolving at the final simulation time, but the overall energy density is mostly in IR modes so the IR part of the density power spectrum will not change dramatically.}  Similar results have been observed in analogous simulations where the axion mass has nontrivial temperature-dependence. This has been studied for the first time in \cite{Vaquero:2018tib}, where a detailed analysis of the power spectrum (at all momenta) has been carried out at small $\log_\star$ for the QCD axion.\footnote{In particular, the power spectrum is also characterized by a peak related to the presence of oscillons.}

It is plausible that the slope of the $k^3$ tail may be similar at large values of $\log_\star$, and if the relic abundance of axions from the destruction of the network is comparable to (or larger than) that from scaling, the isocurvature constraint would be significantly strengthened. For instance, assuming equal DM abundance from the waves produced during the scaling regime and from the destruction of the network, and $C \simeq 5\times 10^{-2}$, using eq.~\eqref{eq:cmbiso} for a fixed $m_a$ the bound on $f_a$ Figure~\ref{fig:param} would strengthen by a factor of $7$, ruling out large parts of the ranges of $f_a$ and $m_a$ that could be detected at SKA. As mentioned in the main text, even if the true $C$ from the destruction of the network at large tension is small, if the DM abundance is larger than that produced during scaling the isocurvature limit can also strengthen, owing to the DM abundance factor in eq.~\eqref{eq:cmbiso}.

\section{GWs from the Nonlinear Transient and Oscillons}\label{app:massGW}

As mentioned, we  refrain from attempting to calculate the contribution to the GW background from the collapse of the system of strings and domain walls at $H\simeq m_a$, since the dynamics of this system is yet not fully understood. One contribution to the GW spectrum from this collapse is expected to lie at frequencies and amplitudes of the same order as the last $e$-folding of the scaling regime, as already pointed out in~\cite{Hiramatsu:2012sc} (where numerical simulations at small scale separations have been carried out). As described in Section~\ref{ss:presentday}, such a contribution has a too low frequency for ultralight axions (for the masses that are not excluded by DM overproduction) and a too small amplitude for the QCD axion, for which $f_a\lesssim10^{10}$ GeV (but in principle at frequencies that are under investigation) to be observed.

As we will now explain, there could be an additional source of observable GWs in this system. As discussed in Section~\ref{ss:relic} and in more detail in~\cite{moreaxions}, the axion waves (accumulated during the scaling regime) experience a period of relativistic redshift after at $H\simeq m_a$ and a small nonlinear transient. During the nonlinear transient, the field is a superposition of waves containing (topologically trivial) domain walls that decay rapidly into axions. After the transient, the axion field is mostly in the linear regime (settling down to $a=0$), except in small regions called oscillons where it oscillates with an amplitude of order $f_a$. If the evolution of the axion waves were purely linear, the axion waves would not produce GWs~\cite{Dufaux:2007pt}. However, the existence of a small nonlinear regime provides a possible source of GWs. Unfortunately, this contribution is again in amplitude and frequency of the same order as the last $e$-folding of the scaling regime, and therefore not observationally relevant both for ultralight axions and for the QCD axion.

\begin{figure}[t]
	\begin{center}
		\includegraphics[width=0.7\textwidth]{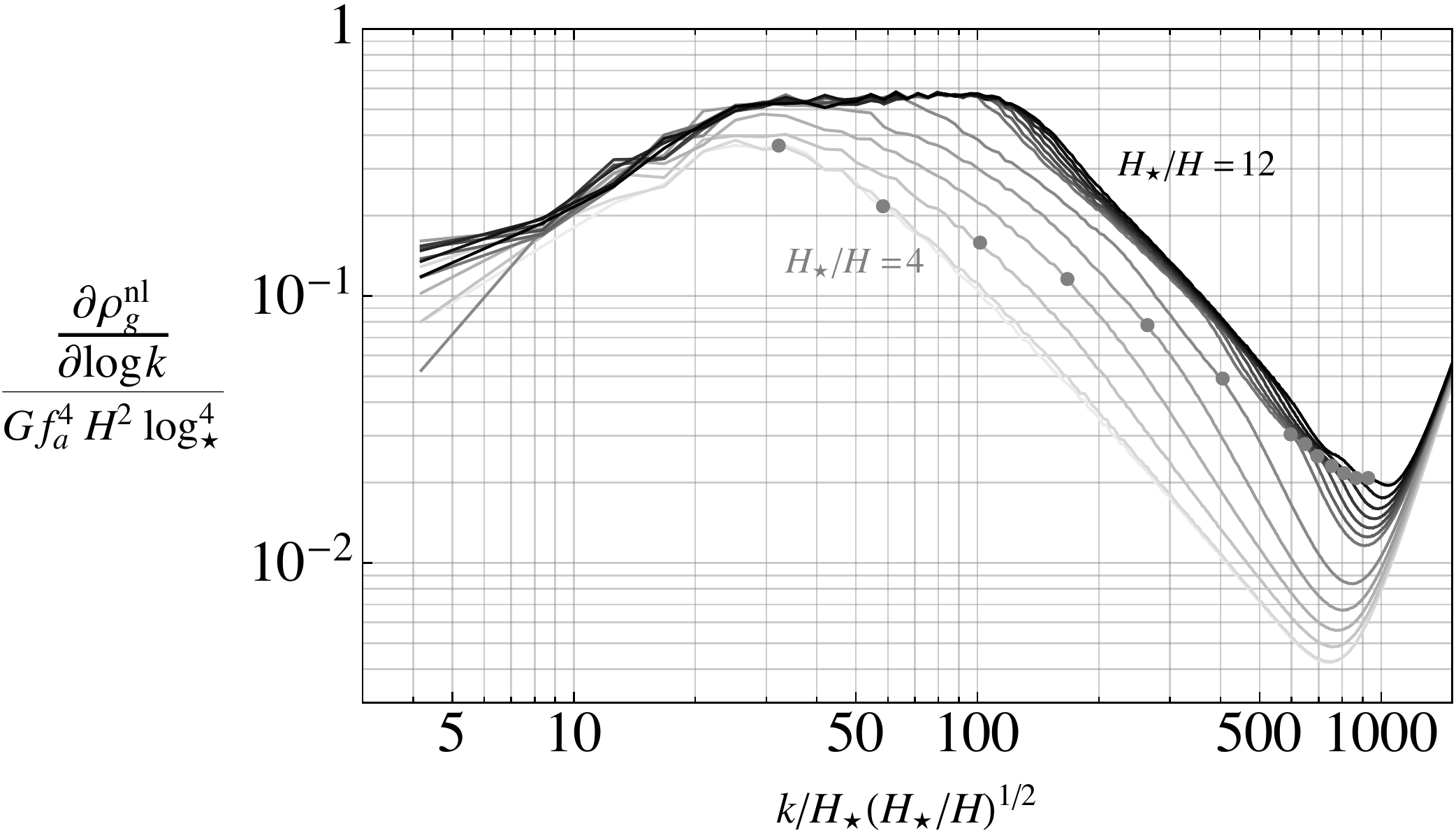}
	\end{center}
	\caption{The evolution of the GW spectrum generated during the nonlinear transient that the axion waves (emitted during the scaling regime) experience at the time $H\lesssim H_\star$, as a function of the (comoving) momentum. Increasing times are labelled by different values of $H_\star/H$. We show with a gray point the momentum corresponding to the axion mass. The overall redshift of the waves has been factored out in the plot by dividing the spectrum by $H^2$.   \label{fig:GWnonlinear}} 
\end{figure}

This conclusion can be easily drawn by estimating the parametric dependence on $f_a$ and $H_\star$ of this contribution via quadrupole formula (valid in the nonrelativistic limit) applied to the topologically trivial domain walls. In any case, in Figure~\ref{fig:GWnonlinear} we show the full spectrum of GWs from the numerical evolution of the axion waves during the nonlinear regime, discussed in Appendix~E of~\cite{moreaxions}. We start with a configuration of waves with the energy density spectrum from the scaling regime with for $\log_\star=65$, and we assume $m_a= R^{\alpha/2}$ with $\alpha=8$, which is the case for the QCD axion (see Section 3.2 in~\cite{moreaxions} for more details on the details of the evolution of these waves). It is immediate to see that the spectrum in Figure~\ref{fig:GWnonlinear} is peaked at momenta a few times larger than $k\simeq x_0 H_\star$ and is of the same order as the spectrum in eq.~\eqref{totalspectrum} evaluated at $H=H_\star$ (and $k=x_0 H_\star$). Moreover, most of the GWs are produced around the time $H/H_\ell\simeq6$, when the the potential energy equals the kinetic energy and the system becomes completely nonlinear (see~\cite{moreaxions}). At the final times the field is in the linear regime except for the presence of oscillons. Being spherical, oscillons do not contribute significantly to the GWs. Indeed, the GWs stop being produced after the nonlinear regime ends (at around $H_\star/H\simeq10$), and are not produced during the subsequent times when oscillons are present.

Finally notice that the for a temperature-independent mass, the nonlinear regime is much milder, and the contribution to the GW spectrum is smaller (and, as mentioned, outside the detectable frequency range). Given the experimental irrelevance of the GWs from the nonlinear transient, we refrain from a more detailed analytical study.

\section{Other bounds} \label{app:otherbounds}

{\bf \emph{Black hole superradiance}} Weakly interacting light particles can spontaneously draw energy out of black holes through the phenomenon of superradiance. The observation of spinning black hole that have not had their angular momentum removed by superradiance therefore constrains the axion parameter space~\cite{Arvanitaki:2010sy,Arvanitaki:2014wva,Stott:2018opm}. Currently, axions in the mass range $10^{-16}\div 10^{-18}~\eV$ are in tension with observations, however sufficiently large self-interactions prevent superradiance so the constraints only apply to $f_a \gtrsim 10^{15}~\GeV$, see \cite{Baryakhtar:2020gao} for a detailed analysis. In the post-inflationary scenario that we consider these limits are less important than that from the axion relic abundance.

{\bf \emph{Other CMB bounds}} Dark matter axions with masses $\sim 10^{-25}~\eV$ are constrained by their effect on CMB observables independently of the presence of strings \cite{Hlozek:2017zzf} (we note that the sensitivity of these observations could increase substantially in the future \cite{Bauer:2020zsj}). In the post-inflationary scenario these bounds are subdominant to the isocurvature constraints even with our most conservative assumption for the power spectrum of density perturbations. Additionally, a new approach to detecting strings
 that exist beyond the time of decoupling and are associated to an axion that interacts with photons  has recently been proposed \cite{Agrawal:2019lkr}. This is probably not relevant for strings that give the observable GW signals, since such long lived strings are likely to be in conflict with CMB anisotropy constraints for $f_a \gtrsim 10^{14}~\GeV$.

\section{Comparison to the Literature}\label{app:compare}
In this Appendix we first comment on the difference between our approach and previous works on GWs from global strings from the scaling regime.\footnote{There has also been some work on GW signals from axions in the pre-inflationary scenario \cite{Machado:2018nqk,Machado:2019xuc}, which can arise if there the an axion is coupled to a light hidden sector gauge boson.}
\begin{itemize}
	\item Refs.~\cite{Chang:2019mza} and~\cite{Gouttenoire:2019kij,Gouttenoire:2019rtn} utilise a particular model of the string evolution (also known as one-scale velocity-dependent model) and the expressions for the rate of energy emission to GWs and axions (in the zero coupling limit, derived in~\cite{Vilenkin:1986ku,Vachaspati:1984gt}) to calculate the GW spectrum from the loops produced during the scaling regime. Such references correctly reproduce the logarithmic deviation of the GW spectrum due to the logarithmic time-dependence of the tension. In particular, the resulting $\log^3$ dependence on the momentum (and the corresponding enhancement of the spectrum) has been already pointed out in~\cite{Gouttenoire:2019kij}. However, as such a model does not seem to always reproduce the logarithmic increase in $\xi$, the corresponding increase in the GW spectrum has not been captured. 

	\item  Refs.~\cite{Figueroa:2012kw,Figueroa:2020lvo} extract the GW spectrum directly from numerical simulations of physical systems similar to that in eq.~\eqref{eq:LPhi} at small log, without any extrapolation. These references claim that the GW spectrum asymptotes to an \emph{exactly} scale invariant form. However, as is clear from Section~\ref{sec:GWs}, this is in contradiction with conservation of energy and effective field theory, and also with our simulation results. Indeed, the spectrum results in~\cite{Figueroa:2020lvo} appear to show a residual time increase (and not an exactly scale invariant form).

\end{itemize}

We also observe that the original analysis of isocurvature perturbation of Section~\ref{ss:iso} has been developed for the QCD axion \cite{Enander:2017ogx} and axion-like particles in \cite{Feix:2019lpo,Feix:2020txt}. However, compared to these works we differ in our expression for the relic abundance, in the power spectrum that we use (which we obtain from simulations of the string network rather than motivated by misalignment production) and in allowing the axion to comprise a subdominant fraction of the dark matter. We also apply constraints on isocurvature from Lyman-$\alpha$ observations. These were derived in the context of primordial black hole dark matter models and extended to axion string scenario in \cite{Irsic:2019iff} (our analysis differs from this work again in our calculation of the relic abundance and in the density power spectrum that we use).

\section{Cosmological Stability of the Axions and Temperature Dependent Masses} \label{app:cosmostable}

In this Appendix we show that all generic axions that lead to observable GW signals  and are not ruled out by the constraints of Section~\ref{sec:Constraints} are cosmologically stable, and that also a temperature-dependent mass forces the axions to be ultralight.

As discussed in Section~\ref{ss:presentday} and visible in Figure~\ref{fig:OmegaGW}, for a temperature-independent axion mass GWs in the observable frequency range are only possible for $f_a\gtrsim10^{14}$ GeV and $m_a \lesssim 100 \keV$ so that the string network is not destroyed before $T_\star\simeq 10^7\GeV$. In this case the axion is always stable on cosmological timescales regardless of its interactions. For example, the axion might have an interaction with photons of the form~\cite{Arias:2012az}
\beq \label{eq:twophoton}
\mathcal{L} \supset - C \frac{\alpha_{\rm EM}}{8 \pi f_a} a F_{\mu\nu} \tilde{F}^{\mu\nu} ~,
\eeq
which allows decays, where $F_{\mu\nu}$ is the electromagnetic field strength with associated coupling constant $\alpha_{\rm EM}$, and the coefficient $C$ is model-dependent and expected to be not much larger than order one for the theory to be perturbative. However, denoting the temperature of the universe when $H =  m_a$ by  $T_{\star}$ the corresponding lifetime
\begin{align}
	\Gamma_{a\rightarrow \gamma \gamma}^{-1} 
	&=  \frac{1}{C^2} \left(\frac{\MeV}{m_a} \right)^3 \left(\frac{f_a}{10^{14}~\GeV}\right)^2  10^{21} ~{\rm s}  \label{eq:lifetime} \\
	&=  \frac{1}{C^2} \left(\frac{6\times 10^7~\GeV}{T_{\star}} \right)^6 \left(\frac{f_a}{10^{14}~\GeV}\right)^2  10^{21} ~{\rm s}~,
\end{align}
exceeds the age of the Universe for all the decay constants of interest and for axion masses $m_a\lesssim \MeV$, and in particular $m_a\lesssim 100$ keV. If the axion is sufficiently heavy and has suitable interactions it could also decay to leptons or hadrons, or hidden sector particles. However, these channels (or similar decays into hidden sector states) are not expected to significantly shorten the axion lifetime compared to that corresponding to the only photon coupling and do not change the conclusion.

As mentioned in Section~\ref{sec:Constraints}, a temperature-dependent mass does not relax the constraints in Figure~\ref{ss:presentday} on $f_a$ and $m_a(0)$ that lead to observable GWs, and still force the axion to be ultralight. To see this, we assume for simplicity that the axion mass dependence on temperature is $m_a(T)\simeq \Lambda^2/f_a\equiv m_a(0)$ for $T<\Lambda$ and $m_a(T)\leq m_a(0)$ for $T>\Lambda$, where $\Lambda$ is the strong coupling scale of a new sector, which happens in typical models (for $T\gg \Lambda$ the dependence is a power law but not relevant for our present argument).

As mentioned above, the conditions for the GWs to be observable are $f_a\gtrsim 10^{14}$ GeV and $T_\star\lesssim 10^{7}$~GeV. If $T_\star <\Lambda$, then as far as the cosmological evolution of the string network and axions are concerned the axion mass is constant and the bounds on dark matter, dark radiation and isocurvature perturbations are those discussed in Section~\ref{sec:Constraints}. On the other hand, $T_\star>\Lambda$ means that $\Lambda\lesssim 10^7$~GeV, which for $f_a\gtrsim10^{14}$ GeV requires $m_a(0)\lesssim$ GeV. If $ \MeV \lesssim m_a(0)\lesssim$ GeV, such values of the axion mass and decay constant are actually ruled out as the axions decay after BBN (from eq.~\eqref{eq:lifetime}) and dominate the energy density of the Universe at the time of BBN (which can be easily seen redshifting back today's would be DM abundance from eq.~\eqref{eq:relic} to $T=$ MeV). If instead $m_a(0)\lesssim \MeV$ the axion is stable and only ultralight axions do not overproduce DM ($T_\star$ in this case is always smaller than the corresponding $T_\star$ if $m_a$ did not depend on the temperature, implying in general a larger DM abundance despite the nonlinear evolution leading to a larger suppression in this case).

We finally note that in such models there are new constraints on the effective number of degrees of freedom in the hidden sector, since $\Lambda$ is far below the scale of BBN for the viable masses. These require the hidden sector is cold relative to the visible sector.

\section{Symmetry Restoration}

In this Appendix we give more details on the ways in which a PQ symmetry with large $f_a$ can be restored in the early Universe described in Section~\ref{sec:restore}.

\subsection{Symmetry restoration during inflation} \label{app:hybrid}

First we further analyse the scenario in which a coupling between the inflaton $\varphi$ and (the radial mode part of) $\phi$ leads to symmetry restoration. In particular, we  discuss the effect that a coupling of the form eq.~\eqref{eq:interaction_potential} has on inflation. 
Such a coupling gives no contribution to $\partial V/\partial \varphi$ (where $V$ is the full potential of the theory) as long as   $\left< \phi \right>=0$. Therefore it has no effect on the inflaton's slow-roll evolution provided that the radial mode's potential energy at this point is small compared to that of the inflaton, i.e. $H_I^2 M_{\rm P}^2 \gg f_a^4$ (which, e.g. for an inflaton with a quadratic potential requires  $4m_\varphi^2 \left<\varphi^2 \right> \gg f_a^4$). This is satisfied in the example theory described in Section~\ref{sec:restore}, which has $m_\varphi \simeq H_\textrm{max}=6\times 10^{13} ~\GeV$, $\left<\varphi\right>\sim M_{\rm P}$, and $f_a\lesssim 10^{15}~\GeV$. Otherwise, if $H_I^2 M_{\rm P}^2 \sim f_a^4$ the potential of the radial mode actually makes slow-roll easier to achieve, which is  the hybrid scenario discussed in Section~\ref{ss:hybrid} and below.

Once the PQ symmetry is broken (which if a string network is to form must happen close to the end of inflation or during reheating), there is a contribution to the inflaton's mass of $\delta m_\varphi^2 \simeq g f_a^2$. For large enough $g$ and $f_a$ this exceeds the inflaton's bare mass, changing the dynamics in a way that we have not analysed. However, for $f_a \sim 10^{15}~\GeV$ and values of $g\in (10^{-6},10^{-2})$ with $\left<\varphi\right>\sim M_{\rm P}$ the PQ symmetry is restored during inflation without this contribution being relevant, assuming that a not too small fraction of the inflaton's potential energy at this time comes from its mass term (i.e. $m_{\varphi}^2 \left<\varphi^2\right> \simeq H_I^2 M_{\rm P}^2$).

Additionally, interactions between the axion sector and the inflation typically lead to radiative corrections in the absence of extra symmetries (e.g. softly broken supersymmetry) or further new physics. Depending on their size, these could require that the inflaton's potential is fine tuned so that the slow roll conditions for inflaton are satisfied. The most dangerous radiative correction is the expected quadratically divergent correction to the mass of the inflaton that is cutoff by a UV scale $\Lambda_{\rm UV}$, which is expected to be of the form $\delta m_\varphi^2= g\Lambda_{\rm UV}^2/(32\pi^2)$ \cite{loop_correction}. For this not to violate the slow roll condition requires  $g/(48\pi^2) \Lambda^2 \left<\varphi\right> M_\textrm{P}<H_I^2 M_\textrm{P}^2$, which for $\Lambda \sim M_{\rm P}$ requires $\sqrt{g \left<\varphi\right> M_\textrm{P}/(48\pi^2)}<H_I$. However, for the interaction to restore the PQ symmetry (with $\left<\varphi\right>\lesssim M_\textrm{P}$) we need $m_r^2<g \left<\varphi^2\right><g\left<\varphi\right> M_\textrm{P}$. Combining these bounds with the observational limit on $H_I$ eq.~\eqref{eq:MaxHI}, the condition that the radiative corrections do not disrupt slow roll  is
\beq 
\frac{m_r}{\sqrt{48 \pi^2}}<H_I< 6\times 10^{13}~\GeV ~ .
\label{radiative corrections}
\eeq
These inequalities can be simultaneously satisfied for values $m_r<1.3\times 10^{15} ~\GeV$ provided $H_I$ is close to its maximum allowed value. For other $m_r$, $g$, and $H_I$ either some tuning of the inflaton's potential is required, or the UV cutoff $\Lambda$ must be below the Planck scale.

\subsubsection*{Hybrid Inflation}

Here we give more details about the hybrid inflation scenario. In particular, we show that hybrid inflation that is compatible with observations of the CMB occurs with a coupling between the radial mode and the inflaton of the form eq.~\eqref{eq:interaction_potential} for a wide range of $g$, when $f_a$ is large enough for observable GW signals.

As an example, we consider a theory with a potential of the form
\beq \label{eq:hybridV}
V=V_\varphi+V_\phi+V_\textrm{int} ~,
\eeq 
where $V_\phi$ is the axion sector potential (which by itself would spontaneously break the PQ symmetry), $V_\varphi$ is the inflaton's potential, and $V_\textrm{int}$ is an interaction between the two sectors.  We take $V_\phi$ and $V_\textrm{int}$ to be given by eqs.~\eqref{eq:LPhi}, \eqref{eq:interaction_potential} respectively, and for the following we assume $V_\varphi$  to be quadratic, although this is not essential (and we no longer fix $\left<\varphi\right>\sim M_\textrm{P}$ and $m_\varphi\sim H_I$). We also assume $f_a\approx m_r$. In combination, the model we consider is a minor modification of the original hybrid inflation theory \cite{hybrid_Linde,hybrid_initial}, with the change that $\phi$ is a complex scalar with a U(1) global symmetry rather than either a real scalar or a complex scalar with a gauge symmetry. 

As described in the main text, $\phi$ is kept at $\left<\phi\right>=0$ during inflation by its interaction with the inflaton, and its potential energy at this point exceeds that of the inflaton. This scenario occurs provided
\beq 
g\left<\varphi^2\right> \gtrsim m_r^2 ~,
\eeq
for symmetry restoration, and
\beq 
m_\varphi^2\left<\varphi^2\right>\ll \frac{f_a^4}{4} ~,
\eeq
for the potential energy of $\phi$ to dominate.

Because of $V_\phi$'s large contribution to the total energy density,  the slow roll condition for inflation is $V^\prime M_\textrm{Pl}/V_\phi\ll 1$, where $V^\prime=dV/d\varphi$, whereas in the absence of $\phi$ it would be $V^\prime M_\textrm{P}/V_\varphi\ll 1$. Thus in the hybrid scenario, inflation can occur with significantly lower values of the inflaton expectation value and mass, and consequently smaller $H_I$, than would otherwise be possible. For the particular realisation of hybrid inflation that we consider, by substituting for the potentials in eq.~\eqref{eq:hybridV}, the slow-roll condition is
\beq 
8 m_\varphi^2\left< \varphi \right> M_\textrm{P}\ll f_a^4 ~.
\label{eq:hybrid_condition}
\eeq

As long as the PQ symmetry remains restored the slow-roll condition above continues to be satisfied and the inflaton expectation value slowly changes (evolving towards $0$ with the potential we consider). Eventually the inflaton expectation value falls below $\varphi_c=m_r/g^{1/2}$ and the PQ symmetry is broken. The potential energy of the Universe then decreases faster (corresponding to the slow roll parameter $V^\prime/(V(\phi)+V_\textrm{int}) M_\textrm{Pl}$ increasing) and inflation will subsequently end at $\left<\varphi\right><\varphi_c$. 
In fact, following \cite{hybrid_Linde,false_vacuum} it can be shown that inflation ends within one $e$-fold of symmetry breaking (and thus the axion string network that forms is not diluted by further inflation) provided
\beq 
f_a^3\ll m_\varphi M_\textrm{P}^2 ~,
\label{eq:waterfall}
\eeq
which is satisfied if the mass of the inflaton is not too small $m_\varphi\gg f_a^3 / M_\textrm{P}^2 \simeq 10^{9} \GeV$ for $f_a \simeq 10^{15}\GeV$.\footnote{In more detail,  eq.~\eqref{eq:waterfall} can be derived by considering the change in the system in the first $e$-fold that follows PQ symmetry breaking. During this time the inflaton's expectation value shifts by 
	$ \Delta \varphi =\Dot{\left<\varphi\right>}/H=8M_\textrm{P}^2 m_\varphi^2 \varphi_c /f_a^4$. Meanwhile, the complex scalar  $\phi$ now has a symmetry breaking minimum at $|\phi|^2=(m_r^2-g\left<\varphi^2\right>)/2$, which is away from the origin but not yet at $f_a/\sqrt{2}$ since the inflaton expectation value is still non-zero. The position of this minimum continues to move away towards $f_a/\sqrt{2}$ as the inflaton rolls toward $0$. As a result, one $e$-fold after symmetry breaking the radial mode part of $\phi$ has an effective mass $m_\phi^2=m_r^2-g\left<\varphi^2\right>\approx 2g \varphi_c \Delta \varphi$. If (\ref{eq:waterfall}) holds, then $m_\phi\gg H$ and the radial mode tracks the minimum of its potential.  The resulting change in the energy density of the Universe is $ \Delta V=32M_\textrm{P}^4 m_\varphi^4/f_a^4+\mathcal{O}(\Delta \varphi^3)$, so the slow roll parameter becomes
	\beq 
	\abs{\frac{V^{\prime\prime}M_\textrm{P}^2}{ V}}=\frac{4 M_\textrm{P}^2}{\varphi_c^2}>1 ~.
	\eeq
	Therefore, for  $\varphi_c\lesssim M_{\rm P}$,  slow roll inflation indeed ends less than one $e$-fold after symmetry breaking, provided eq.~\eqref{eq:waterfall} holds.}

The final constraint on the hybrid scenario comes from the COBE normalisation condition, which fixes the amplitude of density perturbations using observations of CMB anisotropies. This requires
\beq  \label{eq:conditionC}
\frac{V^{3/2}}{M_\textrm{P}^3 V^\prime}=5\times 10^{-4} ~,
\eeq
where the left hand side is evaluated when modes corresponding to the pivot scale $k_\textrm{CMB}=0.05 ~\Mpc^{-1} $ leave the horizon (see \cite{review_inflation} for a review). In the model we consider, eq.~\eqref{eq:conditionC} translates into a relation between $m_\varphi$ and $g$ \cite{hybrid_Linde} 
\beq  \label{eq:cobecond}
m_\varphi=9\sqrt{\frac{g^{1/2} f_a^5}{M_\textrm{P}^3}} ~.
\eeq
Imposing this condition means that (for $\left<\varphi\right>,f_a < M_{\rm P}$) the slow roll condition and eq.~\eqref{eq:waterfall} are automatically satisfied, so inflation will continue for as long as the PQ symmetry is restored and will end immediately after symmetry breaking. Moreover, for $f_a\lesssim M_{\rm P}$ and $g\lesssim 1$, if eq.~\eqref{eq:cobecond} is satisfied the potential of the radial mode dominates the total energy density while $\left<\phi\right>=0$, so our calculation of the slow roll condition in eq.~\eqref{eq:hybrid_condition} is self-consistent.

In summary, the are two relevant conditions that remain in our example model: one from requiring PQ symmetry restoration, and one from the COBE normalisation. For any inflaton expectation value that satisfies $f_a< \left<\varphi \right> \lesssim M_{\rm P}$ there is an allowed range of perturbative $g\in (f_a^2/\left<\varphi \right>^2,1)$ such that hybrid inflation occurs, with the corresponding inflaton mass fixed by eq.~\eqref{eq:cobecond}.

\subsection{Preheating} \label{app:preheat}

Here we give more details of how non-perturbative energy transfer from the inflaton to other states, i.e. preheating, can lead to far higher temperatures after inflation than occur during perturbative reheating. We show that the relatively large mass of the radial mode renders direct preheating to this inefficient (for $m_r\simeq f_a$ and $f_a \gtrsim 10^{14}\GeV$ relevant for observable GWs), but that preheating to another, light, particle can still restore the PQ symmetry.

We consider a general real scalar $\chi$ that interacts with the inflaton through $V_\textrm{int}=\frac{1}{2}g\varphi^2 \chi^2$, and we assume that the inflaton potential  is quadratic in the part of field space that reheating occurs in, $V_\varphi=\frac{1}{2}m_\varphi^2 \varphi^2$. Following the analysis in \cite{Linde_broad}, the momentum modes $\chi_k$ of $\chi$ evolve according to 
\beq \label{eq:modeev}
\Ddot{\chi}_k+3H\Dot{\chi}_k+\left(\frac{k^2}{R(t)^2}+m_\chi^2+g\Tilde{\varphi}^2\sin^2(m_\varphi t) \right)\chi_k=0 ~,
\eeq
where $\Tilde{\varphi}$ denotes the amplitude of the inflaton oscillations. Neglecting the expansion of the Universe, eq.~\eqref{eq:modeev} can be reexpressed as the Mathieu equation
\beq
\chi_k^{\prime\prime}+[A_k-2q\cos(2z)]\chi_k=0 ~,
\label{mathieu}
\eeq
where $q=g\Tilde{\varphi}^2/(4 m_\varphi^2)$; $A_k=((k/R)^2+m_\chi^2)/m_\varphi^2+2q$; $z=m_\varphi t$ and differentiation is with respect to $z$.

The family of solutions of the Mathieu equation has resonant bands at particular momentum $k$, which depend on $q, A_k$ \cite{McLachlan:104546}. The solutions with $k$ inside these bands grow exponentially, as $\chi_k\sim e^{\alpha z}$ where $\alpha$ has a real part   $\Re(\alpha)>0$. Meanwhile, the solutions for $k$ outside these bands oscillate (corresponding to $\Re(\alpha)=0$). Preheating happens when a mode $\chi_k$ has momentum inside one of the resonance bands. The resulting amplification corresponds to an exponentially fast increase in the mode's occupation number, i.e. an extremely fast transfer of energy from the inflaton to $\chi$.

However, a particular mode is only exponentially amplified for a limited time. One  reason for this is that the expansion of the Universe redshifts a mode's momentum, which results in it moving out of a resonance bands.\footnote{The exception to this is if the inflaton's potential has a pure quartic form.} It is shown in \cite{Linde_broad} that this results in the resonance bands being effectively `blurred': modes are mostly amplified inside a broad resonance band at low frequency $\omega\in (0,~\omega_\textrm{max})$,
where $\omega^2\equiv k^2/R^2+m_\chi^2$, and  $\omega_\textrm{max}=\sqrt{g^{1/2}\Tilde{\varphi}m_\varphi/2}$. Outside this broad resonance, the resonance bands are very narrow and modes are quickly moved out of them by the expansion of the Universe, resulting in little energy being transferred. Additionally, if the resonance is efficient enough  that a substantial amount of energy has been transferred into $\chi$, the amplitude of the inflaton oscillations will decrease significantly faster than just due to redshifting. The interaction between the inflaton and $\chi$ also causes a contribution to the mass of the inflaton $m_{\rm{eff}}^2 \simeq m_\varphi^2+g\left<\chi^2\right>$ which can end up dominating. Thus the backreaction of the created particles further modifies (in particular, it decreases) the values of $q, A_k$, which changes the structure of the resonance bands. Overall, these effects mean preheating generally does not transfer all of the inflaton's energy to $\chi$.  Indeed, it is shown in \cite{Linde_broad} that inflation lasts until half of the (redshifted) inflaton energy has been transferred to $\chi$ for $g>10^{-6}$, and this takes roughly $20$ inflaton oscillations, meaning the energy density in $\chi$ at this time is a factor $10^{-4}$ lower than the original energy density at the end of inflation. For lower values of $g$ the fraction of energy transferred decreases very quickly (roughly exponentially with $g^{1/2}$).

From the discussion above, we see that if the mass of $\chi$ is large enough then efficient preheating does not occur. This is simply because the energy of $\chi$ modes is $\omega_k \geq m_\chi$, so if $m_\chi > \omega_\textrm{max}$ no modes are in the broad band (and the remaining narrow resonances are highly inefficient). In the intermediate case  $0< m_\chi < \omega_\textrm{max}$ fewer modes satisfy the condition to be in broad resonance, $\omega_k < \omega_\textrm{max}$. This is expected to reduce the efficiency of preheating, although we do not investigate such a scenario in detail.

We now apply these results to analyse the possibility that the PQ symmetry is restored by preheating  directly to the radial mode of a complex scalar that gives rise to the axion, i.e. we identify $\chi$ with the radial part of $\phi$ of eq.~\eqref{eq:LPhi}, via the interaction  eq.~\eqref{eq:interaction_potential}. The condition for broad resonance $m_\chi \ll \omega_\textrm{max}$ requires $g\Tilde{\varphi}^2 \gg 4m_r^4/m_\varphi^2$. However, for the $m_r \gtrsim 10^{14}\GeV$ as is relevant for GWs $m_r^4/m_\varphi^2 > m_r^2$ (for the inflaton masses that are permitted by the slow roll constraints in typical theories \cite{Martin:2013nzq}).  Consequently, from eq.~\eqref{eq:restore_inflation} efficient preheating requires that the coupling $g$ is large enough that the symmetry is restored directly during inflation anyway, as mentioned in Section~\ref{ss:preh} (or, depending on the sign of the interaction, the complex scalar is displaced to large field values, which might also lead to strings although we do not study this scenario in detail).

We also note that our analysis is consistent with results from simulations of preheating carried out in \cite{khlebnikov_tkachev_1997}, which consider preheating to a real scalar field and include the case that this is massive compared to the inflaton. They consider a quadratic inflaton potential with starting inflaton amplitude $\Tilde{\varphi}\sim M_\textrm{P}$ and inflaton mass $m_\varphi\sim 10^{13}~\GeV$ of same order of magnitude to our case. These papers find that for $m_r\gtrsim 2m_\varphi$, which is the case relevant to the scenario that we are interested in with $m_r\gtrsim 10^{14} ~\GeV$, $q>10^5$ is required for fluctuations created during preheating to be large enough to restore the symmetry. This translates to requiring $g>10^{-4}$ and $\omega_{\rm max}\gtrsim 10^{15} ~\GeV\gg m_r$.

\subsubsection*{Intermediate preheating}

Alternatively, as mentioned, the PQ symmetry could be restored if a new, light, scalar is preheated and this then transfers energy to the sector that gives rise to the axion. As an example in which this happens we consider a theory where the energy transfer to the axion sector happens through an interaction of the form $g_{\chi\phi}\chi^2|\phi|^2$, where $\chi$ is a real scalar that is preheated as before. Since the intermediate field $\chi$ could be effectively massless and a large coupling $g$ does not lead to symmetry restoration during inflation, we will see that in this theory the PQ symmetry can be restored solely thanks to efficient preheating.

Calculating the effective temperatures $\chi$ and $\phi$ reach after this process is complex due to the non-perturbative and out of equilibrium nature of the dynamics. We therefore take a simplified approach in which we analyse the distribution of energy in $\chi$ after preheating and use this to approximate the scattering rate $\Gamma$ of $\chi+\chi\rightarrow \phi+\phi$. We will compare this to the Hubble parameter $H$ at the time when the energy density transferred into $\chi$ and the energy density remaining in the inflaton are equal.  Earlier than this, the energy density in $\chi$ will be significantly lower, leading to a lower effective temperature in the axion sector. However, if preheating lasts beyond this time then backreaction will play an important role in the evolution, which makes the dynamics more complicated and is expected to slow down energy transfer from the inflaton to $\chi$. Assuming preheating ends when half the energy is transferred is enough for an order of magnitude estimate of the maximum effective temperature achievable (indeed, even if the entire energy density of the inflaton is subsequently transferred into $\chi$, this will be at most a factor of $2$ higher than that at the moment of equality). The condition that preheating lasts until this point, and thus that a substantial fraction of the inflaton energy is transferred into $\chi$, is that the coupling $g\gtrsim 10^{-6}$ \cite{Linde_broad, khlebnikov_tkachev_1997}.

We denote the amplitude of the inflaton oscillations at the point where the energy density of the inflaton and of $\chi$ are equal by $\Tilde{\varphi}_\textrm{eq}$. To estimate the typical occupation numbers $n_k$ of $\chi_k$ modes that are inside the resonance band $k\in (0,\omega_\textrm{max})$ we equate the energy density in $\chi$ at this moment
\beq 
\rho_\chi\approx\frac{2\pi}{3} n_k \omega_\textrm{max}^4=\frac{\pi}{6} n_k g \Tilde{\varphi}_\textrm{eq}^2 m_\varphi^2 ~,
\eeq
to the energy density in the inflaton $\frac{1}{2}\Tilde{\varphi}_\textrm{eq}^2 m_\varphi^2$, so $n_k\simeq 3/(\pi g)$ inside the resonance band. Meanwhile, modes outside the resonance band $k>\omega_\textrm{max}$ have not been exponentially amplified, so their occupation numbers are negligible. 

To transfer energy to $\phi$ efficiently there must be $\chi$ modes that are energetic enough  for $\chi+\chi\rightarrow \phi+\phi$ to occur, which requires $\omega_\textrm{max}\gg m_r$. This leads to a condition $g\gg 10^{-5}$ for $m_r\simeq 10^{15}~\GeV$.\footnote{For the lowest value of interest $m_r\simeq 10^{14}~\GeV$ the condition would be $g\gg 10^{-9}$, i.e. weaker than the condition required for efficient preheating.} 
Given the momentum distribution described above, the scattering rate $\Gamma=n\left<\sigma v \right>$ can be calculated, where $\sigma$ is the interaction cross section
\beq 
\sigma=\frac{g_{\chi \phi}^2\sqrt{k^2-m_r^2}}{128\pi k^3} ~,
\label{cross section}
\eeq 
leading to
\beq 
\Gamma=\frac{3 g_{\chi \phi}^2 m_\varphi^{1/2} \Tilde{\varphi}_\textrm{eq}^{1/2}}{32\sqrt{2}\pi g^{3/4}} ~.
\eeq
If this scattering rate is larger than the Hubble at that time preheating ends $\Gamma\gtrsim H$ then $\chi$ and $\phi$ will reach equilibrium at an effective temperature $T\sim \sqrt{H M_\textrm{P}}$. 
As discussed above, preheating lasts for approximately $20$ inflaton oscillations until a significant fraction of the inflaton energy has been transferred \cite{Linde_broad}, so we can estimate the Hubble at the end of preheating $H\simeq 0.025 H_I\lesssim 1.5\times 10^{12}~\GeV~$, leading to an effective temperature $T\simeq 2 \times 10^{15} ~\GeV~$, which is high enough to restore the PQ symmetry for axion decay constants that lead to observable GWs. Meanwhile, if $\Gamma \ll H$ immediately after preheating the energy transferred to the axion sector through this process is not sufficient to restore the PQ symmetry, since the rate of energy transfer by $\chi+\chi\rightarrow \phi+\phi$ will decrease faster than the Hubble parameter drops.

In summary, the conditions for symmetry restoration in this scenario are that preheating is sufficiently efficient and $\chi$ energetic enough to allow scattering, which occur provided $g\gg 10^{-5}$, that the scattering rate is large enough  for thermalisation to be efficient, which requires $g_{\chi \phi}^2>10^{-2} g^{3/4}>10^{-6}$, and that there is sufficient energy at the end of inflation, and subsequently at the end of preheating, $H_I>3\times 10^{13}~\GeV$. For comparison, couplings $g, g_{\chi \phi}$ of such order would lead to a maximum temperature via perturbative reheating $T_{\rm max} \sim 10^{12}~\GeV$ from eq.~\eqref{eq:tmax} (and a far lower final reheating temperature).

\subsection{Symmetry restoration with a light radial mode} \label{app:lightR}

Here we give more details on the scenario where the  symmetry is restored at temperatures $T\ll f_a$ because the radial mode is light. In particular, we justify the parametric dependence for the minimum temperature that leads to symmetry restoration given in eq.~\eqref{eq:Tminmr} and we show that this expression is accurate taking into account the full finite temperature potential.

As in the main text we consider a potential 
\beq \label{eqappM1}
V\left(\phi \right) = \frac{m_r^2}{2 f_a^2} \left(|\phi|^2 - \frac{f_a^2}{2} \right)^2 ~.
\eeq 
Although it will turn out not to restore the PQ symmetry for $T\ll f_a$ it is useful to first consider the finite temperature contribution from $\phi$ to its own thermal potential. In the high temperature limit $T\gg m_r$ this is given by
\beq \label{eqappM2}
V_T \simeq \frac{1}{24} m_r^2\left(\phi\right) T^2 \simeq \frac{1}{16} \frac{m_r^2}{f_a^2}  |\phi|^2 T^2 ~,
\eeq
where $m_{r}\left(\phi\right)^2 \sim \lambda |\phi|^2$ is the mass of the radial mode on the background of its own expectation value (and in the second equality we have dropped a $\phi$ independent term). Comparing eqs.~\eqref{eqappM1} and \eqref{eqappM2}, we immediately see that eq.~\eqref{eqappM2} can only restore the symmetry for $T \gtrsim f_a$ (i.e. $T \gtrsim m_r$ is not sufficient).  It is straightforward to show that the conclusion is unchanged if the full thermal potential is used rather than eq.~\eqref{eqappM2}.

However, the complex scalar could also couple to new fermions. In a QCD axion model these might be the fermions that generate the QCD-PQ anomaly in KSVZ models, but more generally the new fermions need not be charged under the SM gauge group.\footnote{Indeed to avoid a too large axion mass for the parameter space that we are interested in, they must not lead to a QCD induced axion mass.} We consider an interaction of the form
\beq \label{eqappMP}
\mathcal{L} \supset g \phi  \psi^c \psi  + {\rm h.c.}~,
\eeq
where $\psi$ and $\psi^c$ are Weyl fermions that are massless in the absence of a $\phi$ expectation value (and h.c. denotes the Hermitian conjugate).\footnote{Unless $g$ is tiny radiative corrections induced by this term typically require that $\phi$'s mass is fine tuned. We do not worry about this issue, which could be avoided for example if the axion and new fermion sector is supersymmetric.} The dependence of the mass of $\psi$ on $\phi$'s expectation value leads to finite temperature contribution to $\phi$'s potential
\beq \label{eq:finiteTfermion}
V_T = \frac{-n_f T^4}{2\pi^2} \int_0^\infty q^2 \log\left(1+e^{-\sqrt{q^2+g^2|\phi|^2/T^2}} \right) ~ dq ~,
\eeq
where $n_f=4$ if there are a single pair of fermions. In the high temperature limit $T\gg m_\psi = g \left< \phi \right>$ eq.~\eqref{eq:finiteTfermion} is approximately
\beq \label{eqappM3}
V_T \simeq \frac{1}{24} g^2 \phi^2 T^2 ~.
\eeq
Consequently $\left< \phi \right>=0$ is a local minimum of the potential  for any temperature $T \gtrsim m_r/g$. However, we impose a stronger condition,  which is that the thermal potential ensures that the system reaches $\left< \phi \right>=0$ regardless of the initial conditions.\footnote{We could e.g. consider models of inflation such that $\left< \phi \right>=0$ initially, in which case $T \gtrsim m_r/g$ would keep the system at this point. However, in such a theory strings will form anyway, so the thermal potential is not required for this.} This is not automatic given eq.~\eqref{eqappM3}, because this is only valid for $T\gg g \left<\phi\right>$, which is not satisfied around $\left<\phi\right> \sim f_a$ if $T \lesssim g f_a$. Instead, for $T \ll g f_a$ the thermal potential of eq.~\ref{eq:finiteTfermion} is exponentially suppressed at $\left<\phi\right>\sim f_a$. Physically, this happens because $\psi$ decouples from the thermal bath when its mass is greater than the temperature. Therefore, there is a local minimum close to the zero temperature minimum for temperatures in this range.

\begin{figure}[t]
	\begin{center}
		\includegraphics[width=0.55\textwidth]{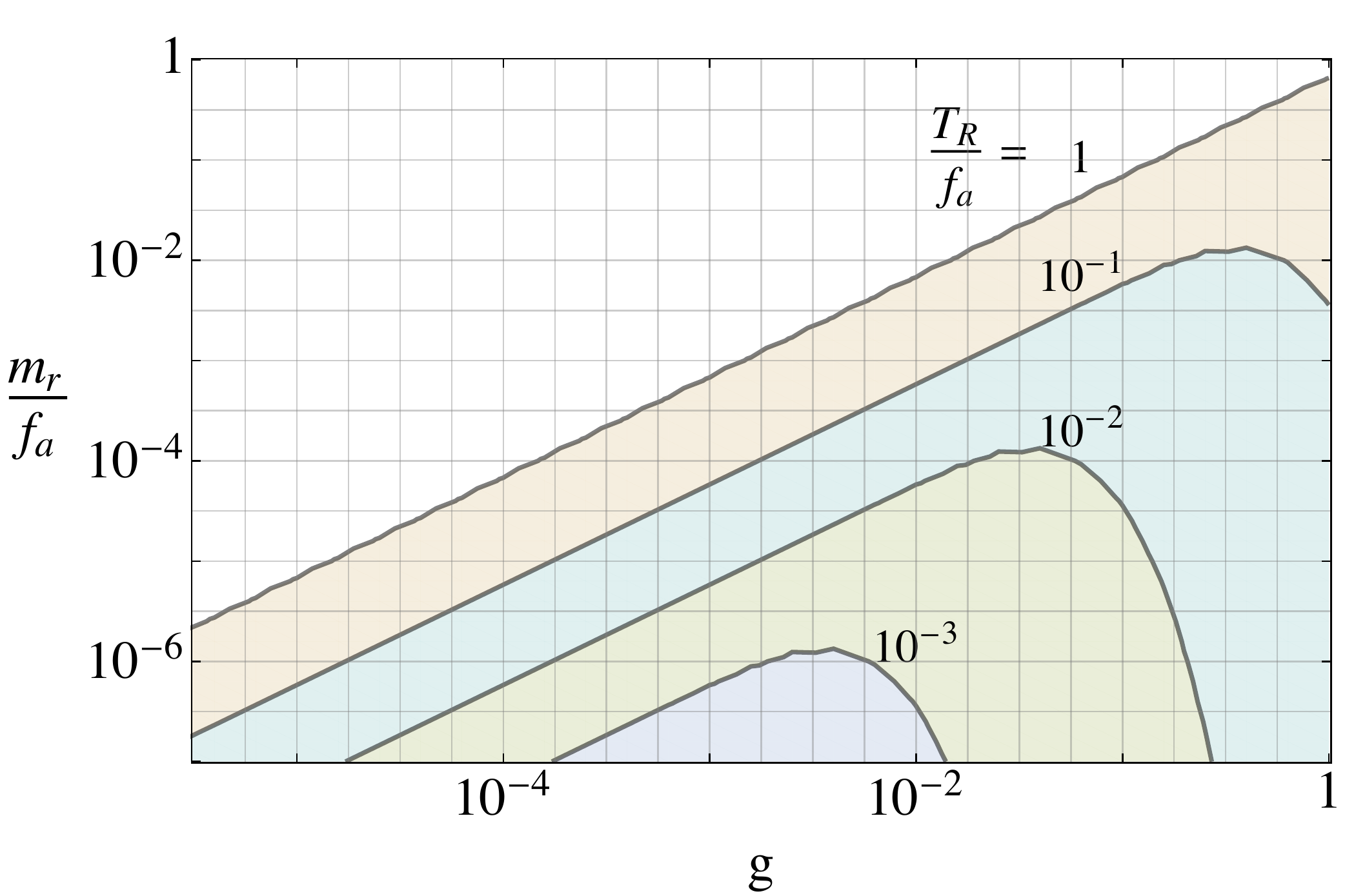}
	\end{center}
	\caption{The minimum reheating temperature required to restore the axion PQ symmetry (regardless of the system's initial conditions) in theories such that the radial mode of the complex scalar $\phi$ that gives rise to the axion has a mass $m_r$ that is significantly smaller than the axion decay constant $f_a$, and $\phi$ has an interaction with fermions with coupling constant $g$ as in eq.~\eqref{eqappMP}. \label{fig:gspace}} 
\end{figure}

Combining the preceding  conditions, the lowest temperature at which the PQ symmetry is restored regardless of the initial condition is parametrically given by eq.~\eqref{eq:Tminmr}. Precise results for the minimum temperature for a given model can easily be obtained by evaluating eq.~\eqref{eq:finiteTfermion} numerically. In Figure~\ref{fig:gspace} we plot the results for the simple model of eq.~\eqref{eq:LPhi} with a single pair of fermions $\psi$ $\psi^c$ as a function of $m_r/f_a$ and the coupling $g$. It can be seen that eq.~\eqref{eq:Tminmr} is quite accurate (although the condition $T> g f_a$ is slightly too strong since the fermions do not decouple from the thermal bath immediately when this condition is violated). If an axion arises from a more complex theory the minimum temperature required will change by order 1 factors, but the main parametric dependence will remain fixed.

We finish our analysis of these models with two additional comments. First, we note that the values of $g$ in our parameter space of interest are small but not tiny, so the fermions $\psi$ are easily in thermal equilibrium (since their typical interaction rate with the thermal bath is $g^2 T \gg H(T)$ where $H(T)$ is the Hubble parameter) and our analysis using the thermal potential is valid. Second, as mentioned in the main text, the GW signal emitted by such a network will be largely unaffected by the small $m_r$, i.e. it will approximately match the predictions of Section~\ref{sec:GWs}. This is because the GW energy depends on the string tension, which is set by $f_a$ not $m_r$, and the GW spectrum is IR dominated so it is unaffected by the UV cutoff at $m_r$ being much smaller than $f_a$. The only effect on the GW spectrum will be through the value of the log $\log(m_r/H)$ being slightly reduced.\footnote{Since the divergence in the energy of the string is cut of by the physical string core size $m_r^{-1}$ rather than $f_a^{-1}$.} This will feed into $\xi$ and the ratio $\Gamma_{\rm GW}/\Gamma_{a}$ as well as the tension. However, the change is not too dramatic as long as $m_r$ is not tiny. For example, taking $m_r=5\times 10^8~\GeV$ and $f_a=5\times 10^{14}~\GeV$ (so that, from Figure~\ref{fig:gspace}, the symmetry can be restored for temperatures $\sim 5\times 10^{11}~\GeV$), the value of the log when the GW emission is relevant to SKA is $\log\sim 60$, as opposed to $\log\sim 75$ if $m_r \sim f_a$. The amplitude of the resulting GW signal is reduced by roughly $50\%$ relative to that plotted in Figure~\ref{fig:OmegaGW}, but it remains detectable by SKA.

\bibliography{axiongw}
\bibliographystyle{JHEP}

\end{document}